\documentclass[journal]{IEEEtran}
\usepackage[utf8]{inputenc}
\usepackage[T1]{fontenc}
\usepackage[strings]{underscore}
\usepackage{graphicx}
\usepackage{booktabs}
\usepackage{amsmath,amssymb}
\usepackage{array}
\usepackage{tabularx}
\usepackage{multirow}
\usepackage{xcolor}
\usepackage{flushend}
\usepackage[font=footnotesize,labelfont=bf]{caption}
\usepackage{enumitem}
\setlist{nosep,leftmargin=1.3em}
\usepackage{listings}
\usepackage{textcomp}

\definecolor{codebg}{HTML}{F6F8FA}
\definecolor{navy}{HTML}{1E3A5F}
\definecolor{kw}{HTML}{1E40AF}
\definecolor{cm}{HTML}{6B7280}
\definecolor{st}{HTML}{B45309}

\lstdefinelanguage{SV}{
  morekeywords={module,endmodule,input,output,logic,wire,reg,always_ff,always_comb,
    posedge,negedge,if,else,begin,end,case,endcase,unique,typedef,enum,struct,packed,
    assign,localparam,parameter,for,int,bit,byte,signed,unsigned,function,endfunction,
    return,default,genvar,generate,endgenerate,always,initial,longint,shortint,break},
  sensitive=true, morecomment=[l]{//}, morecomment=[s]{/*}{*/}, morestring=[b]"}
\usepackage{hyperref}
\hypersetup{colorlinks=true,linkcolor=navy,citecolor=navy,urlcolor=navy}
\graphicspath{{figs/}}

\newcommand{\code}[1]{\texttt{\small #1}}

\begin{document}

\title{Transformer Accelerator (TFA): A Macro-Op INT8\\
Hardware Chip for Transformer Inference\\
and Machine Translation}

\author{Shashank, Independent Researcher, sshashan@alumni.usc.edu, TX%
}

\markboth{Shashank: Transformer Accelerator (TFA) Technical Report, v1}%
{Shashank: A Macro-Op INT8 Transformer-Inference Accelerator Chip}

\maketitle

\begin{abstract}
Transformer neural networks underlie most of modern machine learning, and
deploying them for inference, particularly at the edge, calls for compact,
low-precision hardware. Transformer inference combines a compute-intensive
encoding or prompt-processing pass with an autoregressive generation pass whose
cost is dominated by streaming weights from off-chip memory; a small INT8 engine
that streams weights past a datapath at the bandwidth limit serves both. We
present the \emph{Transformer Accelerator} (TFA), a synthesizable and
parameterizable IP that accelerates transformer inference as a memory-to-memory
INT8 engine. TFA is deliberately architecture-agnostic: it implements the small
set of tensor primitives common to every transformer block (matrix multiply,
softmax, RMSNorm, elementwise, and copy/gather) and exposes them as a macro-op
instruction set of eight 512-bit descriptors, each encoding a complete tensor
operation. The host compiles a model offline into a stored program of these
descriptors, and the engine fetches, exhaustively validates, and dispatches them
onto a single time-multiplexed datapath behind one AXI4 master, one AXI4-Lite
slave, and one interrupt; the same primitives realize self-attention,
cross-attention, and feed-forward blocks, and thus encoder, decoder, and
encoder-decoder models alike. We describe the register-transfer-level (RTL)
microarchitecture in detail: an output-stationary multiply--accumulate array
whose \emph{ping-pong} operand buffers overlap DMA fill with compute across tile
boundaries; serial integer reciprocal-square-root and divide units that keep
RMSNorm and softmax bit-exact; ring-buffer and gather addressing for the
key--value cache and embedding lookup; and an abort-safe write engine that
zero-pads every issued burst so the interconnect cannot deadlock. Because every
operation is specified to the bit, a Universal Verification Methodology
environment built around a bit-exact golden model byte-compares all device
outputs; across a 25-test suite and multi-seed constrained-random regression (34
runs, zero mismatches), it attains 100\,\% functional and 94.96\,\% code
coverage. As a real-world demonstration we target neural machine translation: we
compile a complete pretrained encoder-decoder transformer (t5-small, 60\,M
parameters) and run its full pipeline, the bidirectional encoder, the
cross-attention, and the autoregressive decoder, on the RTL. Selected by the
multi-task model's task prefix, the same compiled weights translate English
into French, German, and Romanian; on a ten-proverb multilingual stress
workload the chip executes 70{,}320 descriptors and the golden model
byte-compares every chip-written output region against its own replay (zero
mismatches over 37.9\,MB of compared data), and the INT8 result reproduces the
floating-point reference token-for-token on half of the sentences (five of
ten), the remainder differing only by an equally valid alternative rendering. We further show that naive per-tensor INT8 fails on
this network owing to large residual-channel outliers, and that an \emph{exact}
randomized-Hadamard reparameterization, a lossless transformation of the model
rather than of the hardware, recovers roughly 11\,dB of per-tensor INT8
signal-to-noise ratio at every layer. The verification-grade configuration
already exceeds a 22-thread CPU baseline by about $20\times$ end-to-end, and
larger configurations are projected to lower energy per token by roughly three
orders of magnitude. We additionally carry the design through a complete
open-source synthesis and place-and-route flow: after recoding the on-chip
memories so they infer as RAM, which removes a comparator-decode artifact and
halves standard-cell logic area to 2.73\,mm\textsuperscript{2}, the design
hardens to a design-rule-clean GDS-II layout on the SkyWater sky130 process,
confirming that the verified RTL is manufacturable and not merely simulatable.
These results indicate that a small macro-op datapath,
paired with a compiler that absorbs quantization difficulty, can run a real
pretrained transformer end-to-end to a bit-exact, fully verified result.
\end{abstract}


\section{Introduction}
\label{sec:intro}

The transformer architecture~\cite{vaswani2017attention} has become the
substrate of modern machine learning and the dominant computational workload of
modern natural-language processing, spanning encoder, decoder, and
encoder--decoder models such as the T5~\cite{raffel2020t5} family and the larger,
billion-parameter-class transformers that have followed. Running this workload
efficiently in low precision at the edge, rather than in large-batch training or
cloud serving, is what governs the cost of practical, interactive inference, and
it makes an INT8-native engine an attractive design point for embedded
deployment. Transformer inference also has a characteristic two-phase structure.
A compute-heavy encoding or prompt-processing pass ingests an input sequence in
parallel and is dominated by dense matrix multiplication, while autoregressive
generation then proceeds one token at a time: at every step the model consumes
the single most recently emitted token, attends to a cached key/value history,
and produces exactly one new token. This incremental, single-batch generation
regime has a defining property: at each step the accelerator must re-read
\emph{all} model weights to produce a single token. With one matrix--vector
product per weight matrix, the arithmetic intensity collapses to roughly one
multiply--accumulate per byte fetched, so steady-state generation is firmly
memory-bandwidth bound rather than compute bound~\cite{pope2023scaling}. An
accelerator that targets autoregressive generation, for example the decoder of a
machine-translation model, is therefore most usefully understood as a machine for
\emph{streaming weights past a datapath} at the bandwidth limit, and is best
served by low-precision weights that shrink the very traffic that bounds it.
INT8 quantization halves or quarters that traffic relative to BF16/FP32 while
remaining bit-reproducible, which makes an INT8-native, streaming engine an
attractive design point for embedded and edge inference.

These observations motivate \emph{TFA}, the general macro-op INT8 transformer
inference accelerator IP presented in this paper. TFA is not a processor: it
exposes no general-purpose instruction set and holds no program counter in the
conventional sense. It is a memory-to-memory engine that executes a short stream
of \emph{macro-operations}, each compiled offline by a host toolchain into a
fixed-size descriptor in an on-chip instruction RAM. The silicon has no notion of
``encoder'' versus ``decoder'': it implements an architecture-agnostic set of
tensor primitives (matrix multiplication, softmax, RMSNorm, elementwise, and
copy/gather) that compose into self-attention, cross-attention, and feed-forward
blocks, and therefore into encoder, decoder, and encoder--decoder transformers
alike, with the host compiler decomposing any model into descriptors. A single
time-multiplexed INT8 datapath (one output-stationary MAC array, one
normalization/vector unit, one softmax unit, and a copy/gather unit) is sequenced
across these macro-ops to realize any transformer block: RMSNorm
pre-normalization, attention (with optional causal masking, sliding-window
masking, and grouped-query (GQA) head sharing available for autoregressive
models), and a SwiGLU feed-forward block.
This organization departs in two directions from prevailing practice. Against
monolithic systolic accelerators such as the TPU~\cite{jouppi2017tpu}, which
expose a wide fixed-function matrix pipeline and push tiling, fusion, and control
into a heavyweight compiler-and-runtime stack, TFA exposes a tiny, explicit
macro-op contract and a deliberately minimal external interface: one AXI4 master
for all data movement, one AXI4-Lite slave for control, and one interrupt. The
hardware validates and sequences; the host compiler owns scheduling. Against
CPU/GPU software execution, TFA replaces per-kernel launch overhead and
floating-point datapaths with a hardware-resident dispatch loop over bit-exact
integer operations, so that the same descriptor stream is reproducible across the
register-transfer model, the simulator, and a software golden model.

The central design claim is that a very small instruction surface, combined with
an exhaustively checked decode contract and a host compiler that pre-resolves all
addressing and tiling, is sufficient to run a complete pretrained transformer
end to end on synthesizable RTL; and that doing so exposes, and lets us close,
the memory-bandwidth ceiling that defines single-batch decode.

This paper makes the following contributions:
\begin{itemize}
  \item \textbf{A compact macro-op ISA with a verifiable decode contract.}
        We define an 8-opcode, 512-bit fixed-width descriptor ISA that expresses
        any transformer as a stream of memory-to-memory macro-ops, paired with
        an exhaustive, overflow-safe decode-validation contract that bounds every
        operand footprint in hardware before any AXI transaction is issued.
  \item \textbf{A streaming RTL microarchitecture with bit-exact INT8
        numerics.} We present a time-multiplexed datapath built around an
        output-stationary MAC array with ping-pong operand buffering that overlaps
        DMA fill with compute across tile boundaries, and define an
        integer-reproducible numeric specification (GEMM/requantization, RMSNorm,
        softmax, and elementwise paths) that the RTL realizes bit-for-bit.
  \item \textbf{A bit-exact golden-model verification methodology.} We build a
        UVM environment whose golden model reconstructs device state purely from
        the control-bus monitor stream and byte-compares every output region,
        reaching 100\% functional coverage and 94.96\% DUT-scoped code coverage
        with reviewed waivers, while surfacing a catalogue of architecture- and
        simulation-level defects.
  \item \textbf{An end-to-end, bit-exact neural machine translation
        demonstration on a pretrained encoder--decoder transformer.} We compile a
        pretrained 60M-parameter T5 encoder--decoder onto the v1 ISA and run its
        full translation pipeline on the RTL, the bidirectional encoder,
        the cross-attention, and the autoregressive decoder, checking the argmax
        of chip-written logits at every decode step against a bit-exact golden
        model. Because the model is multi-task, the same compiled weights
        translate English into French, German, and Romanian by task prefix
        alone; we exercise this on a ten-proverb multilingual stress workload
        ($70{,}320$ operations, zero golden-model mismatches). A lossless
        randomized-Hadamard rotation, folded entirely into weights at compile
        time, restores correct INT8 output with no change to the ISA or silicon.
  \item \textbf{A roofline-honest performance and energy characterization.} We
        report measured RTL throughput and bus/MAC utilization against an explicit
        roofline, attribute read traffic to operation classes, and give
        pre-synthesis energy-per-token estimates for three parameterizations of the
        IP.
\end{itemize}

The remainder of this paper is organized as follows.
Section~\ref{sec:related} positions TFA against prior accelerators and
quantization techniques. Section~\ref{sec:arch} details the macro-op ISA, the
decode-validation contract, and the module hierarchy.
Section~\ref{sec:numerics} specifies the bit-exact INT8 numerics for each
operation. Sections~\ref{sec:uarch} and~\ref{sec:uarch-vec} describe the streaming microarchitecture and its finite-state machines, and Section~\ref{sec:dataflow} develops the dataflow and performance model;
its finite-state machines. Section~\ref{sec:verif} presents the UVM golden-model
methodology and coverage results. Section~\ref{sec:translate} reports the
end-to-end translation case study and the rotation-based quantization result.
Section~\ref{sec:perf} gives the roofline-honest performance and energy analysis,
and Section~\ref{sec:concl} concludes.

\section{Background and Related Work}\label{sec:related}
\label{sec:background}

This section establishes the workload, the numerical regime, and the accelerator
landscape against which TFA is positioned. It is deliberately scoped to prior
art; the TFA contract and microarchitecture are deferred to
Sections~\ref{sec:arch} onward.

\subsection{Transformer Inference}
\label{sec:bg-transformer}
The transformer architecture~\cite{vaswani2017attention} spans three families
that share the same underlying operators: encoder-only models, which apply
bidirectional self-attention over the full input; decoder-only models (for
example the Llama \cite{touvron2023llama} and Mistral \cite{jiang2023mistral}
families), which apply causal self-attention to generate tokens
autoregressively; and encoder-decoder models such as the T5 family
\cite{raffel2020t5}, which pair a bidirectional encoder with an autoregressive
decoder coupled through cross-attention. Neural machine translation is a
canonical encoder-decoder workload: the encoder reads the source sentence in one
bidirectional pass, and the decoder then emits the target sentence one token at a
time, each step attending both causally over its own prefix and, through
cross-attention, over the fixed encoder output. This single application therefore
exercises every block type a transformer can contain.

Despite their differing connectivity, these models are built from a common
pre-norm residual block. Each layer reads a residual stream $x$, applies a
normalization (root-mean-square layer normalization,
RMSNorm~\cite{zhang2019rmsnorm}, which omits the mean subtraction and bias of
LayerNorm and rescales by $1/\sqrt{\operatorname{mean}(x^2)}$ times a learned
gain $\gamma$, in newer designs; a mean-centered LayerNorm in earlier ones), then
an attention sublayer, a second normalization, and a feed-forward (FFN)
sublayer, each wrapped in an additive residual connection. The attention sublayer
realizes self-attention or cross-attention depending on where its keys and values
come from, and the few mechanisms that distinguish modern variants are local to
this sublayer. Grouped-query attention (GQA)~\cite{ainslie2023gqa} shares each
key/value head across a group of query heads, shrinking the key--value (KV) cache
that must be streamed per token while leaving the query projection at full width.
Sliding-window attention~\cite{beltagy2020longformer, jiang2023mistral} bounds
each query's receptive field to a fixed window, capping the per-step attention
work and the resident KV footprint. The FFN may be a plain two-matrix unit or a
gated variant such as SwiGLU~\cite{shazeer2020glu}, in which a SiLU-gated branch
is multiplied elementwise with a linear branch before the down projection; this
introduces a nonlinear elementwise product alongside the two matrix multiplies
and motivates a fused multiply-with-lookup primitive in any supporting datapath.
TFA treats GQA, sliding-window masking, and gated FFNs as optional mechanisms its
ISA supports, not as a fixed target architecture.

The performance behavior of these models splits sharply into two phases.
\emph{Prefill} (equivalently, the encode pass of an encoder or encoder-decoder
model) consumes its input in a single pass: many tokens are processed together,
the matrix multiplies have a large shared dimension, and the operation is compute
bound. \emph{Autoregressive decode} is the opposite: each step generates one
token and, critically, must re-read the entire resident weight set plus the
growing KV cache to do so. A roofline analysis~\cite{williams2009roofline} makes
the consequence explicit: at batch size one the arithmetic intensity of decode is
of order one multiply--accumulate per weight byte fetched, placing the workload
far to the left of the ridge point, so achievable throughput is set by memory
bandwidth rather than by peak compute~\cite{pope2023scaling}. FlashAttention
sharpened this input/output (I/O)-centric view for the attention kernel itself,
showing that tiling the softmax to keep intermediates on chip and minimize
high-bandwidth-memory traffic, rather than reducing floating-point operations,
governs realized speed~\cite{dao2022flashattention}. The autoregressive decoder
of a translation model is precisely this single-batch, memory-bound regime, and
TFA's design choices (streaming operands, overlapping fill with compute, and
minimizing redundant re-reads) are consequences of this memory-bound reality
rather than of a compute ceiling.

\subsection{Low-Precision Inference}
\label{sec:bg-quant}
Because decode is bandwidth bound, the single most effective lever is to shrink
the bytes moved, which makes integer quantization central rather than merely an
efficiency optimization~\cite{nagel2021whitepaper, gholami2022survey}. The W8A8
setting (8-bit integer weights and 8-bit integer activations) is attractive for
hardware because it admits a dense INT8$\times$INT8$\to$INT32 multiply--accumulate
datapath with cheap requantization, and it halves or quarters operand traffic
relative to 16-bit formats. The obstacle is not the weights but the activations:
transformer residual and FFN activations exhibit a small number of channels whose
magnitudes are one to two orders of magnitude larger than the median, and these
outliers dominate the per-tensor dynamic range so that a single symmetric INT8
scale collapses the remaining channels to a handful of
levels~\cite{dettmers2022llmint8, bondarenko2021understanding}. Several families
of methods address this. LLM.int8() isolates the outlier channels into a 16-bit
side computation~\cite{dettmers2022llmint8}; SmoothQuant migrates difficulty from
activations into weights by a per-channel diagonal rescaling that is folded into
adjacent layers~\cite{xiao2023smoothquant}; and weight-only schemes such as GPTQ
and AWQ push weights to very low bit widths using second-order error compensation
or activation-aware scaling~\cite{frantar2023gptq, lin2024awq}. A distinct and,
for this work, pivotal line applies an orthogonal (often randomized-Hadamard)
rotation to render the activation distribution incoherent before quantization,
spreading outlier energy across channels so that a plain per-tensor scale
suffices; QuIP, QuIP\#, and QuaRot develop this incoherence-processing view and,
in the RMSNorm pre-norm setting, exploit that a rotation can be folded into the
surrounding weights as an exact, inference-time-free
reparameterization~\cite{chee2023quip, tseng2024quip, ashkboos2024quarot}. TFA
adopts precisely this stance: the silicon implements only a clean symmetric
per-tensor INT8 datapath, and all outlier handling is pushed into the compiler,
where the rotation is a lossless float identity that leaves the ISA and the
hardware untouched (Section~\ref{sec:casestudy}).

\subsection{Transformer and DNN Accelerators}
\label{sec:bg-accel}
Spatial accelerators for dense linear algebra are well established. The Google
Tensor Processing Unit (TPU) demonstrated a large weight-stationary systolic
array driven by a coarse-grained instruction stream and backed by a software
compiler~\cite{jouppi2017tpu, jouppi2021tpuv4i}, while DianNao established the
case for a small accelerator whose performance is governed by on-chip buffering
and memory traffic rather than by raw multiplier count~\cite{chen2014diannao}.
Eyeriss introduced the row-stationary dataflow and made data reuse and energy
accounting first-class design concerns, showing that the choice of stationary
operand dominates off-chip traffic~\cite{chen2017eyeriss}. More recent work
targets the attention operator specifically, co-designing the dataflow with the
irregular, length-dependent structure of self-attention to keep the array fed
across the softmax boundary~\cite{kao2023flat}. Against this body of work, TFA
occupies a distinct point in the design space. It is a small, host-programmable,
memory-to-memory IP rather than a self-contained machine: it exposes a compact
macro-op ISA whose descriptors map directly onto the layer operators above,
time-multiplexes a single INT8 datapath across them, and, unusually for this
class of accelerator, commits to a \emph{bit-exact} numeric contract, so that
every operation has a single specified integer result that a golden model can
replay and byte-compare. This combination of a host-driven macro-op front end, a
deliberately minimal per-tensor INT8 backend, and a fully specified numeric
contract is what differentiates TFA from both the large systolic engines and the
fixed-function DNN accelerators surveyed here.

\section{System Architecture and Programming Model}
\label{sec:arch}

TFA is a memory-to-memory, macro-op-programmed compute engine rather than a
processor. The host driver places weights, activations, the key/value (KV)
cache, and a \emph{command program} in external memory and in the on-chip
instruction RAM (IRAM); TFA then executes the program (tiled INT8 GEMMs,
masked softmax, RMSNorm, elementwise operations, and DMA copies/gathers) and
raises a level interrupt on completion. The accelerator deliberately omits a
tokenizer, a DRAM/PCIe controller, and any cache or coherence mechanism: its
scratchpads are explicitly software-managed, and its entire external contract
is three wires' worth of standard interfaces. This section defines that
boundary, the macro-op instruction set and 512-bit descriptor that drive it,
how a general transformer layer is expressed as a descriptor sequence, and the
exhaustive decode-time validation contract that guards every descriptor before it is
issued. The bit-exact numeric definitions of each operation are deferred to
Section~\ref{sec:numerics}, the engine micro-architecture to
Section~\ref{sec:uarch-gemm}, and the tiling/performance model to
Section~\ref{sec:dataflow}.

\subsection{IP Boundary and Interfaces}
\label{sec:boundary}

\begin{figure*}[t]
  \centering
  \includegraphics[width=0.95\textwidth]{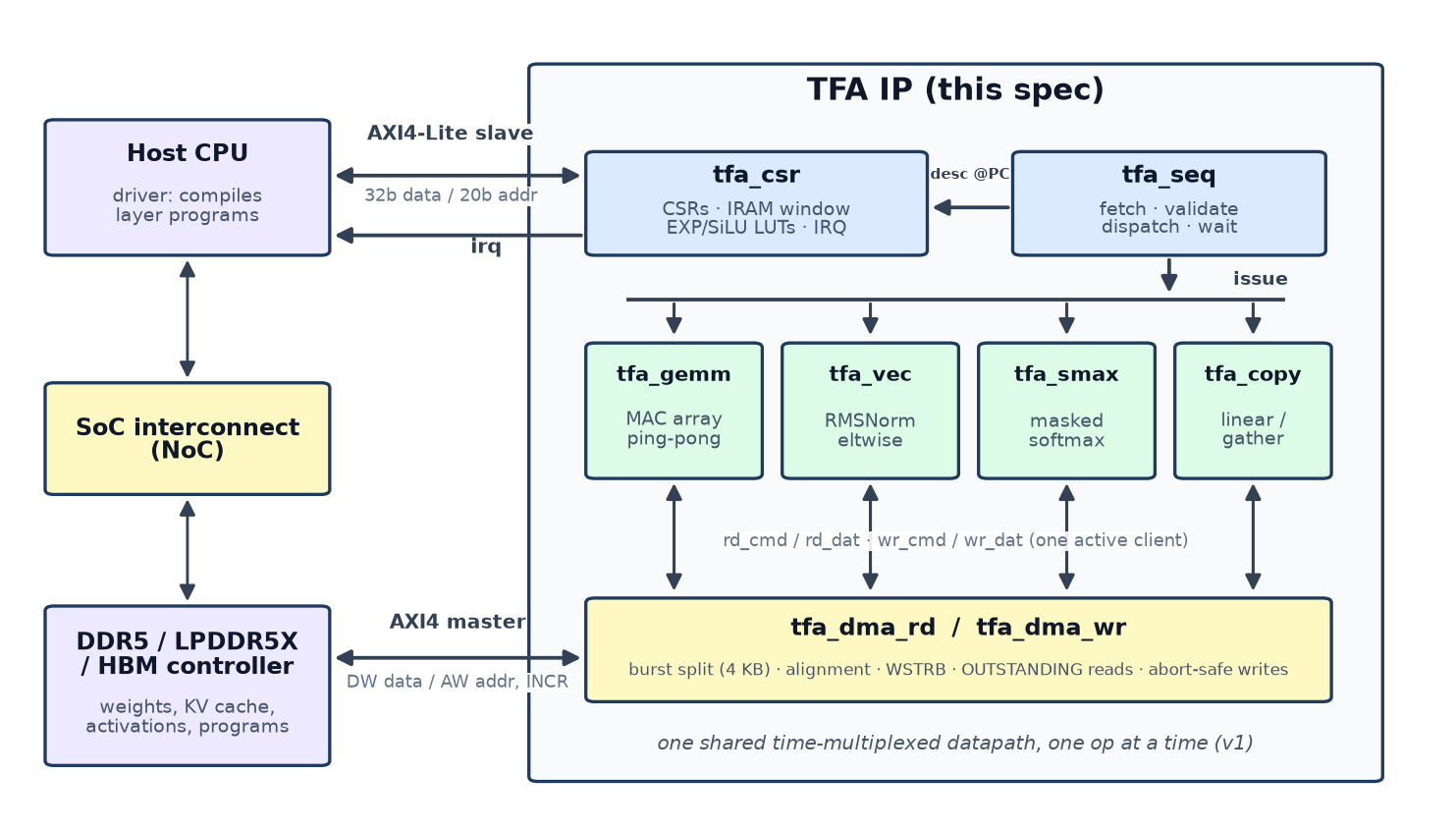}
  \caption{TFA IP boundary: one AXI4 master for all data movement, one
    AXI4-Lite slave exposing the CSR, LUT, and IRAM windows, and one level
    interrupt. A single shared, time-multiplexed datapath dispatches one
    macro-op at a time from the sequencer.}
  \label{fig:boundary}
\end{figure*}

TFA presents a single, technology-agnostic boundary
(Fig.~\ref{fig:boundary}). All bulk data movement traverses \emph{one}
AXI4 master \cite{arm-axi}, on which the engines issue INCR read and write
bursts against external memory; the master is parameterized by a data width
\code{DW} and an address width \code{AW}, uses a single transaction ID per
direction with in-order completion, and emits
\code{AxSIZE}$=\log_2(\code{DW}/8)$ with no narrow transfers. Hardware splits
bursts at 4\,KB boundaries and caps each burst at \code{ABURST\_LEN} beats, so
the IP composes with any DDR5/LPDDR5X/HBM subsystem behind a NoC.
Read-after-write program chaining is made safe by declaring transactions
non-bufferable (\code{AxCACHE}$=$\code{4'b0010}): a \code{B} response is
therefore required to imply global observability, so operation $N{+}1$'s reads
always see operation $N$'s writes. A read or write response carrying
\code{SLVERR}/\code{DECERR} aborts the executing operation with the
corresponding error code (Section~\ref{sec:decval}).

Control and program load arrive over \emph{one} AXI4-Lite slave with 32-bit
data and a 20-bit address, decoded into three address windows. The lower
offsets expose the control/status registers (CSRs): an \code{ID}/%
\code{VERSION} pair, a pulse-style \code{CTRL} (\code{START}, \code{ABORT},
\code{CNT\_CLR}), a sticky \code{STATUS} (\code{BUSY}, \code{DONE},
\code{ERR}, and the opcode of the executing operation), \code{START\_PC},
\code{ERR\_INFO} (error code and faulting PC), interrupt enable/status
registers, and four free-running performance counters (busy cycles, MAC-array
RUN cycles, and AXI read/write beats). A second pair of windows holds the
host-loaded lookup tables: a $256\times16$ exponential table for softmax and
a $256\times8$ signed SiLU table for the FFN gate. The IRAM window, based at
\code{0x10000}, accepts descriptors directly: word $w$ of descriptor $i$ is
written at \code{0x10000}$+64i+4w$. Byte-lane write strobes (\code{WSTRB}) are
honored on every read/write register and window, which lets the host patch
individual descriptor words in place between decode steps without a
read-modify-write. The single \code{irq} output is the level-OR of the
enabled interrupt-status bits.

A small number of race rules complete the contract. \code{START} is accepted
only when \code{BUSY}=0 at the write cycle and is otherwise dropped; a
combined \code{START}+\code{ABORT} resolves to \code{ABORT}; a hardware
interrupt-set wins over a same-cycle write-one-to-clear so no event is lost;
and \code{START} clears \code{DONE}, \code{ERR}, \code{ERR\_INFO}, and the
\code{DONE}/\code{ERR} interrupt bits. The host discipline is thus simply to
poll \code{DONE}/\code{BUSY}, load or patch IRAM, write \code{START\_PC}, and
pulse \code{START}.

\subsection{Macro-Op ISA and Descriptor Format}
\label{sec:isa}

TFA executes a compact macro-op instruction set of eight opcodes
(Table~\ref{tab:isa}), each a coarse-grained tensor operation rather than a
scalar instruction. \code{GEMM} performs a tiled matrix product with either a
fused INT8 requantization (optionally followed by ReLU) or a raw 32-bit
accumulator writeback; \code{SOFTMAX}, \code{RMSNORM}, and \code{ELTWISE} are
row engines; \code{COPY} moves bytes linearly or by index-gather;
\code{SYNC} sets a synchronization flag (optionally interrupting) and
continues; and \code{HALT} terminates the program, sets \code{DONE}, and
interrupts. Opcodes \code{0x8}--\code{0xFF} are rejected as \code{BAD\_OP}.
Because the model's hyper-parameters are program quantities while the silicon
parameters (Section~\ref{sec:dataflow}) only scale the implementation, the
same eight opcodes express every block of a transformer (self-attention,
cross-attention, and feed-forward) and the projection or LM head, independent
of whether the model is an encoder, a decoder, or an encoder-decoder.

\begin{table}[t]
  \centering
  \caption{TFA macro-op instruction set.}
  \label{tab:isa}
  \begin{tabularx}{\columnwidth}{@{}llX@{}}
    \toprule
    Opcode & Val & Function \\
    \midrule
    \code{NOP}     & \code{0x0} & no-op \\
    \code{GEMM}    & \code{0x1} & $C=\mathrm{requant}(A{\cdot}B)$ INT8 or raw INT32; $B$ normal/transposed; $B$/$C$ ring addressing \\
    \code{SOFTMAX} & \code{0x2} & row-wise masked softmax, INT8 in/out \\
    \code{RMSNORM} & \code{0x3} & row-wise RMSNorm with $\gamma$ \\
    \code{ELTWISE} & \code{0x4} & ADD or MUL (optional SiLU LUT on $A$) \\
    \code{COPY}    & \code{0x5} & DMA copy, linear or index-gather \\
    \code{SYNC}    & \code{0x6} & set sync flag, optional IRQ, continue \\
    \code{HALT}    & \code{0x7} & end program, \code{DONE}, IRQ \\
    \bottomrule
  \end{tabularx}
\end{table}

\begin{figure*}[t]
  \centering
  \includegraphics[width=0.95\textwidth]{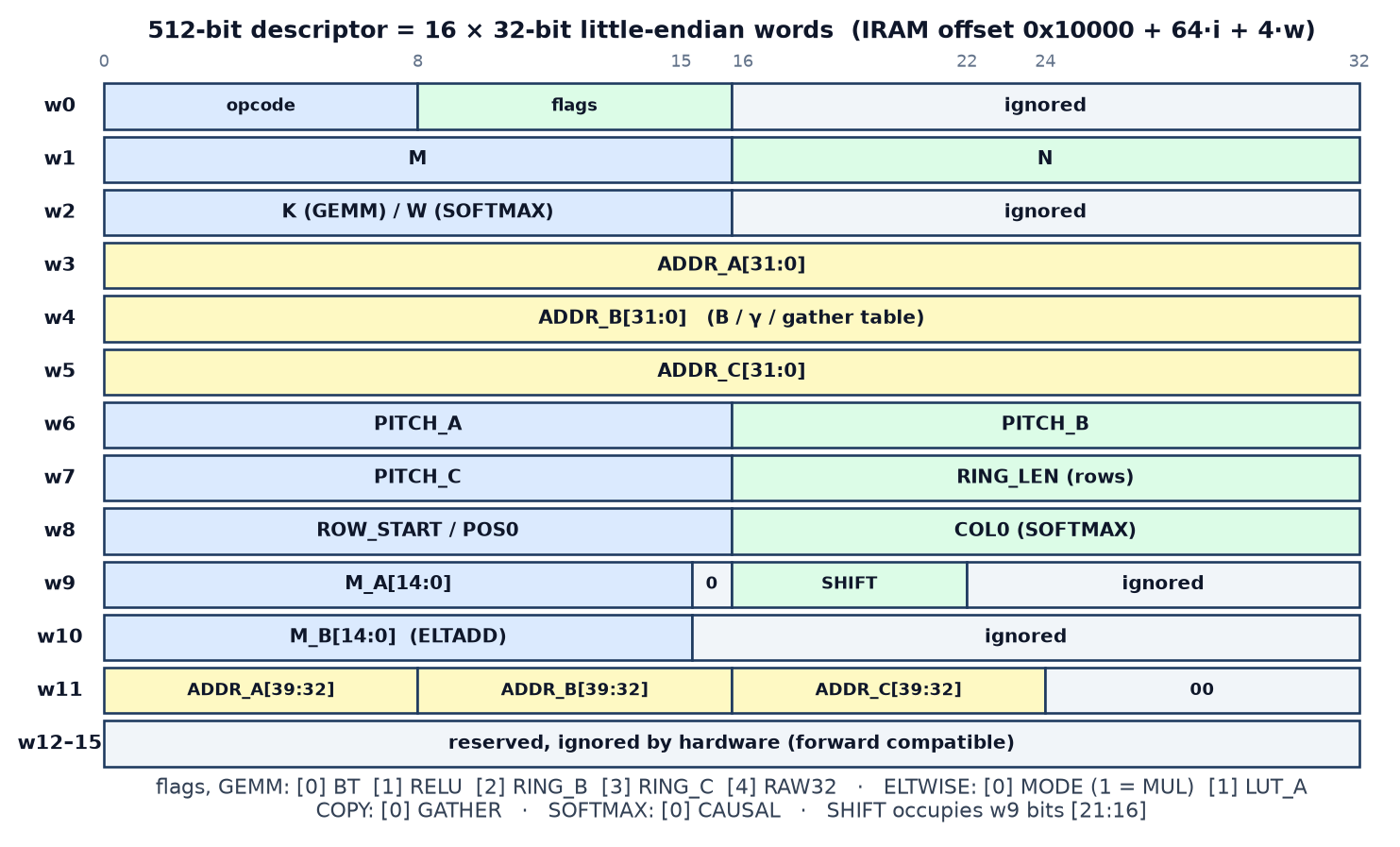}
  \caption{The 512-bit (16$\times$32-bit, little-endian) macro-op descriptor.
    Word~0 holds the opcode and an opcode-specific flag byte; words 1--11
    carry dimensions, base addresses, pitches, ring parameters, and requant
    multipliers; words 12--15 and all unnamed bits are reserved and ignored.}
  \label{fig:desc}
\end{figure*}

Every operation is described by a fixed 512-bit descriptor laid out as
sixteen 32-bit little-endian words (Fig.~\ref{fig:desc}). Word~0 carries the
opcode in its low byte and an opcode-specific flag byte; the remaining words
encode the matrix dimensions $M$, $N$, $K$ (or the softmax window $W$), the
three base addresses \code{ADDR\_A}/\code{ADDR\_B}/\code{ADDR\_C}, their
row pitches, a ring length and start row, the requant multipliers \code{M\_A}/%
\code{M\_B} and right-shift \code{SHIFT}, and, for \code{AW}$>32$ builds, a
word of high address bytes that extends each base to 40 bits. The flag byte is
overloaded per opcode: \code{GEMM} carries
$\{$\code{BT}, \code{RELU}, \code{RING\_B}, \code{RING\_C}, \code{RAW32}$\}$,
selecting a transposed-$B$ layout, fused ReLU, ring addressing on the $B$ and
$C$ operands, and raw 32-bit output respectively; \code{ELTWISE} carries
$\{$\code{MODE}, \code{LUT\_A}$\}$ to pick add versus multiply and to gate the
SiLU table onto operand $A$; \code{COPY} carries \code{GATHER}; and
\code{SOFTMAX} carries \code{CAUSAL}. \code{RELU} is ignored when \code{RAW32}
is set, and \code{RING\_B}/\code{RING\_C} share the single
\code{ROW\_START}/\code{RING\_LEN} pair, since no layer program requires
distinct values.

A deliberate forward-compatibility rule governs the format:
\emph{all unnamed and reserved bits anywhere in words 0--15 are ignored by
hardware and are never an error}. Words 12--15 are entirely reserved, as are
the unused high bits of several packed fields. This lets later revisions
introduce flags or widen fields without invalidating descriptors emitted by
the v1 toolchain, and it bounds the decode-validation surface to exactly the
fields the current ISA reads.

\subsection{Layer Programs}
\label{sec:layerprog}

\begin{figure*}[t]
  \centering
  \includegraphics[width=0.95\textwidth]{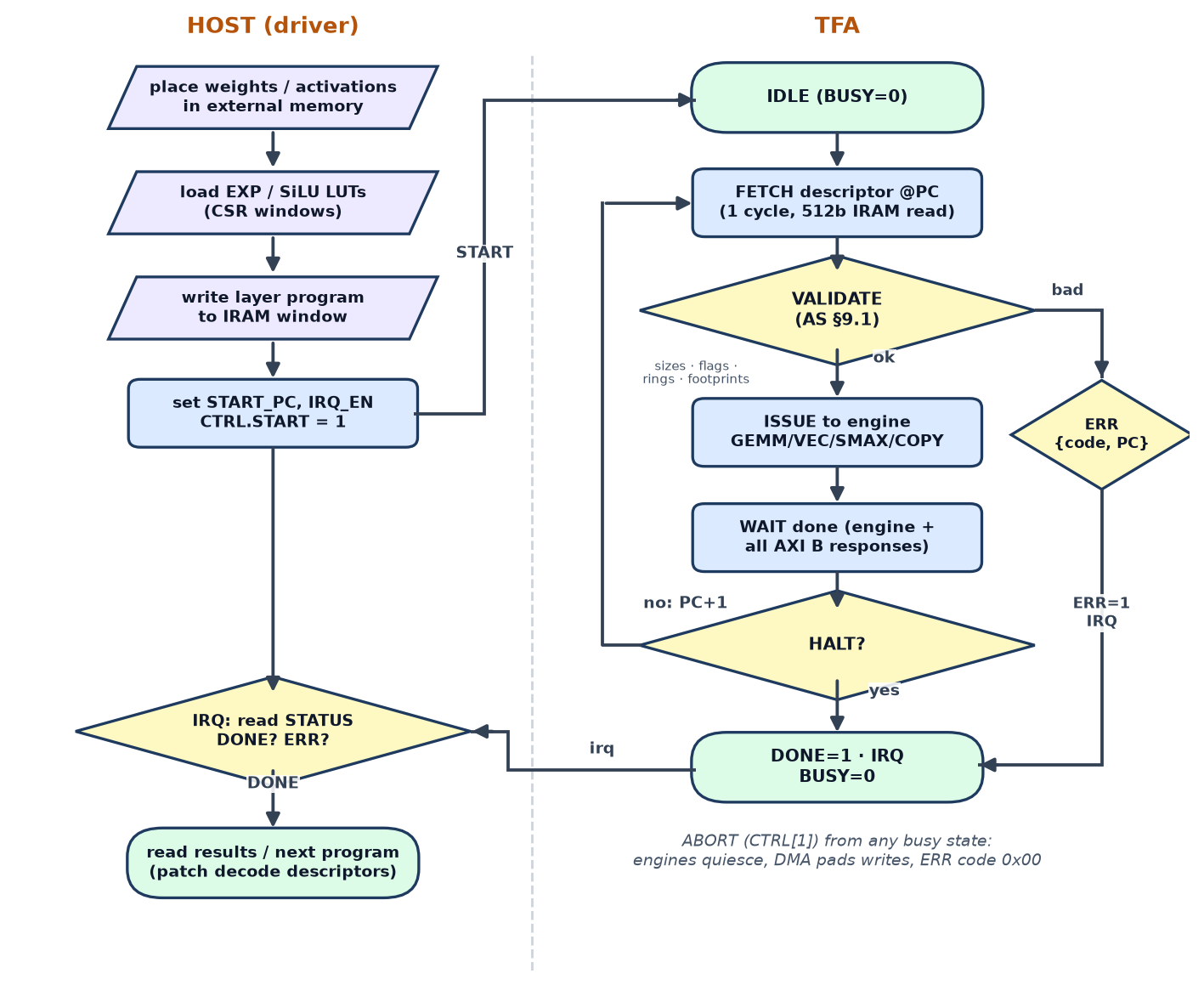}
  \caption{Host/engine program flow: the host loads or patches descriptors
    into IRAM, writes \code{START\_PC}, and pulses \code{START}; the sequencer
    fetches, validates, issues, and waits on one macro-op at a time, signaling
    completion or error through \code{STATUS} and the interrupt.}
  \label{fig:progflow}
\end{figure*}

\begin{figure*}[t]
  \centering
  \includegraphics[width=0.95\textwidth]{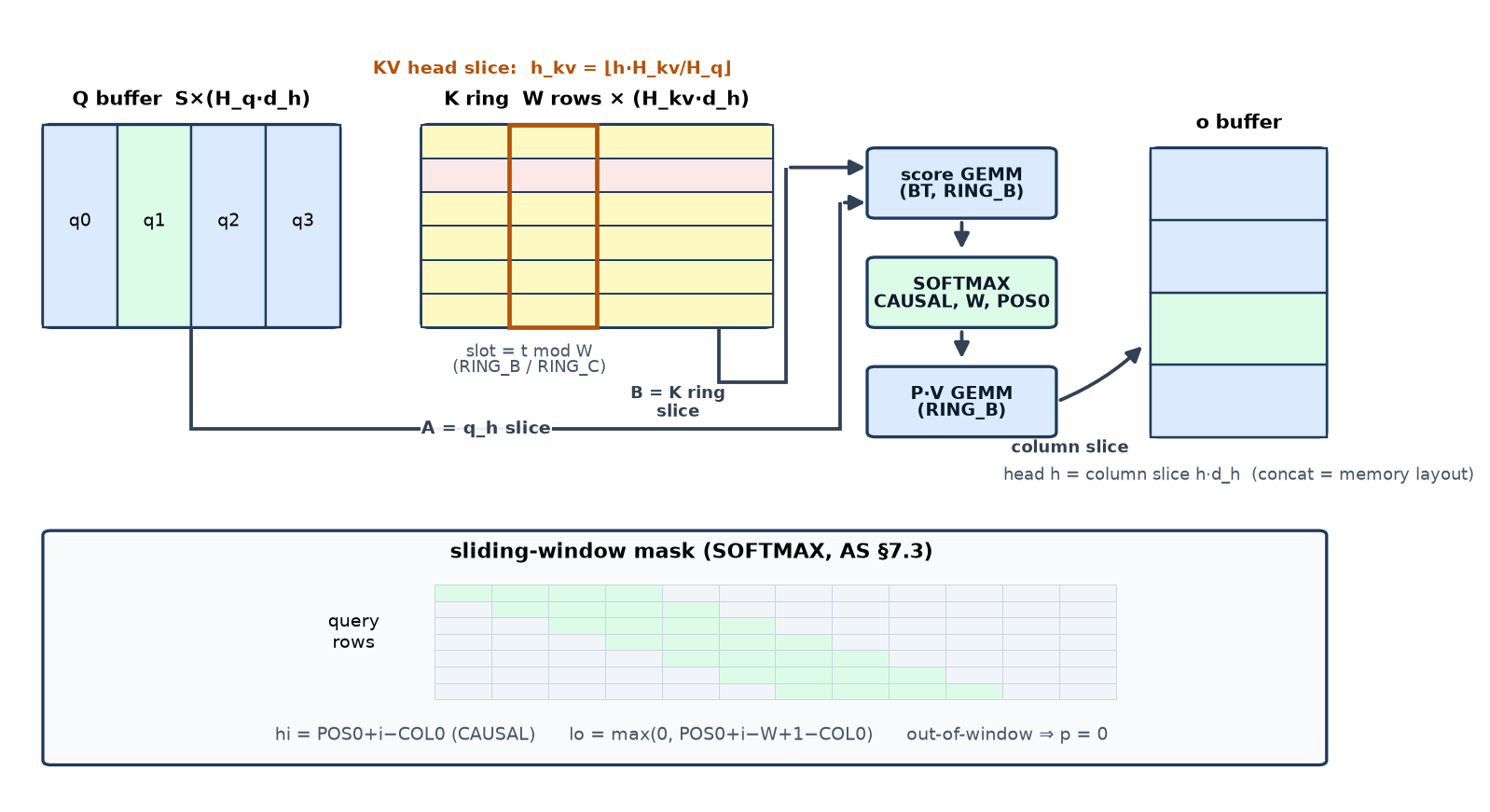}
  \caption{Attention mapping. Grouped-query attention is pure address
    computation ($h_{kv}=\lfloor h\,H_{kv}/H_q\rfloor$); the sliding window of
    depth $W$ is realized as a per-layer KV ring (token $t$ in slot
    $t \bmod W$) read by the score and context GEMMs under ring addressing,
    with the causal/window mask applied in softmax.}
  \label{fig:attn}
\end{figure*}

The ISA expresses a general transformer layer (self-attention, an optional
cross-attention, and a feed-forward block), and the \emph{same} descriptor
vocabulary realizes every layer variant the host compiler may need: an encoder
layer (bidirectional self-attention followed by FFN), a decoder layer (causal
self-attention, then cross-attention into encoder memory, then FFN), and the
cross-attention block itself. The differences among these are entirely
host-side choices of which blocks to emit, which operand a score \code{GEMM}
reads (the layer's own activations for self-attention, the encoder memory for
cross-attention), and whether the softmax mask is enabled. Each layer maps onto
a short, host-generated descriptor sequence that makes the residual-stream
dataflow explicit. The host issues each program by loading (or patching) its
descriptors into the IRAM window, writing \code{START\_PC}, and pulsing
\code{START}; the sequencer then walks the program autonomously, and the host
learns of completion or error only through the interrupt and \code{STATUS}
(Fig.~\ref{fig:progflow}). A pre-norm layer begins with an \code{RMSNORM} of the
residual stream $x$ by $\gamma_1$, whose output feeds three \code{GEMM}s that
project the query, key, and value tensors (all with the \code{BT} flag, since
the natural weight cache layout is transposed). When a KV cache is in use, the K
and V projections additionally set \code{RING\_C} so their output rows append
directly into the per-layer KV ring at slot $t \bmod W$. Attention is then
expressed per query head $h$: a score \code{GEMM} with \code{BT} and (for cached
attention) \code{RING\_B} reads the appropriate K slice, folding the
$1/\sqrt{d_h}$ scale into its requant; a \code{SOFTMAX} normalizes each row; and
a \code{GEMM} forms the per-head context, landing $o_h$ in its column slice of
the output buffer. Three attention mechanisms are available as host-selected
options and do not change the program structure: \emph{grouped-query} head
sharing maps query head $h$ to KV head
$h_{kv}=\lfloor h\,H_{kv}/H_q\rfloor$~\cite{ainslie2023gqa}, a pure
address computation on the ring slice; \emph{causal} masking is requested by the
\code{CAUSAL} flag on \code{SOFTMAX}; and a \emph{sliding window} of depth $W$ is
realized by the KV ring together with the window mask. Bidirectional
(encoder-style) self-attention and full cross-attention simply leave the mask
disabled. Head concatenation is therefore mere memory layout, and the output
projection $W_O$ is an ordinary \code{GEMM} (Fig.~\ref{fig:attn}). An
\code{ELTWISE} add closes the attention residual. The FFN repeats the pattern:
\code{RMSNORM} by $\gamma_2$, two projection \code{GEMM}s for the
SwiGLU~\cite{shazeer2020glu} gate and up paths, an \code{ELTWISE} multiply with
\code{LUT\_A} that applies the SiLU table to the gate before multiplying by the
up tensor, a down-projection \code{GEMM}, and a final \code{ELTWISE} add back
into $x$. Because the row engines process rows strictly in order, the residual
adds operate in place. A representative autoregressive decode program, which
exercises causal masking, GQA, and ring-cached attention at once, is sketched in
Listing~\ref{lst:decode}.

\begin{lstlisting}[language=SV,caption={A decode-step ($M{=}1$) layer
  program in descriptor mnemonics; $h_{kv}=\lfloor h\,H_{kv}/H_q\rfloor$,
  $n=\min(t{+}1,W)$.},label={lst:decode}]
RMSNORM  xn   <- x, g1                  // 1xD
GEMM     q    <- xn.Wq   (BT)           // 1x(Hq.dh)
GEMM     k    <- xn.Wk   (BT, RING_C)   // append slot t%W
GEMM     v    <- xn.Wv   (BT, RING_C)
for each query head h:                  // host-unrolled
  GEMM    s_h <- q_h . K^T (BT, RING_B) // 1 x n
  SOFTMAX p_h <- s_h  (CAUSAL,W,POS0=t,COL0)
  GEMM    o_h <- p_h . V   (RING_B)     // 1 x dh slice
GEMM     ao  <- o.Wo (BT)
ELTWISE  x   <- x + ao (ADD)            // in place
RMSNORM  xn2 <- x, g2
GEMM     g   <- xn2.Wg (BT)
GEMM     u   <- xn2.Wu (BT)
ELTWISE  hdn <- silu(g) * u (MUL, LUT_A)
GEMM     d   <- hdn.Wd (BT)
ELTWISE  x   <- x + d (ADD)             // in place
// ...  HALT terminates the per-token program
\end{lstlisting}

The same program serves prefill and decode. In prefill the sequence runs with
$M{=}S$, so the projections emit $S$ rows and the score/context GEMMs operate
on $S{\times}S$ tiles; in decode it runs with $M{=}1$, the KV ring receives a
single appended row, the score GEMM uses $N=\min(t{+}1,W)$ columns, and
softmax is supplied $\mathtt{POS0}=t$ together with $\mathtt{COL0}$, the
absolute position of the chronologically oldest fetched key. Between tokens
the host patches only the handful of per-token descriptor words through the
IRAM window, an overhead that is a fraction of a percent of DDR5-class token
time. When prefill is longer than the window ($S>W$), a mandatory
linear-then-fold pattern is used: the K/V projections write a linear scratch
region of $S$ rows (no \code{RING\_C}), the score and context GEMMs read it
linearly (no \code{RING\_B}) while the softmax mask enforces the
window~\cite{beltagy2020longformer}, and two \code{COPY} operations afterward
fold the last $W$ rows into the steady-state rings before decoding begins.
This keeps ring aliasing unreachable, because any ring encoding whose row
count exceeds \code{RING\_LEN} is rejected at decode. Finally, the LM head is
a single \code{GEMM} carrying the \code{RAW32} flag, so the 32-bit vocabulary
logits reach the host unsaturated; embedding lookup is expressed as a
\code{COPY} with the \code{GATHER} flag.

\subsection{Decode Validation Contract}
\label{sec:decval}

Before any macro-op is issued, the sequencer validates its descriptor against
an exhaustive, per-opcode contract; a violation aborts the program with
\code{BAD\_SIZE} and never reaches an engine. The checks fall into three
classes. First, \emph{size} checks require the active dimensions to be at
least one and bound the row engines' free dimension by \code{ROW\_MAX}
($1\le N\le\code{ROW\_MAX}$ for \code{SOFTMAX}, \code{RMSNORM}, and
\code{ELTWISE}), which also guarantees the softmax accumulator bound. Second,
\emph{ring} checks apply when \code{RING\_B} or \code{RING\_C} is set: they
require $\code{RING\_LEN}\ge1$, $\code{ROW\_START}<\code{RING\_LEN}$, and that
the number of rows touched in the ring not exceed \code{RING\_LEN}
($N$ if \code{BT} else $K$ for \code{RING\_B}; $M$ for \code{RING\_C}). This
last condition is what makes the aliasing failure mode of an over-long ring
encoding structurally unreachable. Third, a \emph{footprint} check is applied
to every memory operand.

The footprint test verifies that the highest byte an operand can touch lies
within the address space:
\[
  \code{ADDR} + (\text{rows}-1)\cdot\code{PITCH} + \text{row\_bytes}
  \;\le\; 2^{\code{AW}},
\]
with the per-operand \{rows, row\_bytes, pitch\} taken from the descriptor and
the opcode (for example, \code{GEMM}~$C$ uses \code{RING\_LEN} rows of
$N\cdot\mathrm{esz}$ bytes when \code{RING\_C} is set, where
$\mathrm{esz}=4$ under \code{RAW32}). A subtle but essential requirement is
that this sum be evaluated in \emph{at least 41-bit unsigned arithmetic}: at
the 40-bit \code{AW} of the PERF configuration the left-hand side can legally
exceed $2^{40}$, so the carry-out is itself part of the reject condition. A
40-bit modular adder would silently wrap and \emph{false-accept} an
out-of-range descriptor; the wider accumulator makes the comparison exact.
Gather row reads are exempt from the footprint check by host contract, as
their addresses are computed from a host-supplied index list. The error codes
that the contract and the AXI/abort paths can report are summarized in
Table~\ref{tab:errcodes}; the finite-state machine that implements the
fetch--validate--issue--wait flow is described in
Section~\ref{sec:uarch-gemm}.

\begin{table}[t]
  \centering
  \caption{Error codes reported in \code{ERR\_INFO[7:0]}.}
  \label{tab:errcodes}
  \begin{tabularx}{\columnwidth}{@{}llX@{}}
    \toprule
    Code & Name & Meaning \\
    \midrule
    \code{0x00} & \code{ABORT}    & host \code{ABORT} or engine-resolved abort \\
    \code{0x01} & \code{BAD\_OP}  & opcode \code{0x8}--\code{0xFF} \\
    \code{0x02} & \code{BAD\_SIZE}& failed a \S\ref{sec:decval} size/ring/footprint check \\
    \code{0x03} & \code{AXI\_RD}  & \code{SLVERR}/\code{DECERR} on a read response \\
    \code{0x04} & \code{AXI\_WR}  & \code{SLVERR}/\code{DECERR} on a write response \\
    \code{0x05} & \code{BAD\_PC}  & \code{PC}$\ge$\code{IRAM\_DEPTH} without \code{HALT} \\
    \bottomrule
  \end{tabularx}
\end{table}

\section{Bit-Exact INT8 Numerics}
\label{sec:numerics}

The central design principle of TFA's arithmetic is that every operation is
specified to the bit. There is no rounding mode, accumulator width, saturation
boundary, or table-lookup convention left to the implementation: each opcode
denotes a unique integer-valued function of its operands. This is the property
that makes the rest of the work possible. Because the semantics are fully
deterministic, a purely software golden model can be written that is
\emph{byte-identical} to the silicon, and the UVM scoreboard
(Sec.~\ref{sec:verif}) can byte-compare every output region rather than
checking a tolerance. The same determinism lets the host compiler
(Sec.~\ref{sec:translate}) replay each descriptor at compile time and predict
the chip's output exactly, which is how the t5-small case study can
byte-compare every chip output against the golden model and predict the decoded
token at each step. All intermediates are exact in $\le 64$-bit signed
integers; no narrower post-multiply stage is permitted at any point in the
datapath.

\begin{figure*}[t]
\centering
\includegraphics[width=0.95\textwidth]{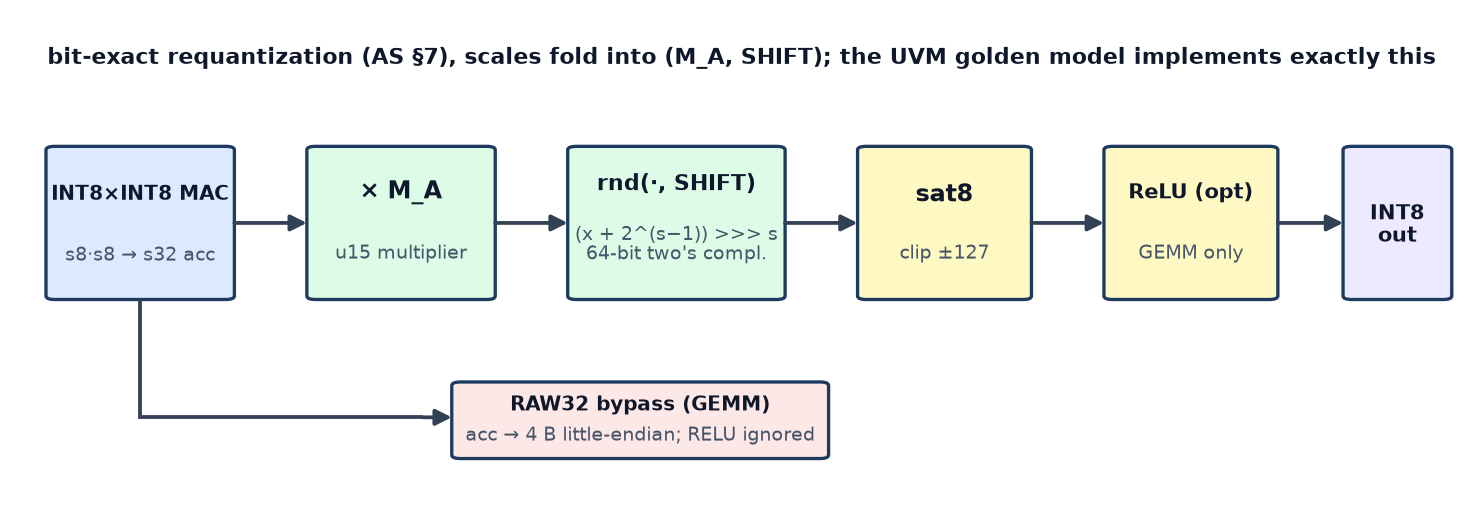}
\caption{Per-op integer pipelines. Each opcode is a fixed integer function from
INT8 (or RAW32) operands to INT8 (or RAW32) outputs; the per-op multiplier
$M_A$ and shift $\mathrm{SHIFT}$ are the only scale-bearing parameters, chosen
by the host compiler. The hardware is scale-agnostic.}
\label{fig:requant}
\end{figure*}

Two primitives recur throughout. Saturation to the signed-byte range is
$\mathrm{sat8}(x)=\mathrm{clip}(x,-128,127)$. Rounded arithmetic right shift is
\begin{equation}
\mathrm{rnd}(x,s)=
\begin{cases}
\big(x+(1\!\ll\!(s\!-\!1))\big)\ggg s, & s>0,\\[2pt]
x, & s=0,
\end{cases}
\end{equation}
i.e.\ round-half-up evaluated in two's complement, with $\ggg$ the arithmetic
shift. The rounding-constant add reaches $\le 2^{62}$ at $\mathrm{SHIFT}=63$ and
the largest product reaches $\le 2^{59}$ (RMSNorm, below), so the sum fits in
$64$ signed bits for every legal operand. A figure of the requantization
pipeline is shown in Fig.~\ref{fig:requant}.

\subsection{GEMM and Requantization}
The MAC array forms an $\mathrm{INT32}$ accumulator from signed-byte products,
\begin{equation}
\mathrm{acc}_{ij}=\textstyle\sum_{k} A_{ik}\,B_{kj},
\qquad |\mathrm{acc}_{ij}|\le 65535\cdot 16384 < 2^{31},
\end{equation}
the bound following from $K\le\mathrm{ROW\_MAX}$ and the $\mathrm{s8}\!\cdot\!
\mathrm{s8}\!\to\!\mathrm{s16}$ product range. Two output paths exist. The
\code{RAW32} path writes \code{acc} directly as a little-endian $4$-byte word
(ignoring \code{RELU}); it is used where full precision is required, notably the
LM-head logits over the $32128$-token vocabulary, whose argmax must be exact.
The default INT8 path requantizes,
\begin{equation}
\begin{aligned}
y_{ij}&=\mathrm{sat8}\!\big(\mathrm{rnd}(\mathrm{acc}_{ij}\cdot M_A,\ \mathrm{SHIFT})\big),\\[1pt]
y_{ij}&\leftarrow\max(y_{ij},0)\quad\text{if \code{RELU}}.
\end{aligned}
\end{equation}
The dyadic pair $(M_A,\mathrm{SHIFT})$ approximates the product of the output
tensor's scale and the reciprocals of the input scales; for attention scores it
additionally folds in the $1/\sqrt{d_h}$ factor. Crucially, the hardware never
sees a floating-point scale: it applies the integer pair the compiler supplies.

\subsection{RMSNorm}
RMSNorm operates on an INT8 row $x[0..N)$ with an INT8 gain $\gamma[0..N)$. The
sum of squares, an integer reciprocal square root via the serial
\code{tfa\_isqrt} unit, and a reciprocal $R$ are formed as
\begin{align}
ss &= \textstyle\sum_i x_i^2 \le 2^{29},\\
q  &= \mathrm{isqrt}(ss\!\ll\!16) \le 2^{24},\\
R  &= \big\lfloor 2^{30}/\max(q,1)\big\rfloor < 2^{31},
\end{align}
after which each output element is
\begin{equation}
y_i=\mathrm{sat8}\!\big(\mathrm{rnd}\big(((x_i\,\gamma_i)\cdot R)\cdot M_A,\ \mathrm{SHIFT}\big)\big).
\end{equation}
The intermediate widens monotonically and provably fits 64 bits:
$x_i\gamma_i$ is $\mathrm{s16}$, the product with $R$ ($\mathrm{u31}$) is
$\mathrm{s47}$, and the product with $M_A$ ($\le\mathrm{u15}$) is $\mathrm{s62}$.
The $1/\sqrt{N}$ factor and the output scale fold into $(M_A,\mathrm{SHIFT})$.
An all-zero row yields $ss=0$, $q=0$, hence $y_i=0$ exactly, with no
division-by-zero or undefined behavior.

\subsection{Softmax}
Softmax is computed per row with masking derived from the descriptor's query
base \code{POS0}, key base \code{COL0}, window \code{W}, and the \code{CAUSAL}
flag. For row $i$ the absolute query position is $h=\mathrm{POS0}+i$, and the
visible key interval $[lo,hi]$ is obtained by clamping
$hi=(\text{CAUSAL})?(h-\mathrm{COL0}):n-1$ to $\le n-1$ and
$lo=(\mathrm{W}{=}0)?0:\max(0,h-\mathrm{W}{+}1-\mathrm{COL0})$ to $\ge 0$, all in
signed arithmetic. If the interval is empty the output row is all zeros.
Otherwise, with $m=\max_{j\in[lo,hi]} s_j$ and a host-provided exponential table,
\begin{align}
e_j &= \mathrm{exp\_lut}[\min(m-s_j,255)],\\
\mathrm{sum} &= \textstyle\sum_{j\in[lo,hi]} e_j < 2^{31},\\
r &= \big\lfloor 2^{31}/\max(\mathrm{sum},1)\big\rfloor,\\
p_j &= \min\!\big(127,\,(e_j\cdot r)\!\gg\!24\big),\quad j\in[lo,hi],
\end{align}
with $p_j=0$ outside the window. The $\max(\cdot,1)$ guard makes the all-zero
row case well-defined; bit-exactness holds for \emph{any} monotone table
contents, so the table is part of the contract rather than an approximation the
hardware must reproduce.

\subsection{Eltwise, Copy, and Gather}
\code{ELTWISE} provides the residual-stream and gating arithmetic. In ADD mode
$z=\mathrm{sat8}(\mathrm{rnd}(a\cdot M_A+b\cdot M_B,\ \mathrm{SHIFT}))$, carrying
two independent operand scales for the residual add. In MUL mode an optional
table applies SiLU, $a'=\mathrm{LUT\_A}?\,\mathrm{silu\_lut}[\mathrm{u8}(a)]:a$
(signed-INT8 entries), then $z=\mathrm{sat8}(\mathrm{rnd}(a'\cdot b\cdot M_A,\
\mathrm{SHIFT}))$, realizing the SwiGLU gate as a single op. \code{COPY} is
byte-exact; its gather variant reads a u32 little-endian index
$\mathrm{idx}[i]$ at $\mathrm{ADDR\_A}+4i$ and copies the $N$-byte row at
$\mathrm{ADDR\_B}+\mathrm{idx}\cdot\mathrm{PITCH\_B}$ to
$\mathrm{ADDR\_C}+i\cdot\mathrm{PITCH\_C}$, the embedding lookup. No arithmetic
is performed, so the copy is reproduced trivially by the golden model.

\subsection{The Scale-Agnostic Seam}
Across every op the only scale-bearing quantities are the per-op dyadic
parameters $(M_A,M_B,\mathrm{SHIFT})$ and the host LUTs. The datapath is
otherwise scale-agnostic: it manipulates integers and never interprets a tensor
scale. This separation is deliberate. It keeps the hardware simple and exactly
specifiable, and it places all numerical policy in the compiler, which is
precisely where quantization-aware transformations such as outlier-tolerant
rotation must live. That seam, and how a randomized-Hadamard reparameterization
exploits it without any change to the ISA or silicon, is the subject of
Sec.~\ref{sec:translate}.

\section{Microarchitecture I: Command Sequencer and GEMM Engine}\label{sec:uarch}\label{sec:gemm}\label{sec:uarch-gemm}
\label{sec:uarch-gemm}

This section describes the control spine of TFA (the macro-op sequencer
\code{tfa\_seq}) and the densest datapath it dispatches to, the GEMM engine
\code{tfa\_gemm}. Together these account for the majority of both the silicon
area and the read traffic of the accelerator, and they establish the internal
conventions (ready/valid engine--DMA channels, single-active-client muxing,
write-idle gating) reused by the vector, softmax, and copy engines of
Sec.~\ref{sec:uarch-vec}. The numeric definitions of the requantization and
RAW32 paths are normative in Sec.~\ref{sec:numerics} and are only referenced
here; the dataflow and roofline analysis appear in Sec.~\ref{sec:perf}.

\subsection{Module Hierarchy and Internal Interfaces}
\label{sec:hier}

\begin{figure*}[t]
\centering
\includegraphics[width=0.95\textwidth]{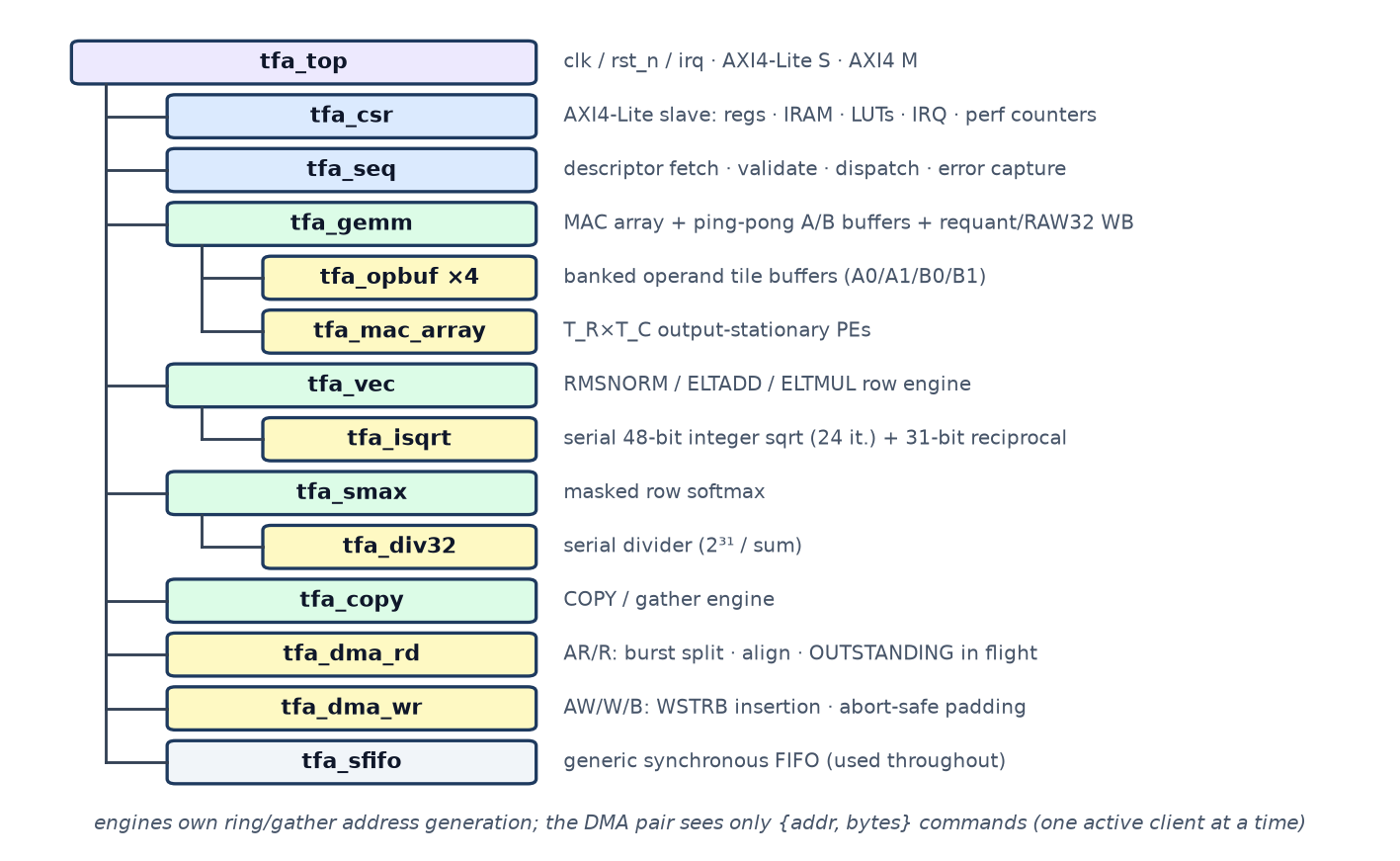}
\caption{Module hierarchy of \code{tfa\_top}. The sequencer fetches and
validates 512-bit descriptors from IRAM and dispatches one compute engine at a
time; the active engine owns the shared read and write DMA clients through the
\code{dma\_owner} one-hot. All engine--DMA traffic uses the four ready/valid
channels \code{rd\_cmd}/\code{rd\_dat}/\code{wr\_cmd}/\code{wr\_dat}.}
\label{fig:hier}
\end{figure*}

TFA is a single-clock, mostly-synchronous design whose top level
\code{tfa\_top} instantiates one control block, four compute engines, two DMA
engines, and the CSR/IRAM block, as shown in Fig.~\ref{fig:hier}. The external
contract is deliberately narrow: one AXI4 master carries all data movement as
\code{INCR} bursts, one AXI4-Lite slave exposes the CSRs, the 64-entry
instruction RAM (IRAM) window, and the host-loaded exponential/SiLU lookup
tables, and one level interrupt signals completion or error. There is no
instruction cache, no register file, and no general-purpose datapath; TFA is a
memory-to-memory engine whose only program state is a 16-bit program counter
and the descriptor currently in flight.

All engines speak to the two DMA blocks over a uniform pair of ready/valid
channels (MAS \S2). The read side is a command/data split: an engine issues a
\code{rd\_cmd} carrying a 40-bit byte address and a 16-bit length (one matrix
row or linear block per command), and \code{tfa\_dma\_rd} returns a packed
\code{rd\_dat} stream of \code{DW}-wide beats with a final partial-keep beat and
a \code{last} marker. The write side is symmetric: \code{wr\_cmd} plus a packed
\code{wr\_dat} stream, with a per-command \code{wr\_rsp} pulse raised only after
all AXI \code{B} responses of that command have returned. Two level signals,
\code{rd\_idle} and \code{wr\_idle}, indicate that no commands, bursts, or
staged bytes remain outstanding. Because v1 executes one macro-op at a time,
the active opcode owns both DMA clients through a one-hot \code{dma\_owner}
vector; an embedded assertion checks \code{\$onehot0(dma\_owner)} every cycle,
and a companion assertion checks that at most one engine reports \code{busy}.

The single most important global ordering rule is that an engine's \code{done}
pulse is gated on \code{wr\_idle} (design rule G-09). An engine that has issued
its last write command has not finished until every \code{B} response is in;
only then does \code{tfa\_seq} advance the program counter. This guarantees that
the reads of macro-op $N{+}1$ observe all writes of op $N$ in DRAM, which is
exactly the producer--consumer dependency a host-compiled transformer layer
relies on (e.g.\ an RMSNorm whose output feeds the next GEMM, or a residual add
that reads a freshly written projection). The rule is enforced uniformly: in
\code{tfa\_gemm} the completion term is
\code{done\_int = (rstt==R\_FIN) \&\& (fst==F\_DONE) \&\& (wst==W\_IDLE) \&\&
rd\_idle \&\& wr\_idle}, so no engine can report completion while any AXI
response is in flight.

\subsection{Command Sequencer}
\label{sec:seq}

\begin{figure}[t]
\centering
\includegraphics[width=\columnwidth]{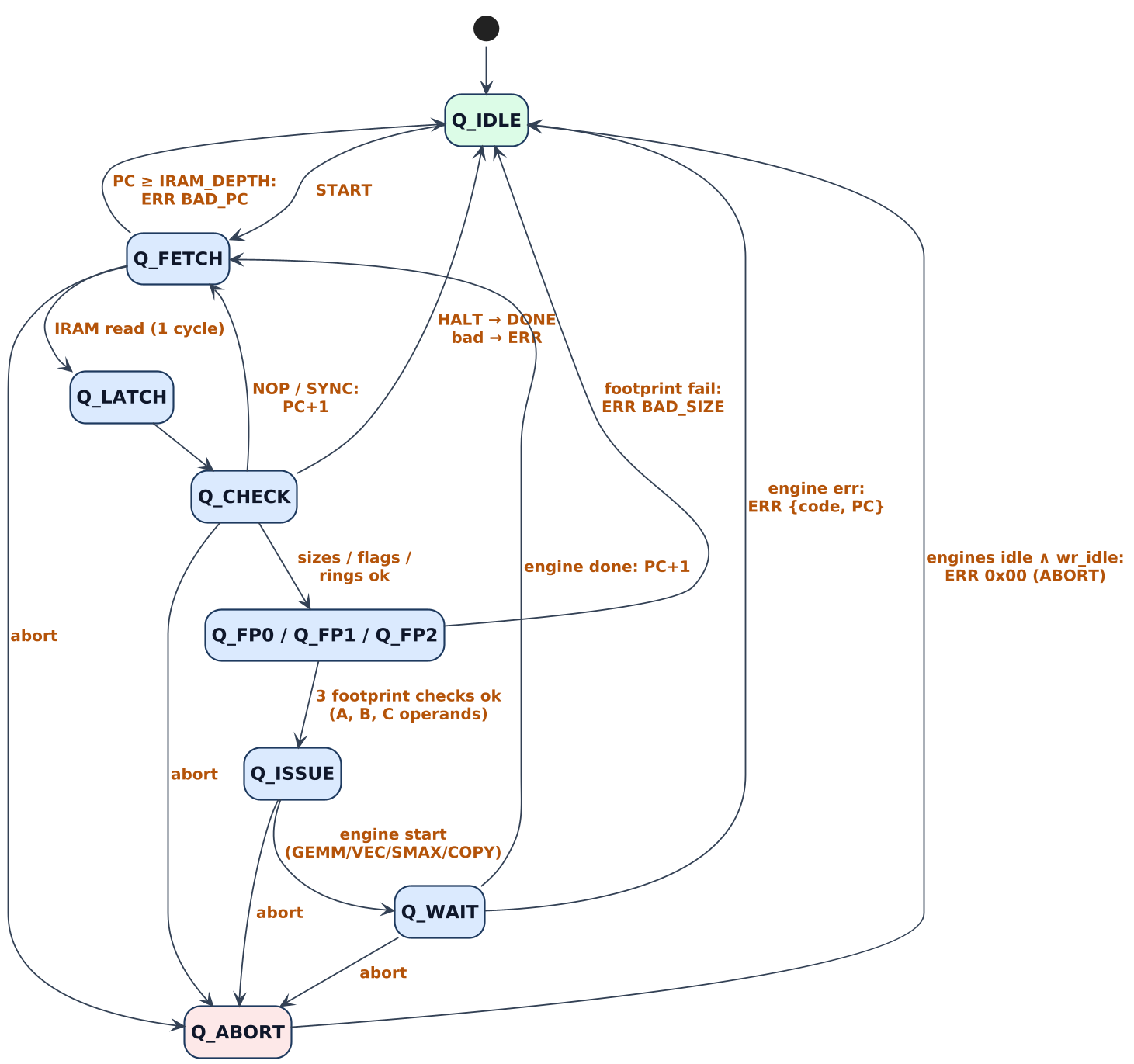}
\caption{Sequencer FSM. Footprint validation is spread across
\code{Q\_FP0}--\code{Q\_FP2}, one operand per cycle, so the two $16{\times}16$
pitch multiplies are never on the timing-critical path. \code{abort\_lvl} is
recognized in every busy state and routes to \code{Q\_ABORT}.}
\label{fig:seqfsm}
\end{figure}

The sequencer \code{tfa\_seq} is not a CPU; it is a
\textsc{fetch}$\rightarrow$\textsc{validate}$\rightarrow$\textsc{issue}$\rightarrow$\textsc{wait}
macro-op dispatcher implemented as the ten-state FSM of Fig.~\ref{fig:seqfsm}.
On a host \code{START} pulse it latches the start PC and enters \code{Q\_FETCH},
which reads all sixteen 32-bit IRAM sub-RAMs in one cycle and, in
\code{Q\_LATCH}, unpacks the 512-bit word into a typed \code{desc\_t} struct.
\code{Q\_CHECK} handles the control opcodes inline: \code{NOP} and \code{SYNC}
advance the PC (\code{SYNC} additionally pulses an interrupt flag), \code{HALT}
sets \code{DONE} and returns to \code{Q\_IDLE}, and an unknown opcode or a
program counter past \code{IRAM\_DEPTH} raises \code{BAD\_OP} or \code{BAD\_PC}.

The distinguishing feature of \code{tfa\_seq} is that it implements the full
decode-validation contract of the architecture spec (AS \S9.1) in hardware,
\emph{exhaustively}, before any DMA traffic is generated. Static checks
(per-opcode size legality and the ring constraints \code{RING\_LEN}$\,\geq 1$,
\code{ROW\_START}$\,<\,$\code{RING\_LEN}, and ring-rows$\,\leq\,$\code{RING\_LEN})
are combinational on the latched descriptor. The footprint checks,
\code{ADDR}$+(\text{rows}{-}1)\cdot\code{PITCH}+\text{row\_bytes}\leq 2^{\code{AW}}$
for each operand, require two $16{\times}16$ multiplies per operand and are
deliberately spread across the dedicated cycles \code{Q\_FP0}, \code{Q\_FP1},
and \code{Q\_FP2}, one operand (A, B, C) per cycle, accumulating a sticky
\code{v\_fail}. This keeps the multipliers off the critical path. Crucially the
sum is evaluated in 48-bit unsigned arithmetic and the comparison includes any
carry-out:

\begin{equation}
\code{fp\_sum} = \code{addr} + (\code{rows}-1)\cdot\code{pitch} + \code{row\_bytes},
\end{equation}

rejected when \code{fp\_sum}$\,>\,2^{\code{AW}}$. A 40-bit modular add would wrap
and \emph{false-accept} an out-of-bounds descriptor; the wider evaluation
(R3-01) makes the carry part of the reject condition. The per-operand row and
row-byte counts are selected by opcode and flags exactly as the spec table
prescribes (ring operands use \code{RING\_LEN} rows, a RAW32 C operand uses
four bytes per element, a gather A operand is one row of $4M$ index bytes),
so the check is meaningful for every macro-op variant.

On any validation failure the FSM writes \code{ERR\_INFO}$\,\leftarrow\,$\{PC,
code\} and returns to idle without issuing; on success \code{Q\_ISSUE} pulses
exactly one engine start, sets \code{dma\_owner}, latches the opcode into
\code{STATUS[15:8]}, and enters \code{Q\_WAIT}. \code{Q\_WAIT} watches for an
engine error (propagating its code into \code{ERR\_INFO}) or an engine
\code{done} (advancing the PC and refetching). Host \code{abort\_lvl} is
honored in \emph{every} busy state (\code{Q\_FETCH}, \code{Q\_CHECK},
\code{Q\_ISSUE}, and \code{Q\_WAIT} all branch to \code{Q\_ABORT}), which
broadcasts \code{eng\_abort}, waits for \code{!any\_busy \&\& rd\_dma\_idle \&\&
wr\_dma\_idle}, and reports \code{ERR\_ABORT} (0x00). The abort therefore never
races a partially issued descriptor and always lands the design in a clean
quiescent state.

\subsection{Output-Stationary MAC Array}
\label{sec:mac}

The arithmetic core is \code{tfa\_mac\_array}, a $T_R\times T_C$ grid of
output-stationary processing elements (with $T_R=T_C=T$ enforced at
elaboration). Each cycle in which \code{en} is asserted, the array consumes one
column of A and one row of B and performs an outer product, accumulating
$\text{acc}[i][j]\mathrel{+}=a[i]\cdot b[j]$. A two-stage input pipe (operand
register, then multiply--accumulate) keeps the array timing-clean and implies a
two-cycle epilogue between the last \code{en} and the first accumulator read.
The accumulators are signed 32-bit, and overflow is impossible by construction:
each INT8$\times$INT8 product has magnitude $\leq 16384$, and with the legal K
bound the running sum satisfies $|\text{acc}|\leq 65535\cdot 16384 < 2^{31}$, so
no saturation logic is needed inside the array.

Two correctness subtleties are visible in the listing below. First, edge tiles
are handled entirely by \emph{zero-masking} at the caller: when fewer than $T$
rows or columns are valid, the GEMM engine drives the corresponding
\code{a\_col}/\code{b\_row} lanes to zero, so the array always runs a full
$T\times T$ outer product and partial tiles need no special datapath. Second,
element selects of a packed \emph{signed} array are unsigned in SystemVerilog;
the products must be explicitly re-signed with \code{\$signed} before the 32-bit
widening multiply.

\begin{lstlisting}[language=SV,caption={Output-stationary PE accumulation with explicit re-signing; edge lanes are zero-masked by the caller.},label={lst:mac}]
// Stage 2: multiply-accumulate grid
logic signed [31:0] acc [T_DIM][T_DIM]; // T_DIM = T = T_R = T_C
always_ff @(posedge clk or negedge rst_n) begin
  if (!rst_n) begin /* clear all acc */ end
  else if (clr) begin /* zero in 1 cycle */ end
  else if (en_q) begin
    // element selects of packed signed arrays are
    // unsigned in SV -- force signed
    for (int i = 0; i < T_DIM; i++)
      for (int j = 0; j < T_DIM; j++)
        acc[i][j] <= acc[i][j]
          + (32'($signed(a_q[i])) * 32'($signed(b_q[j])));
  end
end
\end{lstlisting}

The \code{clr} input zeroes every accumulator in a single cycle, used at the
start of each output tile's first K-tile; a simulation assertion forbids
\code{clr} and a pipelined \code{en} in the same cycle, since accumulating while
clearing would silently corrupt a tile. The accumulator row read is purely
combinational on a \code{drain\_row} select, feeding the requantization lanes
described in Sec.~\ref{sec:wb}.

\subsection{Ping-Pong Operand Buffering}
\label{sec:pingpong}

\begin{figure*}[t]
\centering
\includegraphics[width=0.95\textwidth]{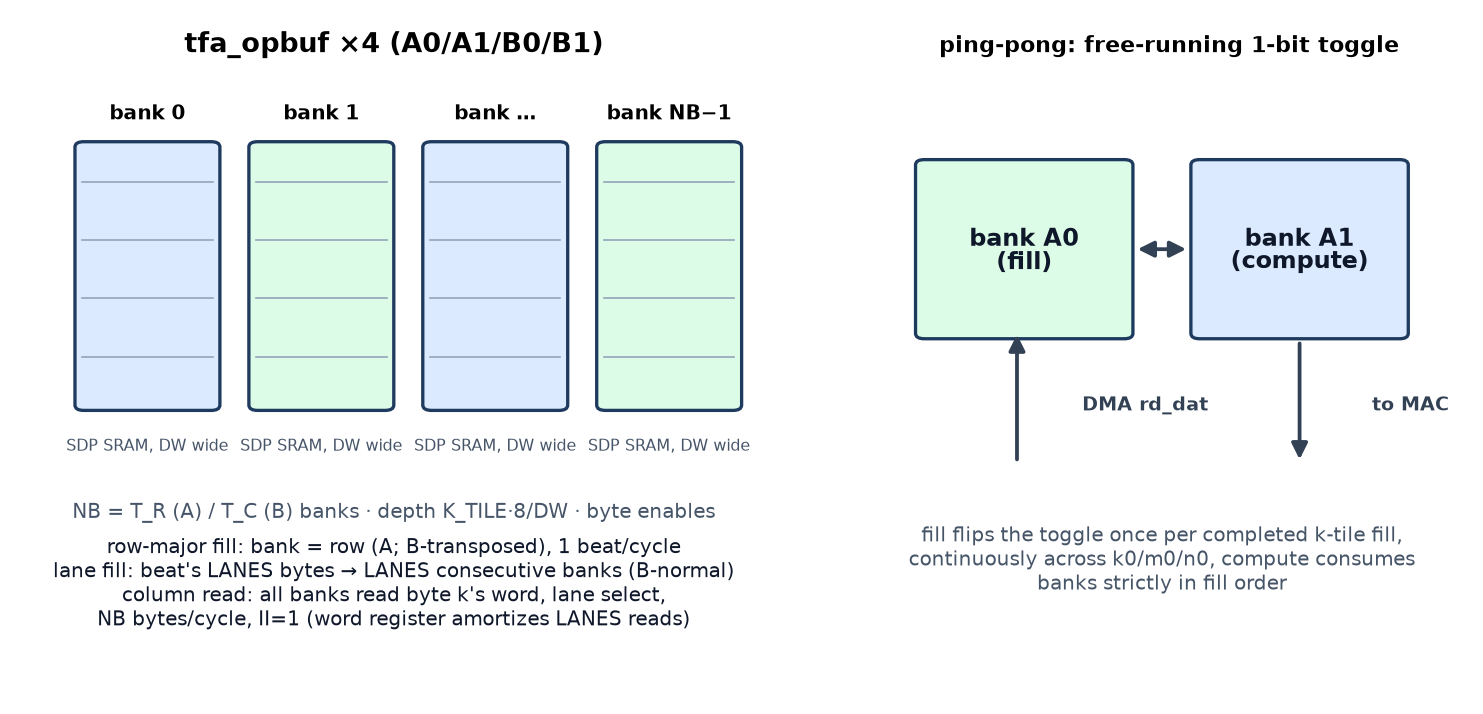}
\caption{Banked operand buffer \code{tfa\_opbuf}. Row-major fill (A, and B under
\code{BT}) writes one bank per row; lane fill (B normal) scatters each beat
across \code{LANES} banks. Column read presents \code{NB} bytes per cycle at
II=1.}
\label{fig:opbuf}
\end{figure*}

Feeding a $T\times T$ array at one outer product per cycle while reading
operands from DRAM is the central throughput problem, and TFA solves it with
double-buffered operand SRAMs (Fig.~\ref{fig:opbuf}). Four instances of
\code{tfa\_opbuf} (A0/A1 and B0/B1) each hold one $K_{\text{TILE}}$-deep
operand tile across \code{NB} banks (\code{NB}$=T$), where a bank is a
simple-dual-port SRAM \code{DW} bits wide with byte write enables. The buffer
supports two fill modes and one read mode. A is always filled row-major, one
bank per row with full-word writes at one beat per cycle; B is filled row-major
in the transposed (\code{BT}) case but \emph{lane-filled} in the normal case,
where each beat scatters its \code{LANES} bytes across \code{LANES} consecutive
banks at word $\lfloor k/\code{LANES}\rfloor$, lane $k \bmod \code{LANES}$. The
compute read is column-oriented: all \code{NB} banks read the same word and the
consumer selects one byte lane, yielding \code{NB} operand bytes per cycle at an
initiation interval of one, with the just-read word held in a register to
amortize the \code{LANES} reads of a single SRAM word.

\begin{figure}[t]
\centering
\includegraphics[width=\columnwidth]{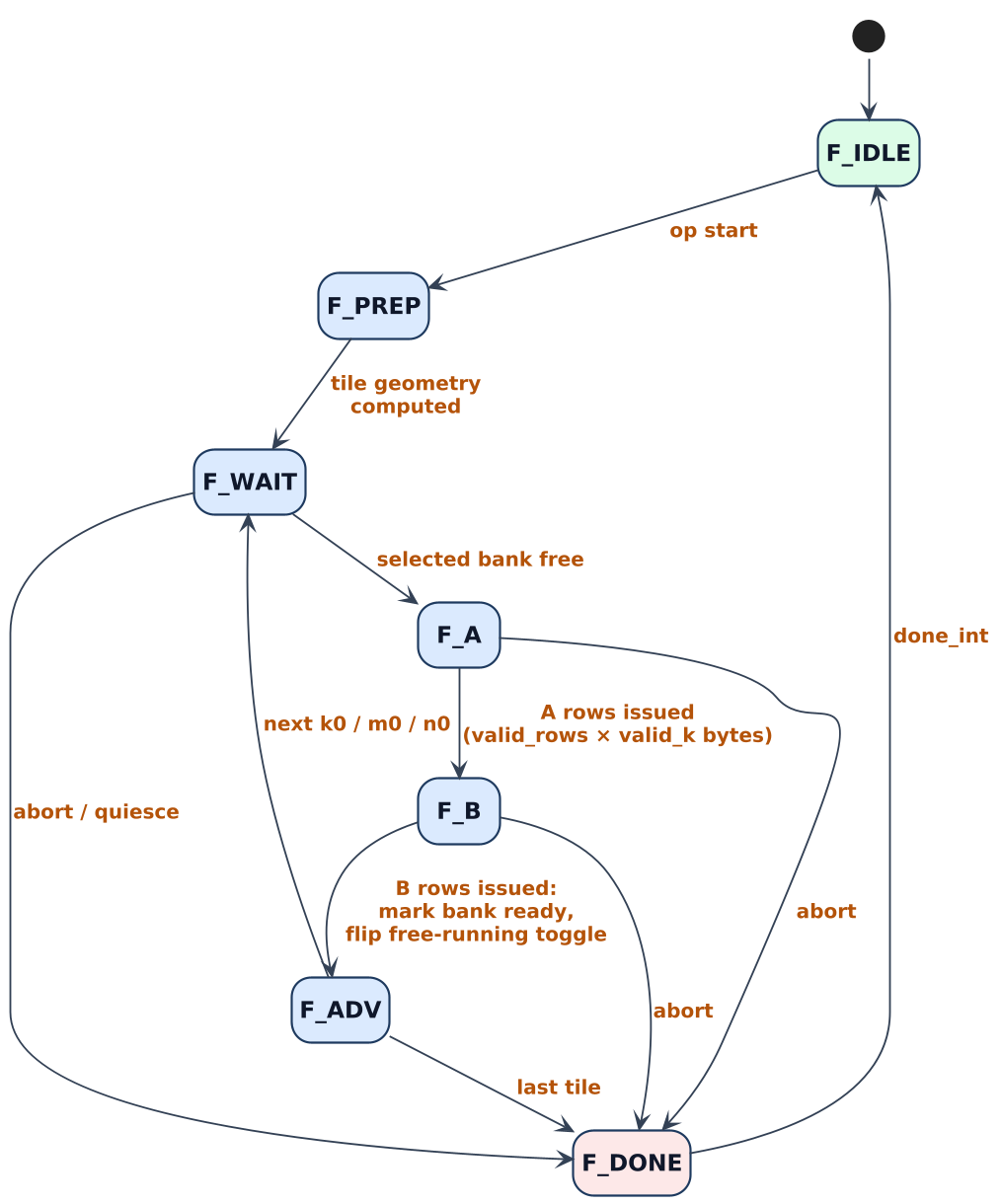}
\caption{GEMM fill FSM. \code{F\_WAIT} claims the next ping-pong bank, captures
its tile geometry and C address, then \code{F\_A}/\code{F\_B} issue the operand
read commands; \code{F\_ADV} toggles the bank and advances $(n_0,m_0,k_0)$.}
\label{fig:fillfsm}
\end{figure}

\begin{figure}[t]
\centering
\includegraphics[width=\columnwidth]{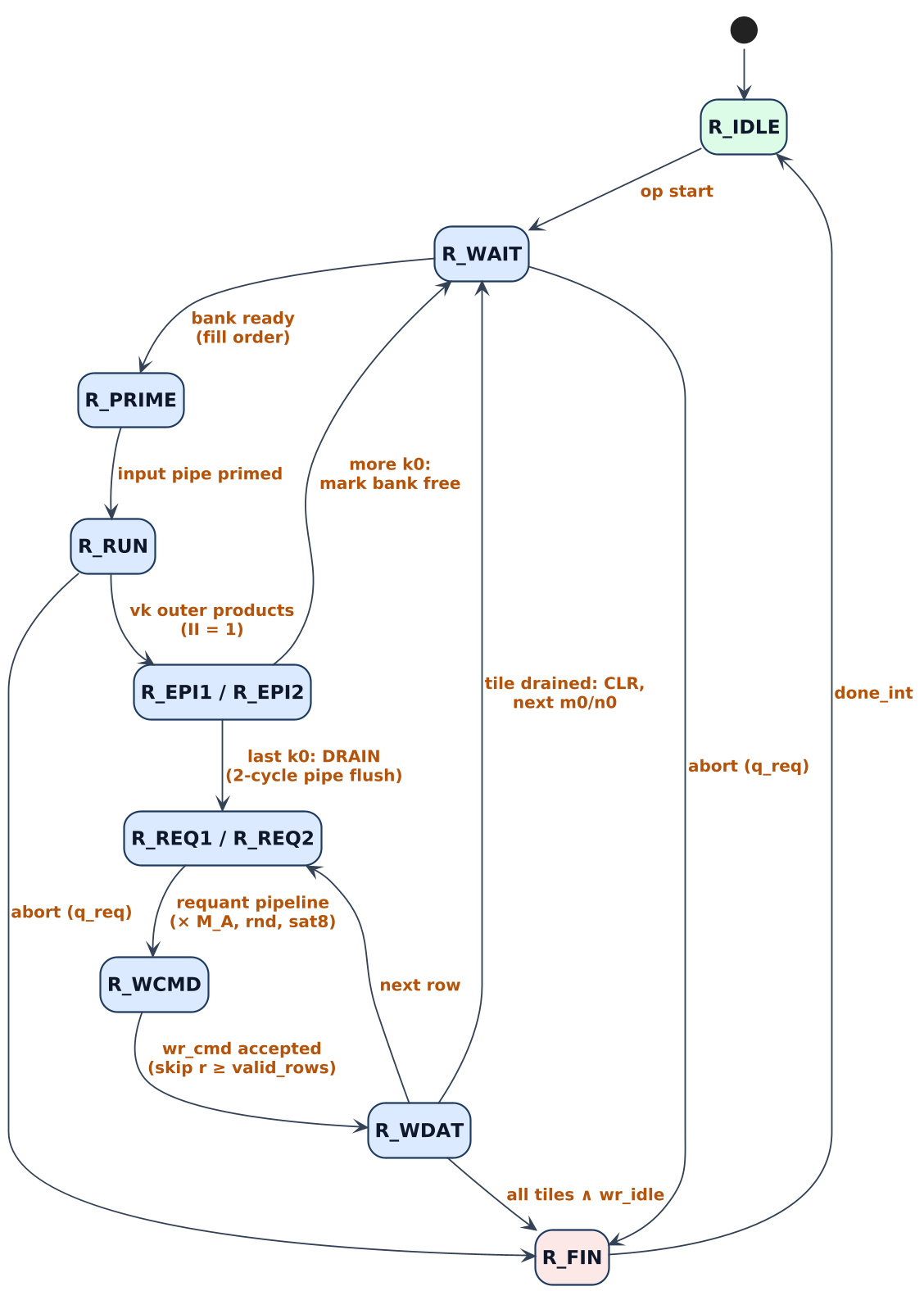}
\caption{GEMM compute/drain FSM. \code{R\_WAIT} blocks on \code{bank\_ready},
\code{R\_RUN} streams \code{valid\_k} outer products, and on the last K-tile the
\code{R\_REQ}/\code{R\_WCMD}/\code{R\_WDAT} states requantize and write C while
the fill process refills the other bank.}
\label{fig:compfsm}
\end{figure}

GEMM control is split into three concurrent FSMs that pass tiles through the
ping-pong buffers. A \emph{fill} process (Fig.~\ref{fig:fillfsm}) owns
\code{rd\_cmd} and walks the loop nest $(n_0,m_0,k_0)$ with $n_0$ outermost and
$k_0$ innermost, issuing the A-row and B-row read commands for each tile. A
separate \emph{writer} FSM routes the returned \code{rd\_dat} stream into the
selected operand buffer. A \emph{compute} process (Fig.~\ref{fig:compfsm})
consumes ready banks, runs the MAC array, and drains C. The two halves are
decoupled by a free-running 1-bit bank toggle, \code{fill\_pp}: it flips once
per completed tile fill and is \emph{continuous across $k_0$, $m_0$, and $n_0$
boundaries}, so the fill of tile $n{+}1$ overlaps the compute of tile $n$ and
the read channel stays busy through every loop-nest transition. The compute
process consumes banks in fill order with its own pointer \code{run\_pp}, and a
pair of per-bank flags (\code{fill\_pend}, commands issued, writer not done;
and \code{bank\_ready}, writer done, run not yet consumed) form the
producer/consumer handshake.

\begin{lstlisting}[language=SV,caption={Bank claim and free-running toggle in the GEMM fill FSM (excerpted).},label={lst:pingpong}]
F_WAIT: begin
  if (q_req) fst <= F_DONE;
  else if (!fill_pend[fill_pp] && !bank_ready[fill_pp]) begin
    tinfo[fill_pp] <= '{vr:vr_c, vc:vc_c, vk:vk_c,
        first_k:(k0==0), last_k:is_last_k, /* ... */
        c_addr0:tile_c_addr0, c_wrap_rows:tile_c_wrap};
    fill_pend[fill_pp] <= 1'b1;
    a_row <= a_m_base + FAW'(k0);
    // ... B iterators initialized here ...
    fst   <= F_A;
  end
end
F_ADV: begin
  fill_pp <= ~fill_pp;   // free-running toggle
  // ... advance k0, then m0, then n0 ...
end
\end{lstlisting}

The compute process clears the array on a tile's first K-tile, streams
\code{valid\_k} outer products in \code{R\_RUN}, and on the final K-tile enters
the drain/writeback states; on any earlier K-tile it simply frees the bank and
returns to wait, so accumulation across K-tiles never crosses a writeback. RUN
and DRAIN never overlap (they share the accumulators), but DRAIN \emph{does}
overlap the next tile's fill, since they use disjoint DMA channels. For the
decode case $M=1$ the A fill is a single row, B streaming dominates, and the
ping-pong keeps the read channel essentially continuously busy: the
arrangement that delivers the measured read-bus utilization of
Sec.~\ref{sec:perf}.

\subsection{Writeback and Ring/RAW32 Addressing}
\label{sec:wb}

Writeback drains accumulator rows through \code{T_C} requantization lanes that
compute $y=\code{sat8}(\code{rnd}(\text{acc}\cdot M_A,\code{SHIFT}))$ with an
optional ReLU (formulas normative in Sec.~\ref{sec:numerics}), packing one INT8
byte per column. The RAW32 flag bypasses the lanes entirely, packing the signed
32-bit accumulators little-endian at four bytes per element and ignoring ReLU;
this path produces, for example, the LM-head logits over the 32128-token
vocabulary. The C-row element size \code{esz} is therefore 1 or 4, and the write
command length is \code{valid\_cols}$\cdot\code{esz}$. A key edge rule is that
the engine issues \emph{no} write command for a row index $r\geq\code{valid\_rows}$
(G-12), so partial output tiles never write garbage rows.

C addressing supports both linear and ring-buffer destinations. The linear row
address is \code{ADDR\_C}$+(m_0+r)\cdot\code{PITCH\_C}+n_0\cdot\code{esz}$; under
\code{RING\_C} the row index is wrapped modulo \code{RING\_LEN}. All ring
iterators are maintained as running wrapped counters initialized from
\code{ROW\_START} and incremented by one with a conditional subtract at
\code{RING\_LEN}, giving true modulo addressing for any walk length without ever
forming a large product; decode guarantees ring-rows$\,\leq\,$\code{RING\_LEN} so
a single GEMM never laps its own ring. The same scheme drives \code{RING\_B}
token-row reads in the fill FSM. All C address arithmetic is carried in 40-bit
registers maintained additively at the loop-nest advance points.

This writeback and ping-pong machinery was the source of two of the most
instructive RTL defects found during verification, both since fixed and
discussed in Sec.~\ref{sec:verif}. The first was a packed-signed
sign-extension error in the MAC array (the very issue the \code{\$signed} cast
in Listing~\ref{lst:mac} guards against), which zero-extended negative operands
and produced systematically wrong negative products. The second was a one-cycle
ping-pong pointer race: the writer toggled \code{wrt\_pp} on the same edge it
pulsed completion, and the fill FSM then cleared \code{fill\_pend} using the
already-toggled pointer, clearing the wrong bank and relaunching the writer on a
phantom tile. The fix latches the completing bank index (pre-toggle) alongside
the pulse and clears exactly that index: a reminder that the value of the
double-buffered structure depends entirely on getting its one-bit handshake
exactly right.

\section{Microarchitecture II: Vector, Softmax, Copy, and DMA Engines}\label{sec:uarch-vec}\label{sec:engines}
\label{sec:uarch-vec-dma}

This section describes the non-GEMM compute engines (\code{tfa\_vec},
\code{tfa\_smax}, \code{tfa\_copy}) and the shared infrastructure that
surrounds every engine: the two AXI4 data movers (\code{tfa\_dma\_rd},
\code{tfa\_dma\_wr}), the configuration block (\code{tfa\_csr}), and the
clocking/parameterization scheme. All engines share a common structural
template (a single per-row finite-state machine that issues read commands
to \code{tfa\_dma\_rd}, performs a fixed numeric transform, and issues write
commands to \code{tfa\_dma\_wr}) and a common completion contract:
\code{done} is asserted only after the write mover reports \code{wr\_idle}
(rule G-09), so no descriptor retires while a $B$ response is still
outstanding. Each engine also exposes a level \code{drain} output asserted
whenever it is quiescing on \code{abort}, a read error, or a write error,
allowing the sequencer to recover deterministically. The GEMM datapath and
the sequencer are covered in Section~\ref{sec:uarch-gemm}; performance
figures are reported in Section~\ref{sec:perf}.

\subsection{Vector Engine: RMSNorm and Elementwise}
\label{sec:vec}

\begin{figure}[t]
\centering
\includegraphics[width=\columnwidth]{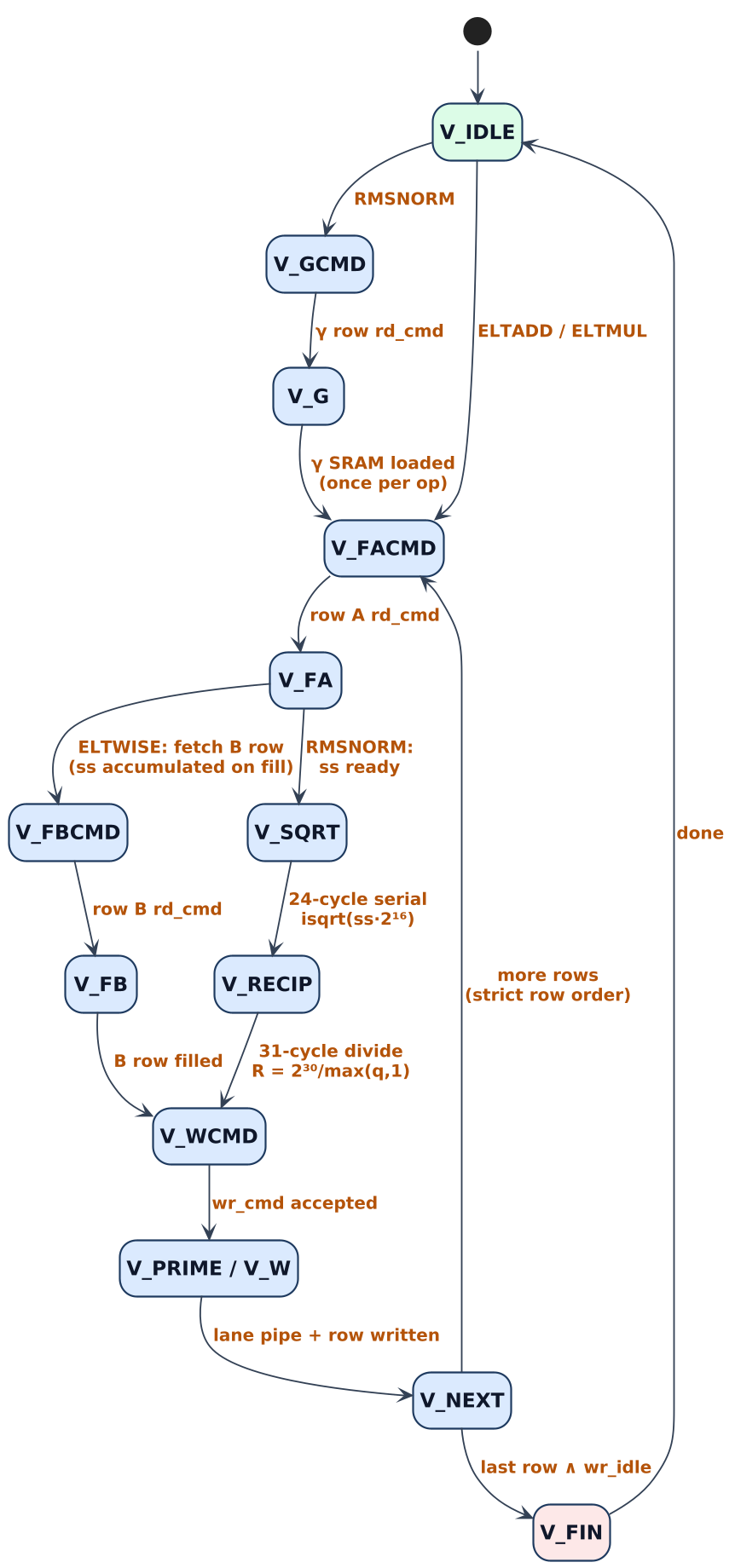}
\caption{\code{tfa\_vec} row FSM. The $\gamma$ row is loaded once
(\code{V\_GCMD}/\code{V\_G}); each output row fills operand buffers, runs the
serial sqrt/reciprocal path for RMSNorm (\code{V\_SQRT}/\code{V\_RECIP}), and
streams the requantized result (\code{V\_W}). Strict read/compute/write
ordering enables in-place $\code{dst}=\code{src}$.}
\label{fig:vecfsm}
\end{figure}

The vector engine \code{tfa\_vec} implements the three row-wise operators of
the residual stream: RMSNorm and the two \code{ELTWISE} modes (\code{ADD},
\code{MUL}). Its FSM (Fig.~\ref{fig:vecfsm}) walks one output row at a time
in the order \emph{read row $\to$ compute $\to$ write row}; because the
write of row $i$ is fully retired before the read of row $i{+}1$ begins, an
in-place transform with destination equal to source ($\code{ADDR\_C} =
\code{ADDR\_A}$) is supported with no aliasing hazard (AS~\S12.1), which the
T5 pipeline relies on for the residual update.

RMSNorm is a two-pass operation. The $\gamma$ table is fetched once per
descriptor into buffer~$B$. For each row, the fill pass (\code{V\_FA})
streams the INT8 input into buffer~$A$ while accumulating the sum of squares
on the fly: each DW-wide beat contributes $\sum_l x_l^2$ over its valid
lanes into a u32 accumulator \code{ss}. After the last beat the engine
launches \code{tfa\_isqrt} on $\code{ss}\ll 16$ to obtain
$q=\lfloor\sqrt{\code{ss}\cdot 2^{16}}\rfloor$, then \code{tfa\_serdiv} to
form the reciprocal scale $R=\lfloor 2^{30}/\max(q,1)\rfloor$. The output
pass (\code{V\_W}) then emits $y=\mathrm{sat8}(\mathrm{rnd}((x\cdot\gamma)
\cdot R\cdot M_A,\ \mathrm{SHIFT}))$, with $x$ and $\gamma$ both INT8 and the
intermediate products widened to 64~bits so the requant is bit-exact with
AS~\S7.2. \code{ELTADD} fills two row buffers sequentially and computes
$z=\mathrm{sat8}(\mathrm{rnd}(a\cdot M_A+b\cdot M_B,\ \mathrm{SHIFT}))$;
\code{ELTMUL} optionally maps the first operand through the SiLU table,
$a'=\code{LUT\_A}\,?\,\code{silu\_lut}[a]:a$, before forming
$z=\mathrm{sat8}(\mathrm{rnd}(a'\cdot b\cdot M_A,\ \mathrm{SHIFT}))$,
realizing the SwiGLU gate in a single \code{ELTWISE} descriptor.

The serial square root is a 24-iteration non-restoring root on the 48-bit
shifted radicand, consuming two radicand MSBs per cycle
(Listing~\ref{lst:isqrt}). The reciprocal is a 32-cycle restoring serial
divide. Both units are invoked once per row and their latency is amortized
against the LANES-byte-per-cycle streaming of the fill and write passes.

\begin{lstlisting}[language=SV,caption={Non-restoring square-root iteration
in \code{tfa\_isqrt}: trial-subtract $(\code{root}\ll2)\,|\,1$ against the
shifted remainder, set the next root bit on success.},label={lst:isqrt}]
assign rem_nxt = {rem[23:0], rad[47:46]};
assign trial   = {root, 2'b01};
// ... per cycle while busy:
rad <= {rad[45:0], 2'b00};
if (rem_nxt >= trial) begin
  rem  <= rem_nxt - trial;
  root <= {root[22:0], 1'b1};
end else begin
  rem  <= rem_nxt;
  root <= {root[22:0], 1'b0};
end
it <= it - 5'd1;
if (it == 5'd1) begin busy <= 1'b0; done <= 1'b1; end
\end{lstlisting}

\subsection{Softmax Engine}
\label{sec:smax}

\begin{figure}[t]
\centering
\includegraphics[width=\columnwidth]{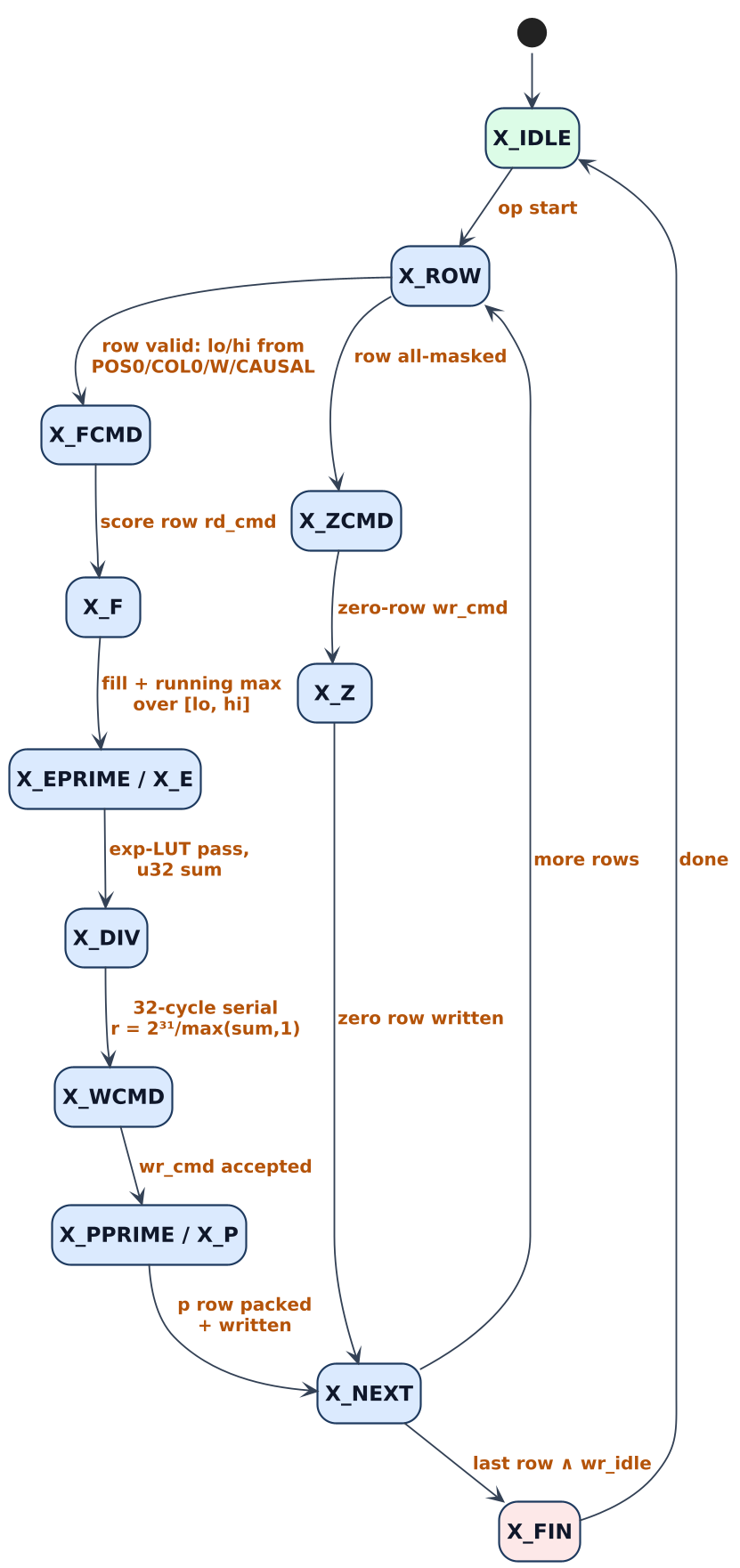}
\caption{\code{tfa\_smax} row FSM. Score fill with running max (\code{X\_F}),
exp-LUT pass with sum accumulation (\code{X\_E}), serial divide
(\code{X\_DIV}), and probability writeback (\code{X\_P}). Fully-masked rows
take the \code{X\_ZCMD}/\code{X\_Z} zero-row short-circuit.}
\label{fig:smaxfsm}
\end{figure}

\code{tfa\_smax} computes a masked row softmax over a sliding window, using a
host-provided 256-entry exponential LUT so the on-chip datapath remains pure
integer. Its FSM (Fig.~\ref{fig:smaxfsm}) processes each row in three passes.
The fill pass (\code{X\_F}) streams the INT8 score row into a buffer while
maintaining a running maximum $m$ over the valid index range $[\code{lo},
\code{hi}]$; the range is derived in signed 18-bit arithmetic from the
descriptor fields \code{POS0}, \code{COL0}, \code{W}, and the \code{CAUSAL}
flag, then clamped to $[0,n{-}1]$. The exponent pass (\code{X\_E}) emits one
u16 word per cycle, $e_j=\code{exp\_lut}[\min(m-s_j,255)]$ inside the window
and $0$ outside, accumulating the u32 sum. A 32-cycle serial divide forms
$r=\lfloor 2^{31}/\max(\code{sum},1)\rfloor$, and the output pass
(\code{X\_P}) writes $p_j=\min(127,(e_j\cdot r)\gg 24)$. Rows whose computed
window is empty (\code{CAUSAL} with negative $\code{hi}$, or
$\code{lo}>\code{hi}$) short-circuit through \code{X\_ZCMD}/\code{X\_Z},
writing a zero row directly without a fill or divide; this both saves
traffic on masked positions and yields well-defined output for fully masked
queries.

\subsection{Copy and Gather}
\label{sec:copy}

\begin{figure}[t]
\centering
\includegraphics[width=\columnwidth]{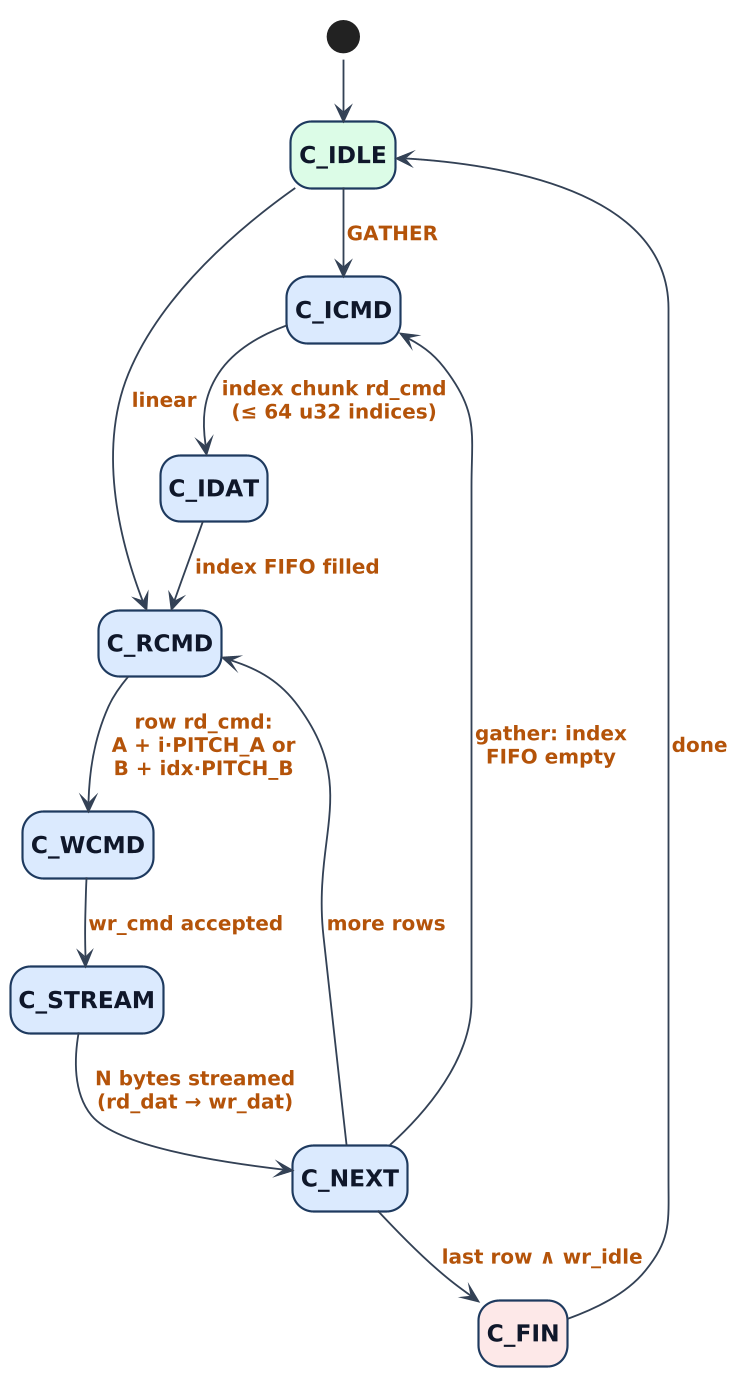}
\caption{\code{tfa\_copy} FSM. Index chunks are prefetched
(\code{C\_ICMD}/\code{C\_IDAT}); per-row source reads
(\code{C\_RCMD}) stream through to the write side (\code{C\_STREAM}).
The index-FIFO write is gated to parse states only.}
\label{fig:copyfsm}
\end{figure}

\code{tfa\_copy} serves both byte-exact linear row copy and index gather
(embedding/KV lookup). Linear copy issues, per row, a read command at
$\code{ADDR\_A}+i\cdot\code{PITCH\_A}$ and a write command at
$\code{ADDR\_C}+i\cdot\code{PITCH\_C}$, passing the packed read stream
straight through to the write stream. Gather (Fig.~\ref{fig:copyfsm}) first
prefetches a chunk of up to 64 u32 indices via a read at
$\code{ADDR\_A}+4i$, parses them serially into an index FIFO, and then reads
each source row from $\code{ADDR\_B}+\code{idx}\cdot\code{PITCH\_B}$ in
40-bit address arithmetic; the index FIFO is refilled in chunks until all
$M$ rows are produced.

Because index data and row data arrive on the same shared \code{rd\_dat}
stream, an in-order command-type discipline demuxes them on last-beat
boundaries: index beats are consumed only in the index-parse states
(\code{C\_IDAT}/\code{C\_RCMD}), and the FIFO write enable is qualified by
both ``a held beat exists'' and ``the current FSM state is a parse state.''
This exact qualification is load-bearing: an earlier revision pushed the
held lane unconditionally, which during \code{C\_WCMD}/\code{C\_STREAM}/
\code{C\_NEXT} flooded the FIFO with duplicate copies of the last parsed
index. The defect surfaced only on an index-FIFO refill beyond 64 indices,
and was caught in constrained-random regression by a $>64$-index gather
sequence.

\begin{lstlisting}[language=SV,caption={Index-FIFO push in \code{tfa\_copy}
qualified by a held beat \emph{and} a parse state: the fix for the
gather duplicate-flood defect.},label={lst:gatherfifo}]
assign ix_wr_valid = ix_beat_v && (ix_lane < ix_cnt)
                   && (st inside {C_IDAT, C_RCMD});
assign ix_wr_data  = ix_beat[32*ix_lane +: 32];
\end{lstlisting}

\subsection{DMA Engines}
\label{sec:dma}

\begin{figure*}[t]
\centering
\includegraphics[width=0.95\textwidth]{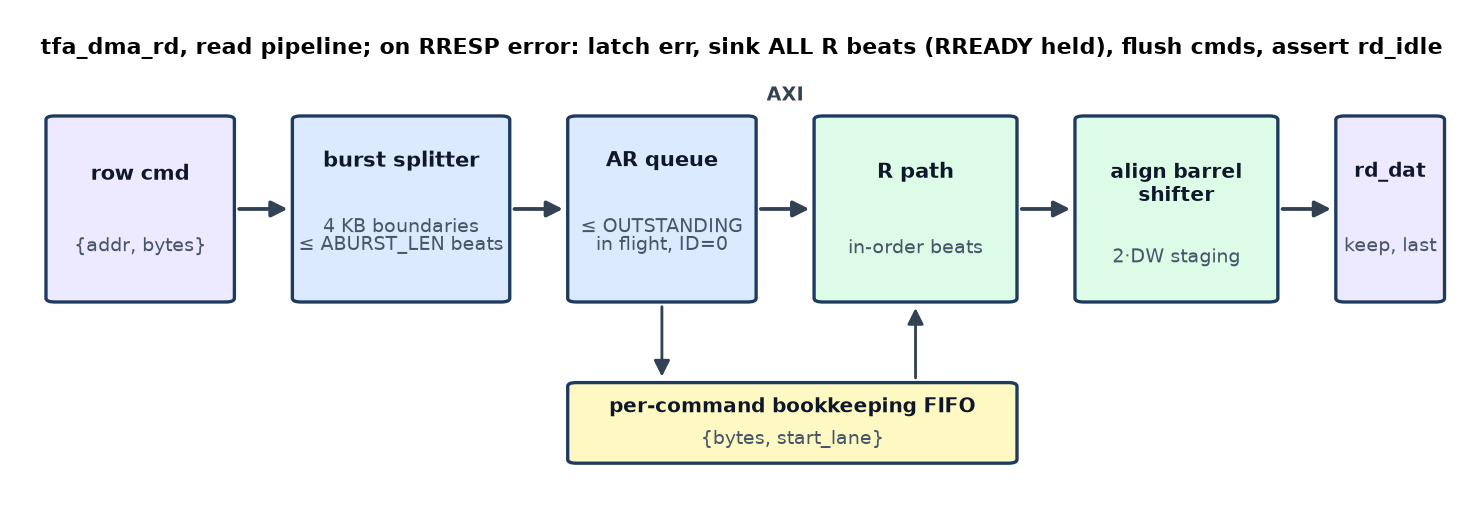}
\caption{\code{tfa\_dma\_rd} datapath: row command $\to$ burst splitter
(4\,KB and \code{ABURST\_LEN}$\cdot$LANES boundaries) $\to$ \code{AR} queue
(up to \code{OUTSTANDING} in flight) $\to$ alignment barrel shifter and
packer producing the contiguous byte stream. Error/abort drives a
sink-and-discard drain.}
\label{fig:dmard}
\end{figure*}

Both data movers convert engine-level row commands $\{\code{addr},
\code{bytes}\}$ into legal AXI4 \code{INCR} bursts and absorb arbitrary byte
alignment, presenting engines a clean packed byte stream. The read mover
\code{tfa\_dma\_rd} (Fig.~\ref{fig:dmard}) splits each command at both 4\,KB
page boundaries and \code{ABURST\_LEN}$\cdot$LANES chunk boundaries, issuing
\code{AR} requests with a single \code{ARID} (in-order) and keeping up to
\code{OUTSTANDING} bursts in flight via an issue/retire credit counter. A
per-burst bookkeeping FIFO records $\{\code{first\_off}, \code{bytes},
\code{last\_of\_cmd}\}$; the $R$ path realigns data through a
$2\!\cdot\!\text{DW}$-byte barrel-shift staging buffer, emitting full
LANES-byte beats with $\code{keep}<\text{LANES}$ only on a command's final
beat. On \code{abort} or any \code{RRESP} error the mover enters a discard
mode: it stops issuing \code{AR}s, holds \code{RREADY} to sink every
in-flight $R$ beat, drops staged data, and only then asserts \code{idle}: a
protocol-clean drain.

\begin{figure}[t]
\centering
\includegraphics[width=\columnwidth]{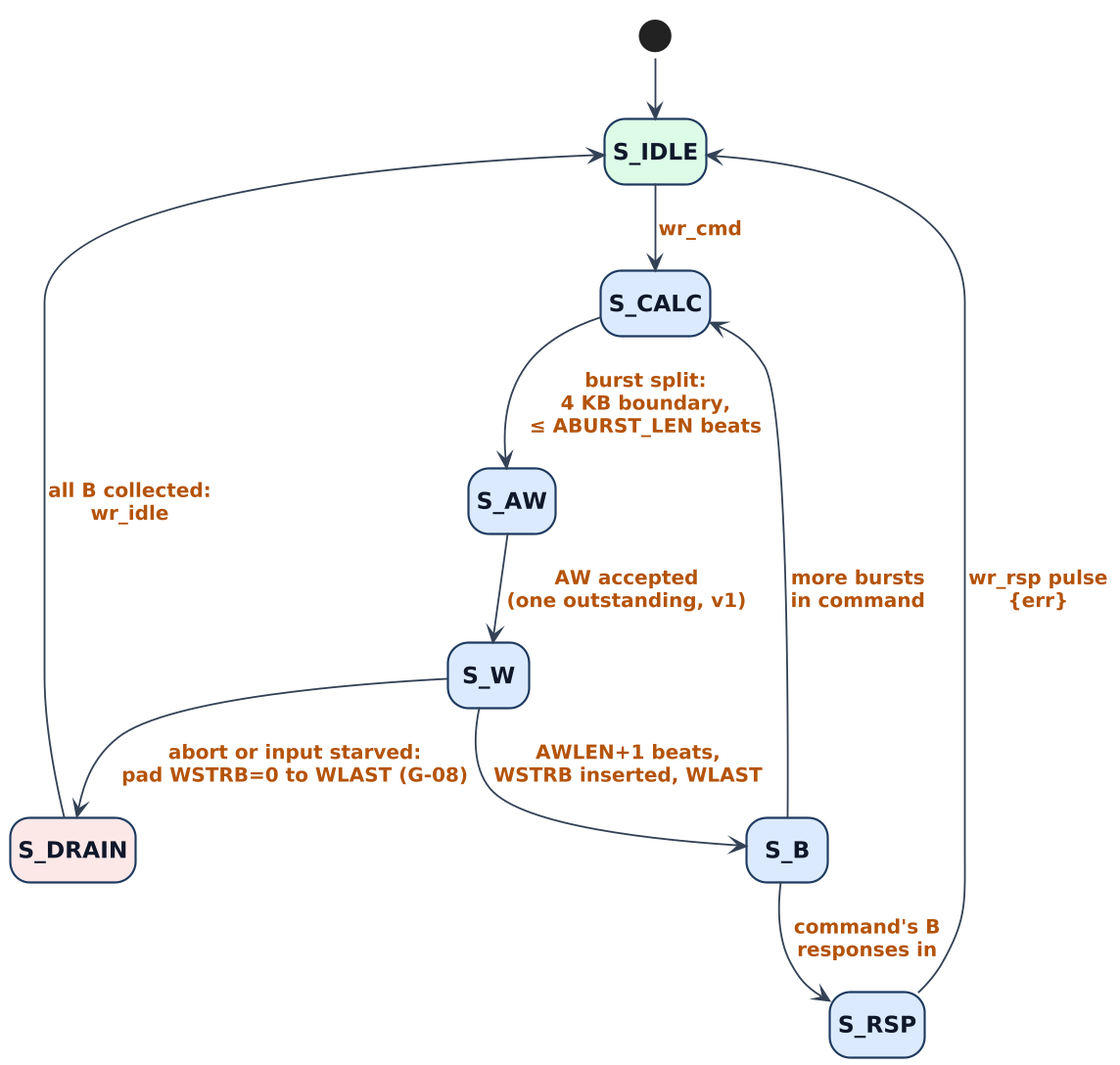}
\caption{\code{tfa\_dma\_wr} FSM. Per burst: compute geometry
(\code{S\_CALC}), issue \code{AW} (\code{S\_AW}), stream $W$ beats
(\code{S\_W}), collect $B$ (\code{S\_B}). The \code{S\_DRAIN} path completes
AW-committed bursts with $\code{WSTRB}=0$ padding and discards staged bytes
on abort.}
\label{fig:dmawrfsm}
\end{figure}

The write mover \code{tfa\_dma\_wr} (Fig.~\ref{fig:dmawrfsm}) mirrors this
structure (burst splitter, alignment staging, \code{WSTRB} insertion for
partial leading/trailing beats) but adds an abort-safe completion rule
(G-08). Once an \code{AW} has been issued for a burst, that burst
\emph{must} be drained on the AXI fabric regardless of whether the upstream
engine still has data. If \code{abort} asserts or the input stream
terminates early, the engine drives the remaining beats of any
AW-committed burst with $\code{WSTRB}=0$ (and correct \code{WLAST}),
cancels un-issued bursts and queued commands, and collects all outstanding
$B$ responses before signaling \code{wr\_idle} (Listing~\ref{lst:pad}). This
guarantees the AXI interconnect never stalls waiting for missing $W$ beats.
The per-beat pad decision is sampled into a registered \code{w\_pad} flag so
\code{WSTRB}/\code{WDATA}/\code{WLAST} remain stable across an asserted
\code{WVALID} even if \code{abort} lands mid-handshake. The contract is
enforced by an SVA stating that every issued \code{AW} is followed by exactly
$\code{AWLEN}{+}1$ $W$ beats containing precisely one \code{WLAST}, including
the abort-padding case:
\begin{verbatim}
w_fire |-> (wlast == (beats_left == 1));
\end{verbatim}
together with a 4\,KB-non-crossing check on every \code{AW}.

\begin{lstlisting}[language=SV,caption={Abort/starvation padding in
\code{tfa\_dma\_wr}: under \code{pad\_mode} all lanes carry zero and
\code{WSTRB} is cleared, while \code{WLAST} still tracks the burst beat
count.},label={lst:pad}]
assign wvalid  = (st == S_W) && beat_avail;
assign wlast   = (beats_left == 9'd1);
always_comb begin
  for (int l = 0; l < LANES; l++)
    wdata[8*l +: 8] = (l >= beat_off &&
        (l - beat_off) < beat_cnt) ? stage[...] : 8'h00;
  wstrb = '0;
  if (!pad_mode)
    for (int l = 0; l < LANES; l++)
      if (l >= beat_off && (l - beat_off) < beat_cnt)
        wstrb[l] = 1'b1;          // real bytes only
end
// w_pad sampled per beat so payload is stable under late abort
if (!wvalid || w_fire) w_pad <= aborted || abort;
\end{lstlisting}

\subsection{CSR, IRAM, and LUTs}
\label{sec:csr}

\code{tfa\_csr} is the AXI4-Lite slave: 32-bit data, 20-bit address, single
outstanding per direction, always responding \code{OKAY}. The 20-bit decode
places registers below \code{0x100}, the exponential and SiLU LUT windows at
\code{0x04000}/\code{0x04400}, and the IRAM descriptor window at
$\code{0x10000}+64i+4w$. The IRAM is implemented as 16 parallel
$32\,\text{b}\times\code{IRAM\_DEPTH}$ simple-dual-port RAMs so the sequencer
reads an entire 512-bit descriptor in one cycle; CSR readback of IRAM shares
that single port at lower priority than an active \code{FETCH}, taking a
bounded one-cycle wait state that the AXI-Lite channel absorbs naturally.
\code{WSTRB} is honored per byte lane on every writable register and window
(G-11).

Control semantics follow AS~\S9. \code{CTRL} bits are write-one-to-pulse:
\code{START} is accepted only when not busy and clears \code{DONE},
\code{ERR}, \code{ERR\_INFO}, and the corresponding IRQ status bits; a
combined \code{START}{+}\code{ABORT} write resolves to \code{ABORT} only;
\code{CTRL} always reads back zero. \code{IRQ\_STAT} is write-one-to-clear,
but a same-cycle hardware set wins over the clear (G-13), so a completion
interrupt is never lost to a racing acknowledgment. Four 32-bit free-running
performance counters, \code{PERF\_CYC} (busy cycles), \code{PERF\_GEMM}
(MAC-array \code{RUN} cycles), and \code{PERF\_RD}/\code{PERF\_WR} (AXI $R$/$W$
beats), are incremented from engine event strobes and zeroed by
\code{CNT\_CLR}, which takes priority over same-cycle increments. These
counters drive the measured utilization figures of
Section~\ref{sec:perf}. The two LUTs live in flop arrays read
combinationally by all lanes of \code{tfa\_vec} and \code{tfa\_smax}; CSR
readback zero-extends.

\subsection{Clocking, Reset, Parameterization}
\label{sec:clk}

The IP is fully synchronous to a single clock. Every state element uses
\code{always\_ff @(posedge clk or negedge rst\_n)} with asynchronous-assert,
synchronous-deassert reset on all control state (SRAM and LUT contents are
not reset). The design contains no latches and instantiates no RTL clock
gates; synthesis is expected to infer integrated clock gating from the
register enables. All module-boundary AXI outputs are registered, and each
FSM carries a default recovery arm. The RTL is a synthesizable SV-2017
subset with no \code{casex}, structs typed in \code{tfa\_pkg}.

A single parameter set scales the whole hierarchy across three design
points. The configurable knobs are the MAC tile $T_R\!\times\!T_C$,
\code{K\_TILE}, the AXI data width DW (hence LANES${}=\text{DW}/8$), address
width AW, \code{ROW\_MAX}, \code{IRAM\_DEPTH}, \code{ABURST\_LEN}, and
\code{OUTSTANDING}. The SIM point uses an $8\!\times\!8$ array with DW${}=64$
and \code{AW}${}=32$; EDGE scales to $64\!\times\!64$ with DW${}=256$; PERF
to $128\!\times\!128$ with DW${}=512$ and \code{AW}${}=40$. A set of
elaboration-time legality assertions (MAS~\S1.1, mirroring AS~\S8.3) reject
illegal combinations at build time (requiring, for example, $T_R=T_C$ a
power of two, $\code{K\_TILE}\cdot 8 \bmod \text{DW}=0$, $\code{ABURST\_LEN}
\cdot\text{LANES}\le 4096$, and $\code{OUTSTANDING}\ge 2$) so that any
elaborable configuration is structurally valid by construction.

\section{Dataflow and Performance Model}
\label{sec:dataflow}

This section develops the dataflow that the \code{tfa\_gemm} engine
implements and the analytic performance model it admits. The treatment is
deliberately structural: it explains why autoregressive, batch-1 generation
(for example the decoder of a translation model emitting one token at a time)
is memory-bound at every silicon configuration, and how the operand
ping-pong organization recovers near-roofline AXI utilization despite a
single shared datapath. Measured rates and energy are reported separately
in Sec.~\ref{sec:results}; the engine finite-state machines are described
in Sec.~\ref{sec:gemm}.

\subsection{Output-Stationary GEMM Tiling}
\label{sec:dataflow-tiling}

\begin{figure*}[t]
  \centering
  \includegraphics[width=0.95\textwidth]{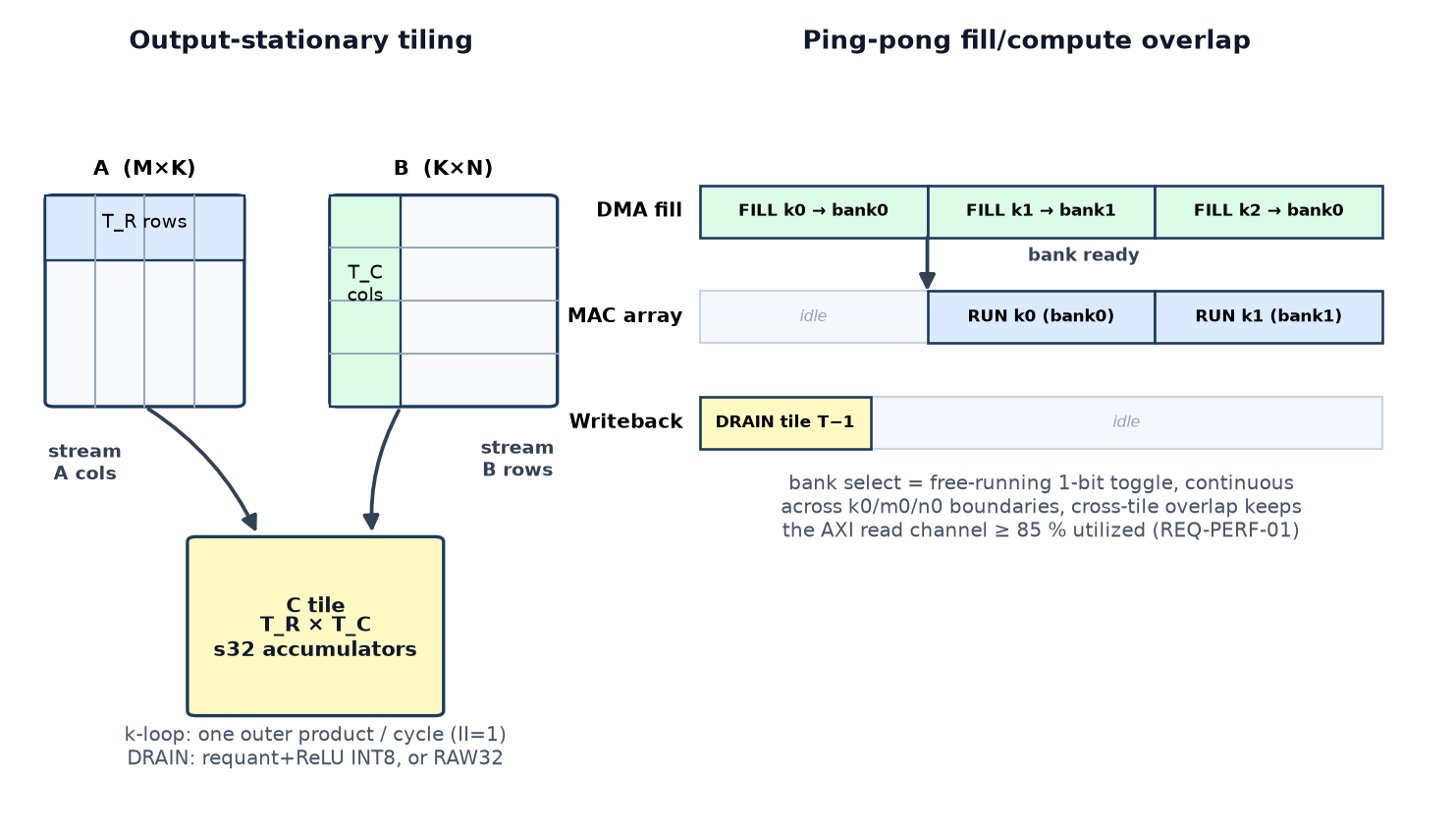}
  \caption{Output-stationary GEMM tiling. The $T_R\times T_C$ accumulator
  tile is held in the MAC array; the $A$ sub-tile
  ($T_R\times K_{\mathrm{TILE}}$) and $B$ sub-tile
  ($K_{\mathrm{TILE}}\times T_C$) are streamed from ping-pong scratchpads as
  outer products over the contraction dimension $K$.}
  \label{fig:tiling}
\end{figure*}

TFA computes $C = A\,B$ with an output-stationary schedule: a
$T_R\times T_C$ tile of $s32$ accumulators resides in the MAC array, while
$A$ columns and $B$ rows stream in as a sequence of outer products. The
loop nest walks the output grid in $(n_0,m_0)$ steps of $(T_C,T_R)$ and,
for each output tile, accumulates over the contraction dimension in
$K_{\mathrm{TILE}}$ chunks, draining the requantized (or RAW32)
result once the $K$ reduction is complete. Each contraction chunk loads an
$A$ sub-tile of $T_R\times K_{\mathrm{TILE}}$ bytes and a $B$ sub-tile of
$K_{\mathrm{TILE}}\times T_C$ bytes into on-chip scratchpads; the array then
consumes them at one outer product per cycle (initiation interval $\mathrm{II}=1$).
Fig.~\ref{fig:tiling} shows the tile geometry and the operand streams.

\begin{figure*}[t]
  \centering
  \includegraphics[width=0.95\textwidth]{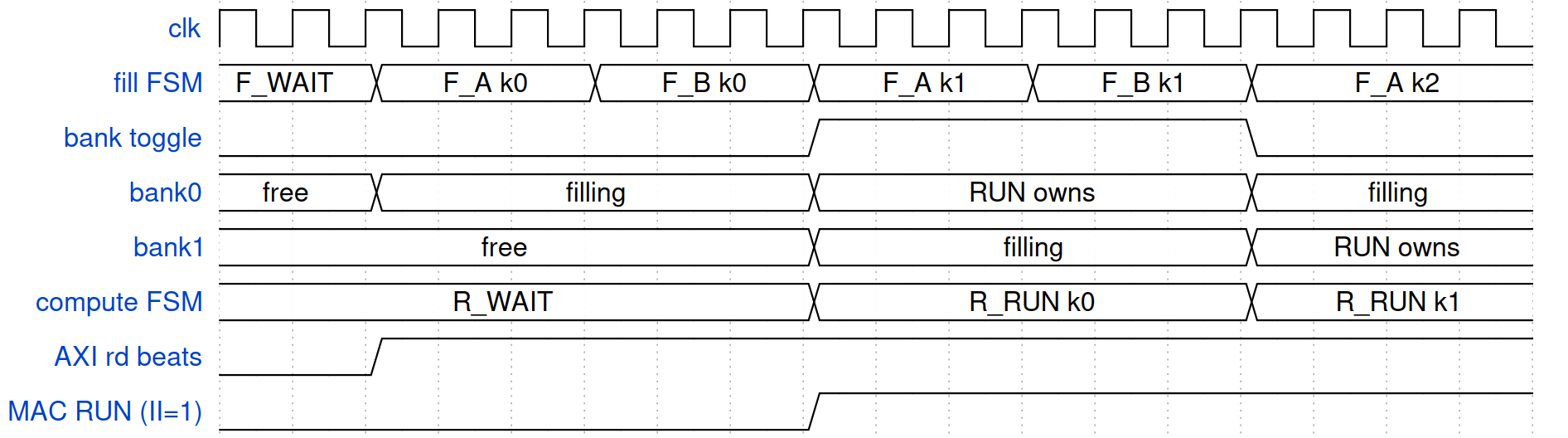}
  \caption{Cross-tile ping-pong overlap. The fill of contraction chunk $n{+}1$
  into the free operand bank runs concurrently with the RUN phase that
  accumulates chunk $n$, so AXI read traffic is continuous across
  $k_0/m_0/n_0$ boundaries. Bank toggling is a single free-running bit; the
  compute pointer follows in fill order.}
  \label{fig:pp}
\end{figure*}

The operand scratchpads are \emph{paired} (A0/A1 and B0/B1), and the bank
select is a free-running one-bit toggle that flips once per completed
contraction-chunk fill. The fill of chunk $n{+}1$ therefore proceeds into the
free bank while the array computes (RUN) over chunk $n$, and this overlap is
continuous across $k_0$, $m_0$ and $n_0$ boundaries: the first fill of output
tile $T{+}1$ is launched into the free bank during the RUN/DRAIN of tile $T$'s
final chunk. The compute pointer consumes banks strictly in fill order from the
same initial value, so no explicit producer/consumer handshake counter is
required. Fig.~\ref{fig:pp} is a waveform of this cross-tile overlap: the
DMA-driven fill of contraction chunk $n{+}1$ is concurrent with the RUN phase
that drains chunk $n$, hiding fill latency behind compute whenever the two
durations are comparable.

\subsection{GEMM Bus-Utilization Model}
\label{sec:dataflow-util}

Because the array sustains $\mathrm{II}=1$ once an operand tile is resident, the
throughput-limiting resource per contraction chunk is the larger of the fill
time (AXI read beats) and the run time (cycles of accumulation). For a single
chunk the read-channel utilization is
\begin{equation}
  U \;=\; \frac{\text{fill beats}}{\max(\text{fill},\ \text{run})}
  \;+\; \varepsilon_{\text{epi}},
  \label{eq:util}
\end{equation}
where $\varepsilon_{\text{epi}}$ is a small epilogue term for tile draining and
for the first (unoverlapped) fill. During single-token (autoregressive) steps the
activation dimension collapses to $M=1$: the $A$ stream is a single row, so the
$B$ weight stream dominates the fill, and the schedule is essentially weight
streaming with the accumulator tile spanning $T_C$ output columns. The cross-tile
overlap of Sec.~\ref{sec:dataflow-tiling} is precisely what makes \eqref{eq:util}
hold for GEMMs whose contraction fits in a single chunk ($K\le K_{\mathrm{TILE}}$),
which is the common single-token case after KV-cache slicing.

The structural ceiling depends on how $T_C$ relates to the AXI lane count
$\mathrm{LANES}=\mathrm{DW}/8$. At the EDGE and PERF geometries
$T_C \ge 2\cdot\mathrm{LANES}$, so the $B$ fill of one chunk takes more cycles
than its RUN, the read channel is the bottleneck, and \eqref{eq:util} approaches
$\approx 95\%$ (EDGE) and $\approx 93\%$ (PERF). At the SIM geometry
$T_C=\mathrm{LANES}$, which forces fill $=$ run for the $B$ stream; the channel
is then idle for roughly the duration of every RUN that is not overlapped at the
chunk granularity, capping utilization at $\approx 80\%$. This SIM ceiling is
informative, not a requirement; the $\ge 85\%$ utilization requirement applies
only to the product geometries.

\subsection{Attention as Address Computation}
\label{sec:dataflow-attn}

Attention of any form, self- or cross-attention, bidirectional or causal,
maps onto the same GEMM and softmax engines without a dedicated datapath; the
variant is selected purely by descriptor address arithmetic and softmax
parameters. The supported attention options include full (bidirectional) and
causal masking, optional sliding-window masking, and grouped-query (GQA) head
sharing; all are available but none is wired into the silicon as a fixed mode.
Grouped-query head sharing, for instance, is realized entirely by descriptor
address arithmetic: query head $h$ occupies the column slice $h\cdot d_h$ of the
$S\times(H_q\cdot d_h)$ activation; it reads KV head
$h_{\mathit{kv}}=\lfloor h\cdot H_{\mathit{kv}}/H_q\rfloor$ by offsetting
\code{ADDR\_B} to slice $h_{\mathit{kv}}\cdot d_h$ of the per-layer KV rings
(with $H_{\mathit{kv}}=H_q$ recovering ordinary multi-head attention). The score
GEMM uses a transposed-$B$, ring-addressed read of the $K$ ring; the
$P\!\cdot\!V$ GEMM uses a normal-layout, ring-addressed read of the $V$ ring; in
both the residual stream $x$ and head concatenation are pure memory layout.
Cross-attention is the same schedule with \code{ADDR\_B} pointing at the
encoder-produced KV instead of the per-layer ring. Sliding-window and causal
masking are not handled in the GEMM at all: the score GEMM computes the full
row and the masking is applied bit-exactly inside softmax from the
host-supplied $(\textit{POS0},\textit{COL0})$ pair and window $W$
(Sec.~\ref{sec:engines}); leaving the window unset yields unmasked
(bidirectional) attention. Out-of-window score computation is not skipped in
v1; compute-skip for fully masked tiles is a documented v1.1 item.

\subsection{Roofline and the Memory-Bound Regime}
\label{sec:dataflow-roofline}

Batch-1 autoregressive generation reads the entire resident weight set once per
token, so token throughput is set by effective read bandwidth rather than by
peak compute:
\begin{equation}
  \text{tok/s} \;=\; \frac{\mathrm{BW}_{\mathrm{eff}}}{B_{\mathrm{token}}},
  \label{eq:roofline}
\end{equation}
where $B_{\mathrm{token}}$ is the bytes read per token. With INT8 weights the
arithmetic intensity of single-token generation is approximately one MAC per
byte read: each weight byte is multiplied into a single $M=1$ activation and
never reused within the token. Comparing this to the per-configuration balance
point (MACs per cycle divided by read bytes per cycle), the generation step
falls below the ridge for SIM, EDGE and PERF alike, so \eqref{eq:roofline}
governs and adding MACs does not move the rate. The same argument extends to
larger, billion-parameter-class transformers: their resident weight sets are
larger, but each autoregressive token still streams them once, so they sit
deeper in the same memory-bound regime. This is why TFA states its performance requirements as efficiency
against the roofline rather than as absolute throughput. Table~\ref{tab:configs}
lists the three reference configurations and their peak compute, MAC count and
on-chip SRAM.

\begin{table}[t]
  \centering
  \caption{Reference silicon configurations (pre-synthesis, informative;
  rates at the 1\,GHz design point).}
  \label{tab:configs}
  \small
  \begin{tabularx}{\columnwidth}{@{}lXXX@{}}
    \toprule
    Parameter & SIM & EDGE & PERF \\
    \midrule
    $T_R\times T_C$        & $8\times8$ & $64\times64$ & $128\times128$ \\
    $K_{\mathrm{TILE}}$    & 64   & 1024 & 4096 \\
    DW (bits)              & 64   & 256  & 512  \\
    AW (bits)              & 32   & 32   & 40   \\
    OUTSTANDING            & 4    & 8    & 16   \\
    INT8 MACs              & 64   & 4096 & 16384 \\
    Peak TOPS @1\,GHz      & 0.13 & 8.2  & 32.8 \\
    SRAM total             & $\sim$28\,KB & $\sim$1.4\,MB & $\sim$7.3\,MB \\
    Read-util.\ ceiling    & $\sim$80\% & $\sim$95\% & $\sim$93\% \\
    \bottomrule
  \end{tabularx}
\end{table}

\subsection{Prefill Limitation and the v1.1 Path}
\label{sec:dataflow-prefill}

The streaming dataflow above is bandwidth-efficient for single-token
(autoregressive) steps but arithmetic-intensity-limited for prefill, the
parallel pass over a full input context (encoder prompts or the decoder's
priming sequence). With the v1 schedule, $A$ and $B$
sub-tiles are streamed per output tile without inter-tile weight reuse, so the
prefill arithmetic intensity is
\begin{equation}
  I \;=\; \frac{T_R\cdot T_C}{T_R + T_C}\ \text{MAC/byte},
  \label{eq:intensity}
\end{equation}
which lies below the array's MAC-per-bus-byte balance at every configuration;
prefill GEMM is therefore bus-bound rather than compute-bound even when
$M\gg1$. The mitigation is structural and does not touch the ISA or numerics: a
weight-stationary (B-stationary) loop order that holds each $B$ tile resident
across a strip of output columns, combined with a C-strip accumulator buffer,
amortizes the weight fetch over many activation rows and lifts $I$ above the
ridge. This B-stationary $+$ C-strip path is the documented v1.1 route to
compute-bound prefill; the v1 RTL is correct and roofline-honest as specified,
and the prefill limitation is an efficiency, not a functional, gap.

\section{Verification Methodology}
\label{sec:verif}

TFA is verified at the SIM configuration ($T_R{\times}T_C{=}8{\times}8$,
$\mathrm{K\_TILE}{=}64$, $\mathrm{DW}{=}64$, $\mathrm{AW}{=}32$) in a
UVM-1.2~\cite{uvm12} environment on Synopsys VCS W-2024.09. The verification
contract is deliberately black-box for \emph{checking} and white-box for
\emph{coverage}: every result is judged against the architectural numerics of
AS~\textsection7 and the AXI4~\cite{arm-axi} protocol, while code coverage and
assertions reach into the DUT hierarchy. Exit criteria are functional
correctness (bit-exact per-op behavior), protocol correctness (no hangs, no
orphan beats under backpressure, error and abort), robustness of decode
validation and reset, and coverage closure. The methodology rests on four
pillars developed below: an active/reactive agent set, a reconstruct-from-bus
bit-exact scoreboard, a constraint-solver stimulus library, and an
exclusion-audited coverage flow. Their value is then argued empirically from
the bugs they caught.

\subsection{UVM Environment}
\label{sec:verif:env}

\begin{figure*}[t]
  \centering
  \includegraphics[width=0.95\textwidth]{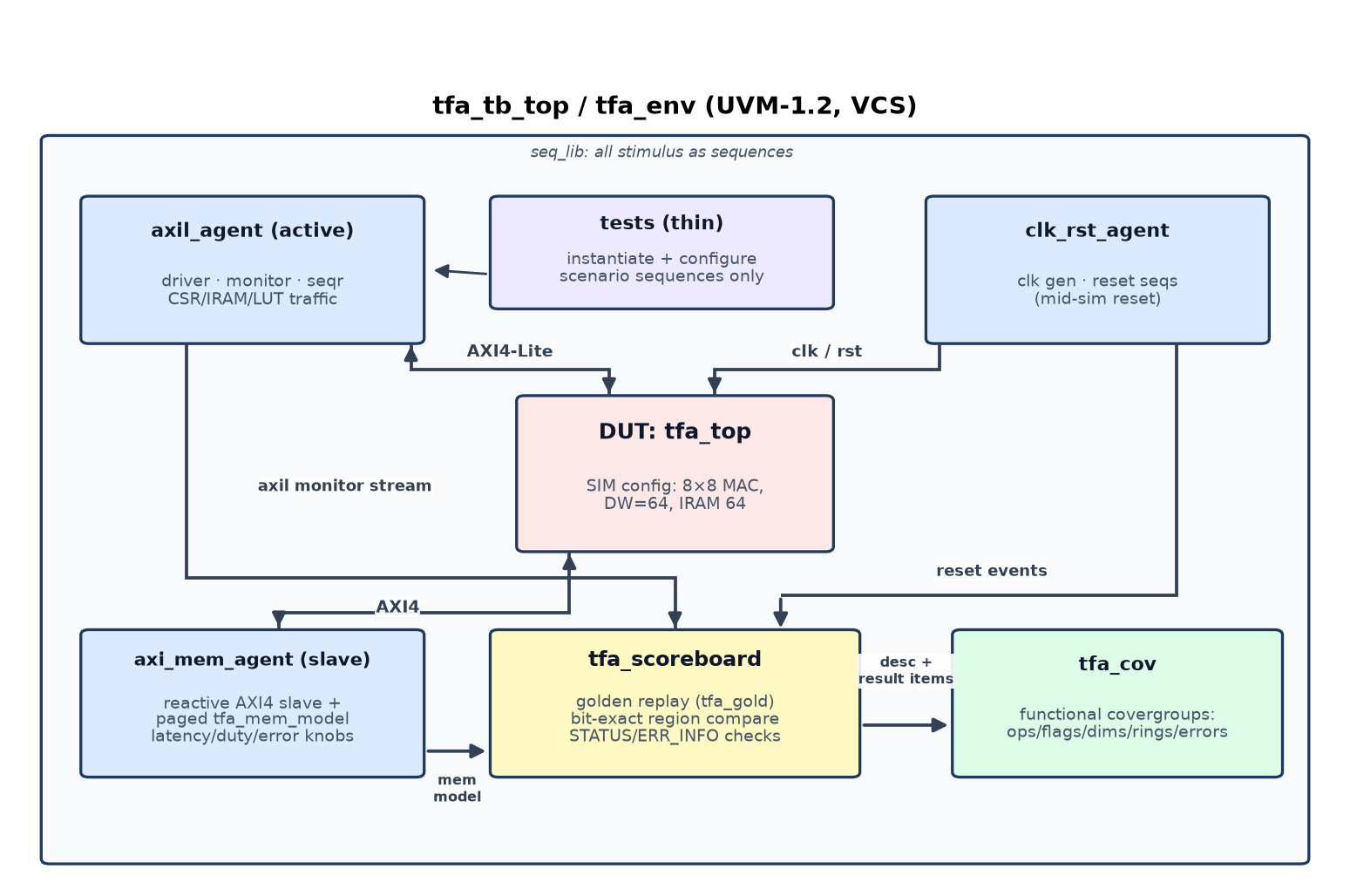}
  \caption{UVM-1.2 testbench architecture. The \code{axil\_agent} is the only
  active master (CSR/IRAM/LUT); the \code{axi\_mem\_agent} is a reactive
  AXI4 slave backed by a paged sparse memory model with randomized
  latency, \code{READY}-duty throttling, and \code{SLVERR}/\code{DECERR}
  injection; the \code{clk\_rst\_agent} drives clock and mid-simulation
  reset. The scoreboard reconstructs device state from the AXI-Lite
  monitor stream alone and checks every output byte against the golden
  model.}
  \label{fig:tb}
\end{figure*}

The environment (Fig.~\ref{fig:tb}) wraps the DUT in three agents, a
scoreboard carrying the golden model, and a functional-coverage subscriber.
The \code{axil\_agent} is the only active master: a driver/monitor/sequencer
triple that drives every CSR, IRAM-window, and exp/SiLU LUT-window access on
the AXI4-Lite slave port, with explicit WSTRB control and both back-to-back
and delayed traffic. The \code{axi\_mem\_agent} is reactive: it models the
single AXI4 master's view of external memory as a \emph{paged sparse} store
(4\,KB pages) so that the same model scales from kilobyte directed tests to
the multi-megabyte application image without allocating dense DRAM. Crucially,
this slave is the principal source of adversarial timing: it randomizes
\code{AR}/\code{R}/\code{B} latencies independently, throttles channel
\code{READY} by a programmable duty cycle to inject backpressure, and injects
\code{SLVERR}/\code{DECERR} responses by address window or transaction count.
The randomized \code{B}-channel latency is not cosmetic: it is what
exercises the AS~\textsection12 rule that an engine may signal \code{done}
only after all write responses are collected (the G-09 read-after-write race).
The \code{clk\_rst\_agent} generates the core clock and issues
reset sequences, including mid-simulation reset pulses at programmable offsets
that drive the recovery sweeps of Sec.~\ref{sec:verif:cov}. Protocol legality
is enforced by 30 SystemVerilog assertions~\cite{ieee1800} bound into the DUT
and the bus interface: handshake stability (\code{VALID} held and payload
stable until \code{READY}), the 4\,KB boundary rule, burst accounting (every
\code{AW} followed by exactly $\mathrm{AWLEN}{+}1$ \code{W} beats with one
\code{WLAST}, \emph{including abort padding}), FSM safety, the
accumulator-never-overflows invariant, and operand-buffer bank-conflict
absence.

\subsection{Bit-Exact Scoreboard}
\label{sec:verif:sb}

\begin{figure*}[t]
  \centering
  \includegraphics[width=0.95\textwidth]{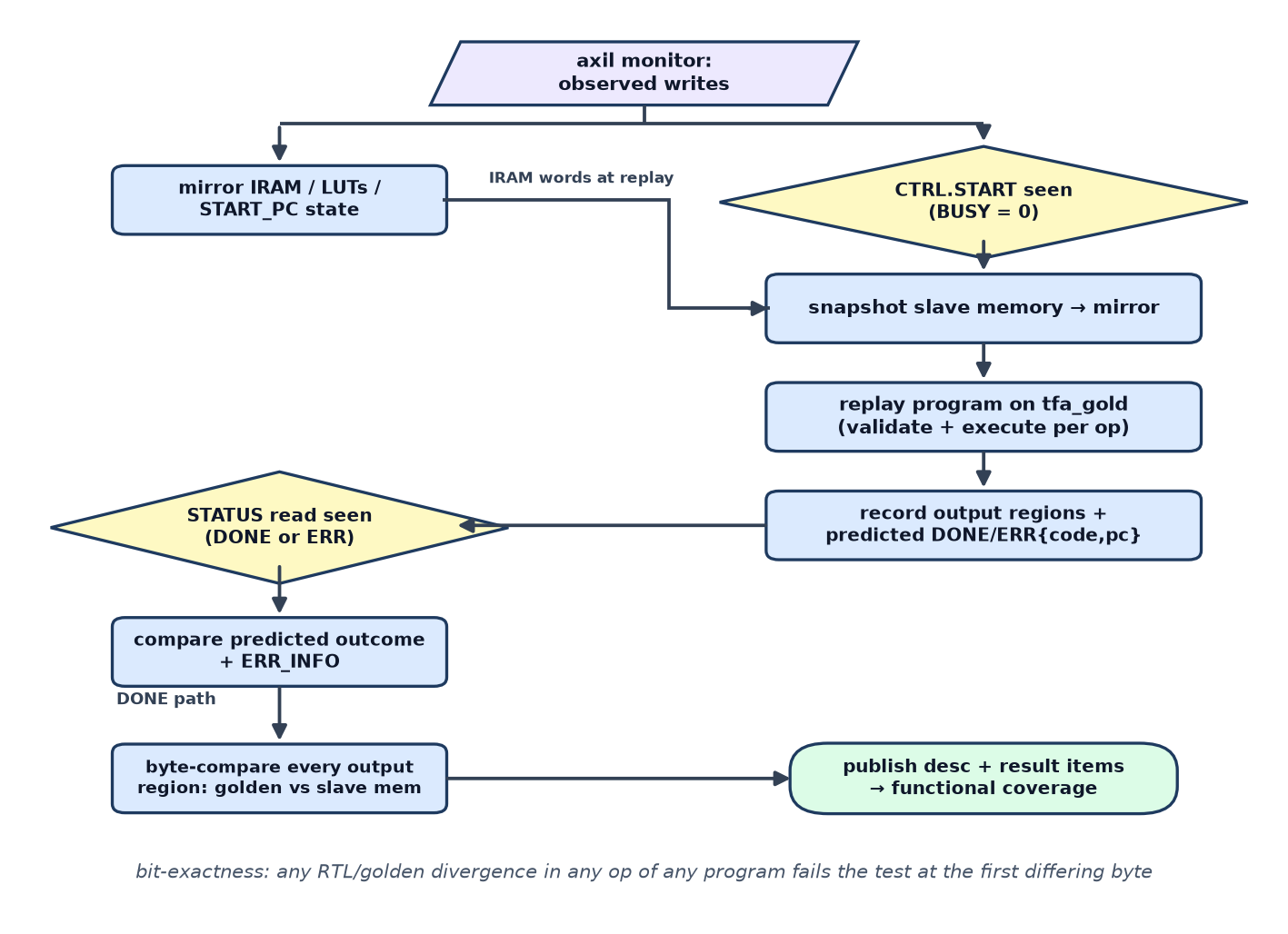}
  \caption{Bit-exact scoreboard dataflow. State is reconstructed purely from
  the AXI-Lite monitor stream; on each observed \code{START} the slave memory
  is snapshotted and the program is replayed through the AS~\textsection7
  golden model \code{tfa\_gold}; every output region is byte-compared and
  \code{STATUS}/\code{ERR\_INFO} are predicted. No test-side back-channel
  carries expected data.}
  \label{fig:sb}
\end{figure*}

The scoreboard (Fig.~\ref{fig:sb}) is the primary data checker, and its
defining property is that it has \emph{no test-side back-channel}. It does not
receive the descriptor object the test built, the expected output, or any
hint of intent; it reconstructs the entire device state from the
\emph{observed} AXI-Lite monitor stream (the IRAM descriptor writes, the
exp/SiLU LUT writes, the \code{START\_PC} and \code{CTRL} writes) exactly
as silicon would see them. On each observed \code{START}, it snapshots the
reactive slave's memory into a mirror, then replays the whole program through
\code{tfa\_gold}, a golden model that implements AS~\textsection7 literally:
the same integer arithmetic, the same $\mathrm{rnd}(x,s){=}(x{+}(1{\ll}(s{-}1))){\ggg}s$
rounding and $\mathrm{sat8}(x){=}\mathrm{clip}(x,-128,127)$ saturation order,
the same per-descriptor validate-then-execute sequence. For each descriptor it
records the precise output region and predicts the program outcome: \code{DONE},
or an error \{code, pc\} drawn from the AS~\textsection9.1 reject classes.
When the test polls \code{STATUS}, the scoreboard checks its prediction; on
\code{DONE} it byte-compares \emph{every} output region golden-versus-actual;
\code{ERR\_INFO} reads are checked against the predicted \{code, pc\}. Data
checking is disabled only where the result is architecturally unpredictable
(an in-flight abort, an injected AXI error), and there the expectation is a
test-driven override rather than a relaxation of the model. Because the model
is reconstructed from the bus, a single mechanism simultaneously validates the
numerics, the addressing, and the DMA engine: any corruption in any of the
three surfaces as a byte mismatch. Executed descriptors and outcomes are
published to the coverage subscriber, closing the loop between what was
actually exercised and what is credited.

\subsection{Constrained-Random Stimulus}
\label{sec:verif:stim}

\begin{figure*}[t]
  \centering
  \includegraphics[width=0.95\textwidth]{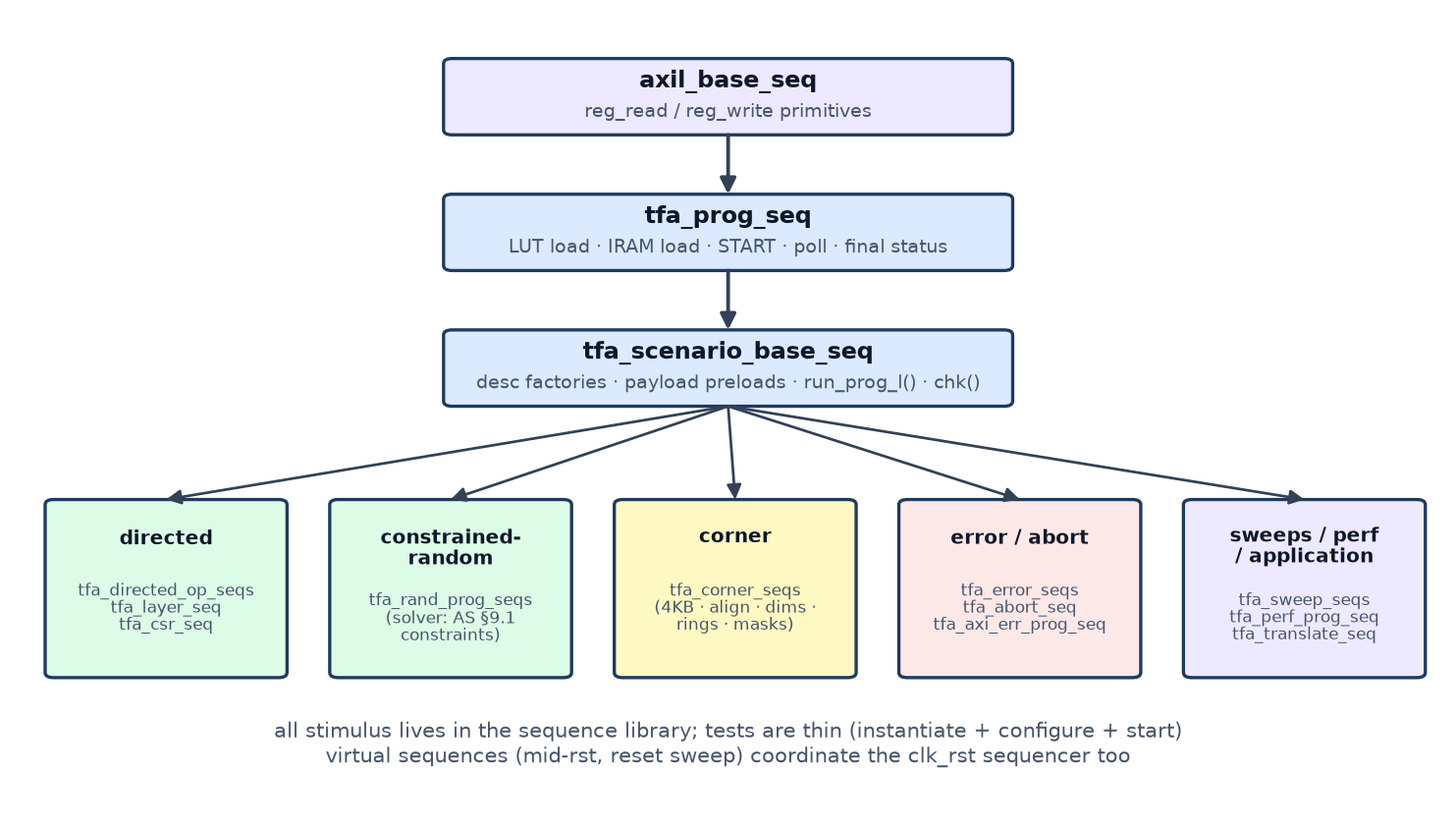}
  \caption{Sequence-library architecture. All stimulus descends from
  \code{tfa\_scenario\_base\_seq}; \code{tfa\_desc} carries the AS~\textsection9.1
  legality constraints so randomized programs are accepted-by-construction,
  while error sequences construct illegal descriptors directly, bypassing the
  solver. Abort/reset sweep virtual sequences drive coverage closure for the
  transient FSM states.}
  \label{fig:seqlib}
\end{figure*}

All stimulus lives in per-agent sequence libraries; tests are thin
configurators that select and parameterize sequences (Fig.~\ref{fig:seqlib}).
The class hierarchy descends \code{axil\_base\_seq} (register primitives)
$\rightarrow$ \code{tfa\_prog\_seq} (LUT/IRAM load, \code{START}, poll, result
capture) $\rightarrow$ \code{tfa\_scenario\_base\_seq} (descriptor factories,
saturation-weighted payload preloads, multi-program orchestration)
$\rightarrow$ the scenario sequences for register, directed-op,
constrained-random, corner, error, abort/race, sweep, and performance classes
(18 sequences in the AXI-Lite library). Randomization is constraint-solver
driven end to end. The descriptor class \code{tfa\_desc} carries the
AS~\textsection9.1 legality predicates as hard SystemVerilog constraints:
per-opcode size bounds, the ring rules ($\mathrm{RING\_LEN}{\ge}1$,
$\mathrm{ROW\_START}{<}\mathrm{RING\_LEN}$, ring-rows ${\le}\mathrm{RING\_LEN}$),
and the per-operand footprint $\mathrm{ADDR}{+}(\text{rows}{-}1)\cdot
\mathrm{PITCH}{+}\text{row\_bytes}\le 2^{\mathrm{AW}}$. So every randomized
program is \emph{accepted-by-construction}: the solver cannot emit a
descriptor the hardware should reject, and a solver failure is itself a
signal. Error stimulus deliberately takes the opposite path: illegal
descriptors are constructed directly, \emph{bypassing} the solver, so that
\code{tfa\_decode\_err\_test} can drive every reject class, including the
41-bit carry attack where a 40-bit modular footprint add would false-accept a
4\,GB scribble. Two virtual sequences coordinate the AXI-Lite and clk/rst
sequencers for coverage closure: \code{tfa\_abort\_sweep\_seq} fires an abort
at single-cycle granularity across every engine FSM window, and
\code{tfa\_reset\_sweep\_vseq} injects reset during both descriptor load and
mid-execution, walking the transient states that no functional program dwells
in long enough to hit by chance.

Listing~\ref{lst:translate} shows the program-driving idiom shared by every
scenario sequence: the LUT and IRAM windows are written word-by-word over
AXI-Lite, \code{START\_PC} and \code{CTRL.START} launch the program, and the
sequence polls \code{STATUS} to completion, the same path the scoreboard
silently observes.

\begin{lstlisting}[language=SV,caption={Program launch and poll idiom from the AXI-Lite sequence library (\code{tfa\_translate\_seq}).},label={lst:translate}]
// load 256x16b exp LUT, then the descriptor program into IRAM
foreach (lut[i]) write_reg(EXP_LUT_BASE + 4*i, lut[i]);
foreach (prog[w]) write_reg(IRAM_BASE   + 4*w, prog[w]);
// launch: point PC at program 0 and pulse START
write_reg(START_PC, 0);
write_reg(CTRL, CTRL_START);
// poll STATUS.BUSY (bit 0) down; scoreboard observes the same reads
do begin
  read_reg(STATUS, st);
end while (st[0]);
if (st[`ERR_BIT]) read_reg(ERR_INFO, ei); // {code, pc}
// ...
\end{lstlisting}

\subsection{Coverage Closure}
\label{sec:verif:cov}

\begin{figure*}[t]
  \centering
  \includegraphics[width=0.95\textwidth]{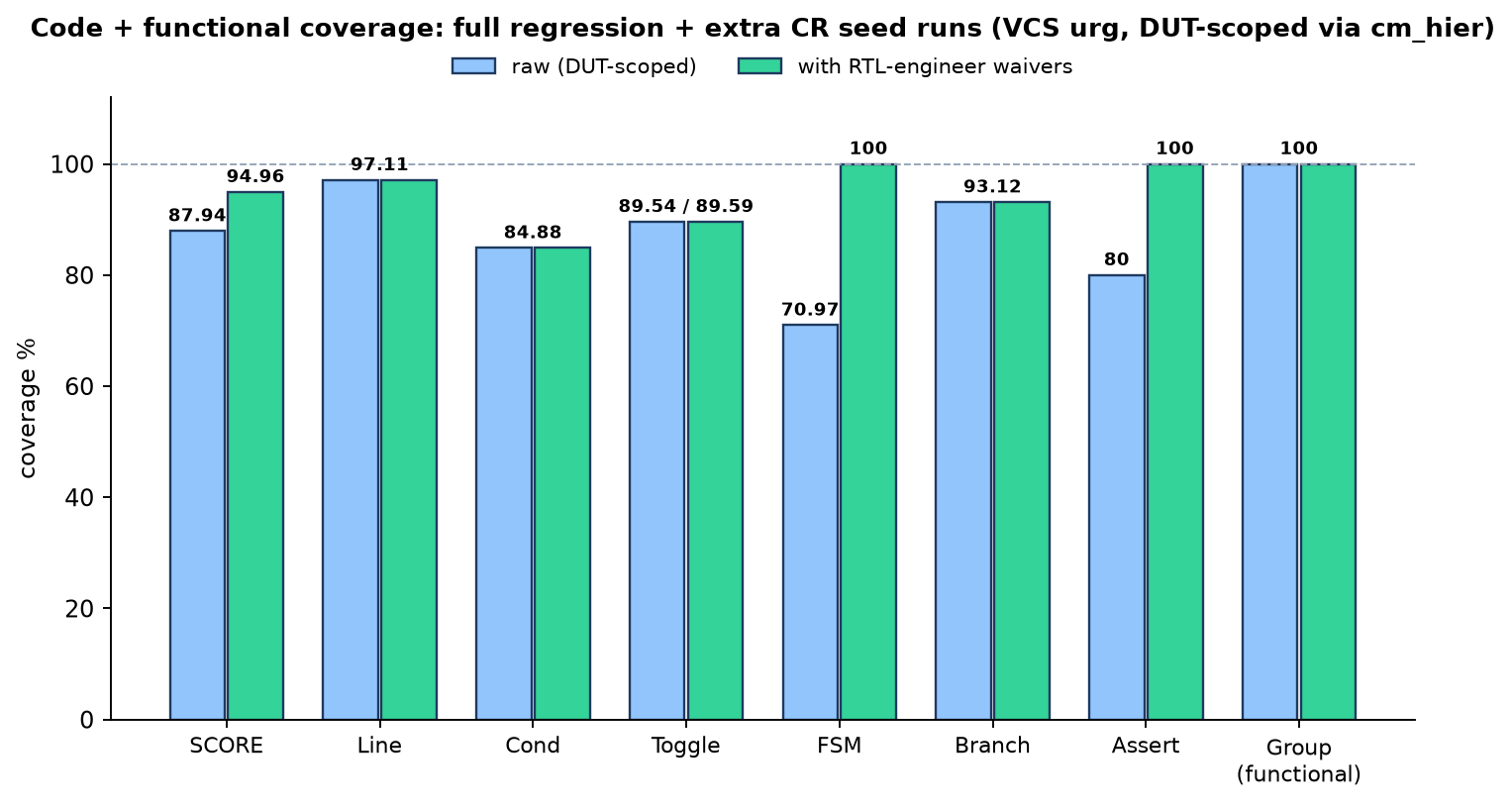}
  \caption{Coverage closure across 34 merged runs (0~\code{UVM\_ERROR}).
  Functional coverage is 100\%; DUT-scoped code coverage rises from 87.94\% raw
  to 94.96\% with engineer-reviewed \code{.el} waivers (FSM 70.97$\rightarrow$100,
  assert 80$\rightarrow$100). Waiver classes (async-reset arcs, 1-cycle abort
  windows, static AXI attributes) are reachability judgments, not hidden
  holes.}
  \label{fig:cov}
\end{figure*}

Closure is measured over 34 merged runs (25 tests at seed~1 plus 9
additional seeds of the three constrained-random stress tests), all passing
with 0~\code{UVM\_ERROR} and 0~\code{UVM\_FATAL} (Fig.~\ref{fig:cov}).
Functional coverage is \textbf{100\%}: covergroups subscribed to the
scoreboard-published descriptors and outcomes hit every opcode$\times$flag
cross (\code{BT}/\code{RELU}/\code{RING\_B}/\code{RING\_C}/\code{RAW32},
\code{MODE}/\code{LUT\_A}, \code{GATHER}, \code{CAUSAL}), every dimension bin,
ring geometry, softmax mask class, requant extreme, error code, abort phase,
and AXI response bin. Code coverage is collected with VCS
\code{line{+}tgl{+}cond{+}branch{+}fsm{+}assert} \emph{scoped to the DUT}
via \code{-cm\_hier} so that testbench and UVM-library code do not dilute the
metric. The raw merged score is 87.94\%; with engineer-reviewed,
tool-native \code{.el} exclusions it is \textbf{94.96\%}, with FSM and assert
both reaching 100\%. The exclusions are audited, not convenient, and fall into
three named classes whose rationale is recorded per item:

\begin{itemize}
  \item \emph{Async-reset arcs (FSM, 54 transitions).} VCS credits a
  $\langle\text{state}\rangle{\rightarrow}\code{IDLE}$ reset arc only when reset
  lands in that exact state in a covered run. The reset sweep covers every
  multi-cycle state; the residual arcs are 1-cycle transient states (e.g.
  \code{*\_CMD}, \code{*\_PRIME}) whose reset-coincidence is not controllable.
  Functional reset behavior in every engine is nonetheless proven by
  \code{tfa\_mid\_rst\_test} and \code{tfa\_reset\_sweep\_test}.
  \item \emph{1-cycle abort windows (FSM).} The same transient states under the
  abort path, with functional abort behavior proven by
  \code{tfa\_abort\_sweep\_test} at single-cycle granularity.
  \item \emph{Static attributes / tied-off bits (toggle, 8 signals;
  assert, 6).} \code{AWSIZE}/\code{AWBURST}/\code{AWCACHE}/\code{AWPROT} are
  architectural constants per AS~\textsection8.1, and address bits $[39{:}32]$
  are tied off at $\mathrm{AW}{=}32$; the 6 assertion waivers are 4 non-DUT
  checks leaking past the \code{cm\_hier} scoping and 2 CSR response-stability
  checks whose antecedent is unreachable because the verification master holds
  \code{READY} high (204k vacuous attempts, 0 failures; the checkers remain
  armed for any backpressuring master).
\end{itemize}

No line, condition, or branch exclusions are taken: that residue is
enumerable defensive code (default arms, parameter-guard terms that cannot
toggle at the SIM parameterization) left deliberately visible. The waivers
narrow the analysis to genuine engineering judgments about reachability rather
than papering over functional holes.

\subsection{Representative Bugs}
\label{sec:verif:bugs}

The methodology earns its keep in the defects it surfaced; the following are
representative of the two simulation-adjacent rounds (full disposition in the
review log). The pre-simulation adversarial RTL review of the DMA path found
\textbf{six} bugs, all functional hangs or protocol violations: an
\code{AWLEN} driven from a never-assigned register (every write burst
advertised 0xFF beats), three distinct abort-path deadlocks in the read and
write engines (a burst generator that never cleared, a stage flush skipped on
a final-burst abort, and queued commands frozen under abort), and an \code{AR}
\code{VALID} that could deassert without \code{ARREADY}, an AXI A3.2.1
violation. Bit-exact simulation against the golden model then exposed
\textbf{eight} more that no lint or directed smoke test would catch:

\begin{itemize}
  \item \emph{Packed-signed sign-extension (MAC array).} Element selects of a
  packed \emph{signed} array are unsigned in SystemVerilog, so the operands
  zero-extended and every negative INT8 product computed wrong, caught by the
  scoreboard observing $+127$ where $-128$ was expected.
  \item \emph{Ping-pong pointer/handshake races (two).} A fill pointer reset
  only by hard reset deadlocked the second GEMM of any program after an
  odd-tile-count op; a one-cycle bookkeeping race cleared the wrong operand
  bank and re-launched the writer for a phantom tile.
  \item \emph{Stale ring write base.} A non-blocking assignment captured the
  ring base one cycle late, so the first \code{RING\_C} tile wrote to a stale
  address.
  \item \emph{Engine-error write-DMA starvation.} A \code{BRESP} error during a
  \code{COPY} quiesced the engine but never asserted the abort/padding path,
  starving the write DMA mid-burst forever, the exact G-08 scenario, fixed by
  exporting an engine \code{drain} into the DMA abort inputs.
  \item \emph{Gather index-FIFO duplicate flood.} A combinational, state-blind
  index-push flooded the FIFO with duplicates outside the parse states. It was
  masked for the first 64 indices and surfaced \emph{only} by the $>$64-index
  gather refill test, a defect invisible to every smaller gather and a direct
  payoff of coverage-driven stimulus.
  \item \emph{Dead FSM state and a TB responder reset-abandon bug.} A
  \code{tfa\_gemm} state with no inbound arc was removed; the reactive slave
  was hardened to abandon in-flight bursts on mid-simulation reset rather than
  leak a stale beat into the next program.
\end{itemize}

Two further testbench defects were found by the reset sweeps (a responder that
kept phantom in-flight responses across a mid-simulation reset), confirming
that the closure sweeps stress the environment as hard as the DUT. The pattern
across both rounds is consistent: the failures cluster at exactly the surfaces
the methodology was built to attack: signed arithmetic checked bit-exactly,
ping-pong overlap and abort exercised at cycle granularity, and rare refill
and reset paths reached only by coverage-targeted stimulus.

\section{Case Study: End-to-End Multilingual Translation and a Quantization Ablation}\label{sec:translate}
\label{sec:casestudy}

The preceding sections establish that the TFA ISA is bit-exact and that the
RTL matches its golden model. Neither result, on its own, demonstrates that
the eight macro-ops are \emph{expressive enough} to run a real pretrained
transformer, nor that the per-tensor INT8 requant contract of
Section~\ref{sec:isa} is numerically adequate for one. This section closes
both questions with a single end-to-end experiment, scaled up here into a
ten-sentence multilingual stress test. A HuggingFace \textsf{t5-small} (60\,M
parameters) is compiled onto the v1 ISA and executed on the RTL on ten tricky,
idiomatic English proverbs, each translated into a \emph{different} target
language. Because the bundled \textsf{t5-small} is the multi-task checkpoint,
French, German, and Romanian are all reachable from the \emph{same} weights,
selected only by the task prefix (\code{translate English to
French/German/Romanian:}); no additional model or retraining is needed. Every
one of the resulting operations is checked bit-exactly against the golden
model, and we report, per sentence, whether the INT8 chip additionally
reproduces the floating-point reference token-for-token. Along the way we
expose why naive per-tensor quantization fails on this model, and that the fix
is an \emph{exact} change of basis in the compiler that needs no ISA or silicon
change. The expanded workload is chosen to stress the datapath far harder than
the original short declaratives: idiomatic inputs broaden the activation range,
and the longest proverb decodes to $21$ tokens.

\subsection{Compiling a Pretrained Transformer onto the ISA}
\label{sec:casestudy:compile}

The offline flow lives in \code{build\_t5.py}. We deliberately choose
\textsf{t5-small}, a complete pretrained encoder--decoder translation model,
as the demonstrator: its full pipeline exercises every transformer block
type, bidirectional encoder self-attention, causal decoder self-attention,
encoder--decoder cross-attention, and the position-wise FFN, which is a strict
superset of what a decoder-only model requires. Each of these constituent
tensor operations, RMSNorm pre-norm, multi-head attention with a learned
relative position bias, the cross-attention $K$/$V$ projection over the encoder
output, and a ReLU FFN, maps onto exactly the same eight macro-ops. Compiling
it whole is therefore the most demanding portability test of the ISA, and a
direct demonstration that the operator set spans general transformer inference
rather than any one model family.

The compiler proceeds in five stages. First, a float \texttt{numpy} greedy
decode produces the golden token stream for each input sentence. Second, a
calibration pass observes the float op graph and records, at $194$ activation
sites, the maximum absolute value used to derive a symmetric max-abs INT8
scale $s=\max|t|/127$. Third (and this is the load-bearing idea), the
compiler does not merely \emph{emit} a descriptor for every tensor op; it also
\emph{executes} that descriptor bit-exactly, against a mirror of the AS
section~\ref{sec:isa} arithmetic (\code{exec\_desc} reproduces
$\mathrm{sat8}(\mathrm{rnd}(\mathrm{acc}\cdot M_A,\,\mathrm{SHIFT}))$, the
RMSNorm \code{isqrt}/reciprocal path, the masked softmax LUT, and the gather
\code{COPY}). Because the compiler holds the device's live memory image, it
knows every intermediate value the chip will produce and selects each next
token by \texttt{argmax} over the chip's \emph{own} RAW32 logits. Fourth, the
emitted descriptors are grouped into IRAM-sized programs; the LM head is a
single \code{GEMM} with the \code{RAW32} flag set, producing $32{,}128$
$32$-bit logits over the full vocabulary (\texttt{argmax} is scale-invariant,
so the $d_{\mathrm{model}}^{-1/2}$ logit scaling is dropped). Decode is
incremental with a KV cache: each step processes a single query row ($M=1$),
appending its $K$/$V$ projections into per-layer cache buffers with the
single-row \code{GEMM} writer \code{gemm\_w\_into}.

\begin{figure*}[t]
  \centering
  \includegraphics[width=0.95\textwidth]{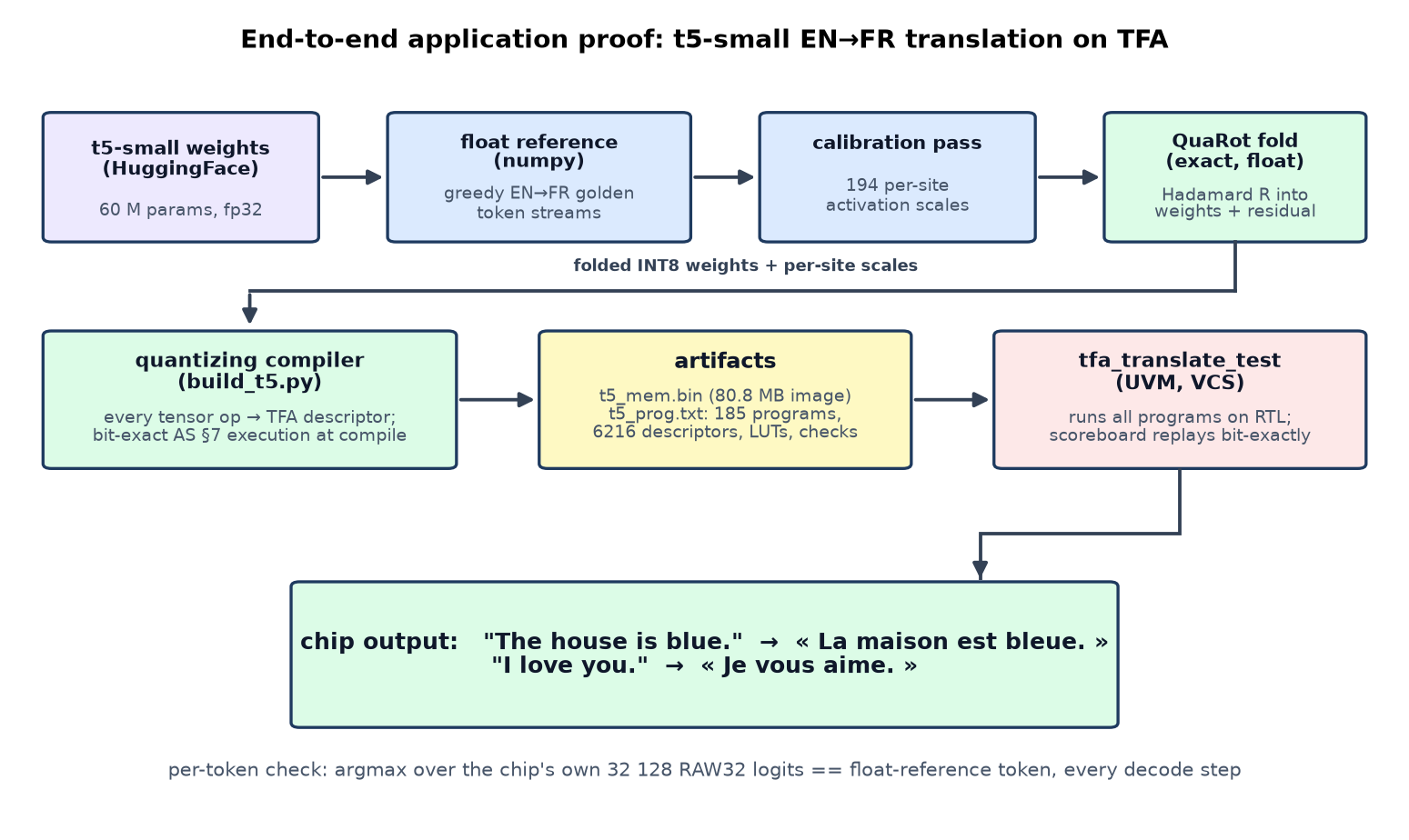}
  \caption{End-to-end translation flow. A pretrained \textsf{t5-small} is
  calibrated and compiled into descriptor programs and a memory image
  ($2{,}095$ programs / $70{,}320$ descriptors / $77.6$\,MB for the ten-sentence
  multilingual workload); each descriptor is executed bit-exactly at compile
  time so the per-step next token is chosen by \texttt{argmax} over the chip's
  own RAW32 LM-head logits. The same image and programs are then replayed on
  the RTL.}
  \label{fig:xlate}
\end{figure*}

For the ten-sentence workload the result is $2{,}095$ descriptor programs,
$70{,}320$ descriptors, and a $77.6$\,MB memory image (\code{t5\_mem.bin})
accompanied by a program script (\code{t5\_prog.txt}) of LUT loads,
descriptors, run directives, and $155$ per-token \texttt{argmax} checks. The
image stays weight-sized despite the ten sentences because the compiler places
the INT8 weights and per-sentence inputs in one arena, serialized into the
preload image, and the chip-computed intermediates (always written before they
are read) in a second arena that is recycled between sentences; the resulting
$77.6$\,MB is in fact smaller than the original two-sentence $80.8$\,MB image.
Fig.~\ref{fig:xlate} traces a decode step through this compiled program.
Because the compiler executes every descriptor bit-exactly \emph{before any RTL
is involved}, it knows the chip's exact token stream a priori, drives the
per-step \texttt{argmax} from the chip's own logits, and records whether that
stream matches the float reference; the compiler is thus itself a complete,
self-checking implementation of the ISA.

\subsection{The Per-Tensor INT8 Failure}
\label{sec:casestudy:fail}

A direct application of the ISA's per-tensor W8A8 contract to
\textsf{t5-small} produces garbage: the decoder emits incoherent tokens and
the \texttt{argmax} stream diverges immediately from the float reference. The
cause is structural, not an implementation defect. T5's residual stream
contains a small number of channels whose magnitude dwarfs the rest: at a mid
encoder layer the per-channel max-abs magnitude reaches $\sim\!9531$ against a
median of $\sim\!72$ ($133\times$), and globally the outliers run up to
$\sim\!24{,}500$ against a median near $124$. A single symmetric per-tensor
scale must be sized to the largest channel, so $s$ is set by the outlier;
every typical channel is then quantized with a step far larger than its own
range and collapses below one LSB. Fig.~\ref{fig:ablation}(a) plots the
sorted per-channel residual magnitude with the per-tensor INT8 step overlaid:
the entire body of the distribution falls beneath the step and is destroyed,
which is fatal because RMSNorm reads the residual to compute its scale.

This is the well-documented activation-outlier problem of large transformers
\cite{dettmers2022llmint8,bondarenko2021understanding,xiao2023smoothquant}.
Crucially, it is a property of the model's representation and not of the
hardware contract: no per-tensor requant pipeline, in TFA or elsewhere, can
recover channels that the chosen scale has already annihilated.

\subsection{An Exact Rotation Fix}
\label{sec:casestudy:rotation}

The fix is to change the basis in which the residual stream is represented so
that the outlier energy is spread across all channels, lowering the
per-tensor step until the body of the distribution is once again resolvable.
TFA adopts a randomized-Hadamard reparameterization in the spirit of QuaRot
and QuIP \cite{ashkboos2024quarot,chee2023quip,tseng2024quip}. A dense
orthogonal rotation $R\in\mathbb{R}^{D\times D}$ is constructed as a Hadamard
matrix with random sign flips, normalized to $1/\sqrt{D}$
(Listing~\ref{lst:quarot}). The residual stream is carried in the rotated
basis $x' = xR$; every weight that \emph{reads} the residual is pre-multiplied
by $R$ on its input side, and every weight that \emph{writes} the residual is
pre-multiplied by $R^{\mathsf{T}}$ on its output side
(\code{(W*g)@R} for projections-in, \code{R.T@W} for projections-out in
\code{\_plant\_weights}).

The decisive property is that this is a \emph{lossless float identity}, not a
lossy approximation. Pre-norm RMSNorm without a bias term is rotation
invariant, $\mathrm{rms}(xR)=\mathrm{rms}(x)$ for orthogonal $R$, so inserting
$R$ and $R^{\mathsf{T}}$ around the residual leaves the float computation
exactly unchanged \cite{zhang2019rmsnorm}. The rotation is therefore folded
entirely into the weights at compile time; the chip is never aware of it. The
same fold absorbs each RMSNorm $\gamma$ vector into its consumer weights
(\code{W*g[None,:]}), so the on-chip $\gamma$ table is a constant $127$ and
its quantization error vanishes exactly.

\begin{lstlisting}[language=Python,caption={Randomized-Hadamard rotation and
the exact weight fold that carries the residual in the rotated basis (from
\code{build\_t5.py}). $R$ is orthogonal, so it is a lossless float identity;
RMSNorm $\gamma$ is folded into consumers, leaving a constant on-chip
table.},label={lst:quarot}]
def make_rotation(D, seed=0x75):
    h = np.array([[1.0]])
    while h.shape[0] < D:               # build D x D Hadamard
        h = np.block([[h, h], [h, -h]])
    signs = np.where(np.random.default_rng(seed)
                     .random(D) < 0.5, -1.0, 1.0)
    return (h * signs[None, :]) / np.sqrt(D)

# in _plant_weights: residual basis rotated by R
#   read residual:  W' = (W * gamma) @ R
#   write residual: W' = R.T @ W
self._plant("lm_head", (emb * g_dfin[None, :]) @ R)
self._plant(f"{base}.0.SelfAttention.o.weight",
            R.T @ w[f"{base}.0.SelfAttention.o.weight"])
\end{lstlisting}

\begin{figure*}[t]
  \centering
  \includegraphics[width=0.95\textwidth]{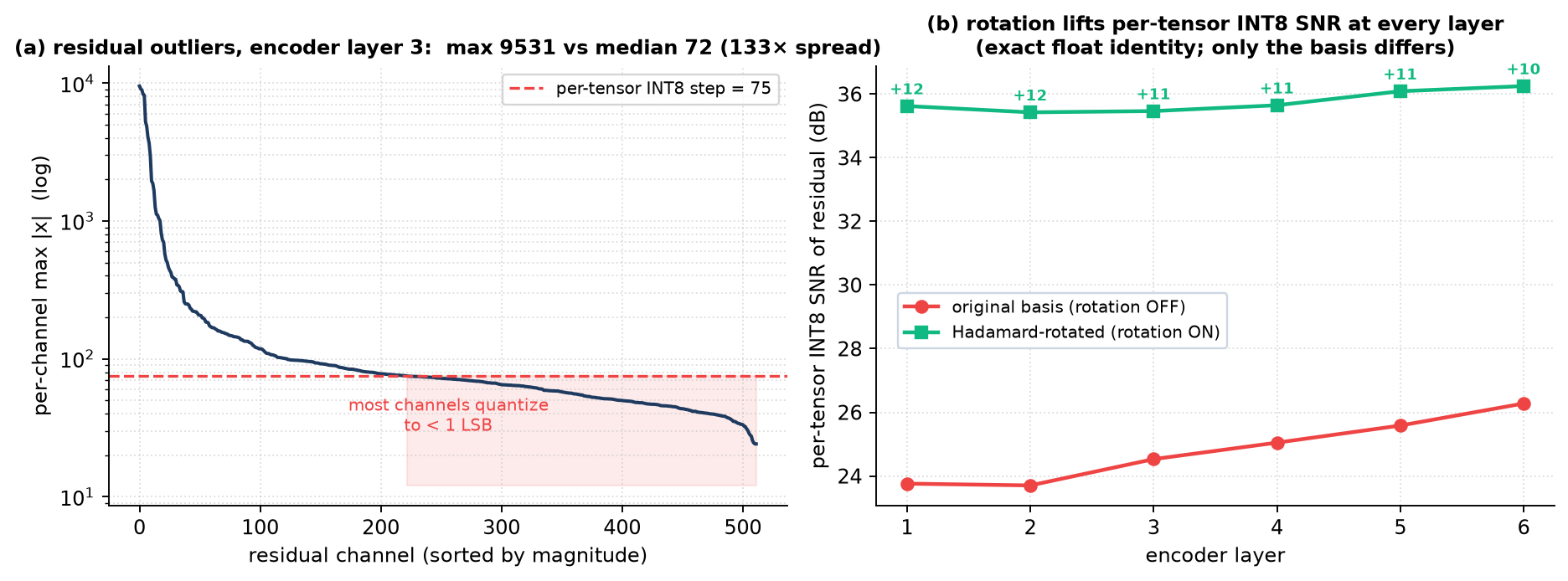}
  \caption{Quantization ablation. (a)~Sorted per-channel max-abs magnitude of
  the float residual at a mid encoder layer (log scale) with the symmetric
  per-tensor INT8 step overlaid; the body of the distribution (outliers up to
  $\sim\!9531$ against a $\sim\!72$ median, $133\times$) is crushed below one
  LSB, which is why naive W8A8 collapses. (b)~Per-tensor INT8 SNR of the
  residual at each encoder-layer output, original basis versus
  randomized-Hadamard rotated basis. The exact (lossless-float) rotation lifts
  the SNR by $\sim\!11$\,dB ($\sim\!24\!\to\!36$\,dB, $\sim\!2$ effective bits)
  at every layer with no ISA or silicon change.}
  \label{fig:ablation}
\end{figure*}

The numerical payoff is quantified in Fig.~\ref{fig:ablation}(b), which
reports the symmetric per-tensor INT8 signal-to-quantization-noise ratio of
the residual at the output of each encoder layer, in the original basis versus
the rotated basis. Rotation lifts the residual SNR from $\sim\!24$\,dB to
$\sim\!36$\,dB at \emph{every} encoder layer, a uniform gain of
$\sim\!11$\,dB, roughly two effective bits of precision, which is precisely
the margin needed to flip the end-to-end output from garbage back to the exact
reference translation. The entire correction lives in the offline compiler;
the RTL, the ISA, and the silicon are untouched.

\subsection{Running on the RTL}
\label{sec:casestudy:rtl}

The compiled artifacts are exercised on the DUT by \code{tfa\_translate\_test}
through the application sequence \code{tfa\_translate\_seq}. The sequence
backdoor-loads the $80.8$\,MB image, then walks \code{t5\_prog.txt}: \texttt{L}
lines drive each program's calibrated EXP LUT through the AXI-Lite LUT window,
\texttt{D} lines stage descriptors into a list, \texttt{R} loads the IRAM,
issues \code{START}, and polls the CSR \code{STATUS} to \code{DONE}, and
\texttt{K} lines invoke the per-token check. \code{do\_argmax\_check}
recomputes the \texttt{argmax} over the chip-written little-endian RAW32
logits and asserts it equals the float-reference token
(Listing~\ref{lst:argmax}); a first-max-wins tie rule matches
\texttt{numpy.argmax}. In parallel, the UVM scoreboard, driven only by the
AXI-Lite monitor stream, independently replays every started program
bit-exactly and byte-compares all device-written outputs against its own
golden execution.

\begin{lstlisting}[language=SV,caption={Per-token check in
\code{tfa\_translate\_seq}: \texttt{argmax} over the chip's RAW32 LM-head
logits must equal the token the bit-exact compiler predicted at every decode
step.},label={lst:argmax}]
for (int unsigned i = 0; i < count; i++) begin
  int v = {mem.read8(addr + 4*i + 3),
           mem.read8(addr + 4*i + 2),
           mem.read8(addr + 4*i + 1),
           mem.read8(addr + 4*i)};   // LE int32
  if (i == 0 || v > best) begin      // first-max-wins
    best = v; bidx = i;
  end
end
chk(bidx == exp_tok, $sformatf(
    "token %0d: chip argmax %0d != expected %0d",
    n_tok, bidx, exp_tok));
\end{lstlisting}

The outcome is summarized in Table~\ref{tab:xlate}. Each of the ten proverbs
was run as its own RTL simulation (a single sentence per run, since one
LM-head \code{GEMM} over the $32{,}128$-entry vocabulary is emitted at every
decode step). Across the ten runs the engine executes the full $2{,}095$
programs, the sequence performs all $155$ per-token \texttt{argmax} checks, and
the scoreboard byte-compares $37{,}905{,}664$ bytes of chip-written output
against its independent golden replay, with \textbf{zero mismatches and zero}
\code{UVM\_ERROR}/\code{UVM\_FATAL}. The bit-exact chain
$\text{RTL}=\text{golden}=\text{compiler}$ therefore holds for every one of the
$70{,}320$ operations, in all three target languages.

A separate, weaker question is whether the INT8 chip reproduces the
\emph{floating-point} reference token-for-token. Five of the ten outputs do
(the ``$=$float'' $\checkmark$ rows: all three German cases plus two Romanian);
the other five (marked $\approx$) differ from the float greedy stream only by a
curly-versus-straight apostrophe (two cases), an added connective (one), or a
single differing word choice (two), and remain valid translations. The
divergences fall on the French outputs, where residual per-tensor INT8 noise
occasionally flips a single near-tie \texttt{argmax} decision early in the
decode and the autoregressive continuation then follows the alternative; all
three German cases, and two of the three Romanian, reproduce the float
reference outright. This is the expected boundary of static per-tensor W8A8 on
long idiomatic generation,
and it does not affect the bit-exact guarantee above: chip-versus-float
agreement is a quantization-quality metric, whereas chip-versus-golden equality
is the correctness guarantee, and the latter is exact for all ten.

\begin{table*}[t]
  \centering
  \caption{End-to-end multilingual RTL translation stress test. Ten idiomatic
  English proverbs compiled onto the ISA and executed on the cycle-accurate
  RTL, one sentence per simulation. Every operation is bit-exact against the
  golden model; the ``$=$float'' column reports whether the INT8 chip
  \emph{additionally} reproduces the floating-point greedy reference
  token-for-token ($\checkmark$) or yields an equally valid alternative
  rendering ($\approx$).}
  \label{tab:xlate}
  \renewcommand{\arraystretch}{1.2}
  \footnotesize
  \begin{tabularx}{\textwidth}{@{}l X X c@{}}
    \toprule
    \textbf{Lang} & \textbf{English proverb} & \textbf{Chip output} & \textbf{$=$float} \\
    \midrule
    FR & Don't count your chickens before they hatch. & \textit{Ne pas compter vos poulets avant qu'ils ne soient \'eclos.} & $\approx$ \\
    FR & The grass is always greener on the other side. & \textit{L'herbe est toujours plus verte de l'autre c\^ot\'e.} & $\approx$ \\
    FR & All that glitters is not gold. & \textit{Tout ce qui glitzera n'est pas l'or.} & $\approx$ \\
    FR & Necessity is the mother of invention. & \textit{La n\'ecessit\'e est la m\`ere de l'invention.} & $\approx$ \\
    DE & The early bird catches the worm. & \textit{Der fr\"uhe Vogel f\"angt den Wurm.} & $\checkmark$ \\
    DE & Where there is a will there is a way. & \textit{Wo es einen Willen gibt, gibt es einen Weg.} & $\checkmark$ \\
    DE & The pen is mightier than the sword. & \textit{Der Stift ist m\"achtiger als das Schwert.} & $\checkmark$ \\
    RO & When in Rome, do as the Romans do. & \textit{C\^and se afl\u{a} la Roma, face\c{t}i a\c{s}a cum fac \c{s}i romanii.} & $\approx$ \\
    RO & Better late than never. & \textit{Mai bine mai t\^arziu dec\^at niciodat\u{a}.} & $\checkmark$ \\
    RO & A friend in need is a friend indeed. & \textit{Un prieten \^in nevoie este \^intr-adev\u{a}r un prieten.} & $\checkmark$ \\
    \midrule
    \multicolumn{4}{@{}p{\textwidth}}{\footnotesize Totals: $2{,}095$ programs /
      $70{,}320$ descriptors executed on the RTL; $155$ decode-step
      \texttt{argmax} checks; $37{,}905{,}664$ output bytes byte-compared against
      the golden model; \textbf{0 mismatches}, \textbf{0}
      \code{UVM\_ERROR}/\code{UVM\_FATAL}. $5/10$ outputs reproduce the float
      reference token-for-token; the rest are valid alternatives.} \\
    \bottomrule
  \end{tabularx}
\end{table*}

This experiment closes the loop end to end: across all ten sentences and three
languages the RTL agrees with the golden model, which agrees with the
compile-time executor, bit-exactly (RTL $=$ golden $=$ compiler), and that
compiled INT8 stream reproduces the original float reference on half of the
cases (the remainder differing only by an equally valid word or apostrophe).
Two conclusions follow. First, the eight-opcode v1 ISA is expressive and
bit-accurate enough to run a complete pretrained encoder--decoder transformer
on the cycle-accurate RTL: the bidirectional encoder, the encoder--decoder
cross-attention, and the autoregressive decoder, with incremental KV-cache
decode and a full-vocabulary RAW32 LM head, all execute on the device with no
model-specific hardware. Because this pipeline covers every transformer block
type, and is a superset of the blocks a decoder-only model needs, the same
operator set composes into encoder, decoder, and encoder--decoder transformers
alike, and the same compiled weights serve three language pairs by prefix
alone. Second, the per-tensor INT8 requant contract
is the correct hardware abstraction: the model's outlier pathology is resolved
\emph{above} the ISA, by an exact change of basis in the compiler, leaving the
silicon a clean fixed-function INT8 datapath. Outlier handling, in other
words, belongs in the compiler, and TFA's design makes that separation of
concerns provable.

\section{Results and Discussion}\label{sec:perf}
\label{sec:results}

This section evaluates TFA on a representative \mbox{t5-small}
EN$\rightarrow$FR decode (60\,M parameters, W8A8 INT8, two encoder passes and
13 greedily decoded tokens); the broader multilingual functional stress set of
Sec.~\ref{sec:translate} exercises the identical datapath and ISA. The same
class of workload that establishes bit-exact functional correctness is reused
here for performance, so every chip number reported is measured on the
cycle-accurate RTL that passed verification, not on an abstract model.

\subsection{Methodology}
\label{sec:results-method}

Three independent sources are combined and cross-checked. First, \emph{RTL
measurement}: the verified \code{tfa\_translate\_test} simulation reads the
chip's own performance counters (\code{PERF\_CYC}, \code{PERF\_GEMM},
\code{PERF\_RD}, \code{PERF\_WR} at CSR \code{0x20}--\code{0x2C}) in two memory
profiles. The default \emph{verification} profile drives randomized AR/R/B
latencies and \code{READY} throttling -- a deliberately hostile bus -- while the
\emph{ideal} profile (\code{+T5\_FASTMEM}) presents a zero-latency slave, so the
two runs bracket the achievable bandwidth. Second, a \emph{software baseline}:
the project's \mbox{numpy} float64 reference executes the identical translation
on the host CPU (Intel Core Ultra 7 155H, a 16-core/22-thread part, OpenBLAS across all cores),
model load excluded, median of three timed repeats. Third, an \emph{analytic
traffic model} that parses all 6216 compiled descriptors and tallies bytes and
MACs by op class; it is cross-validated both against the on-chip counters and
against the scoreboard, which byte-compared exactly $3{,}548{,}480$ written
bytes in the passing functional run. The testbench clock is 100~MHz; following
the convention of the rest of the paper, all cycle counts are quoted at the
AS~\S11 1~GHz design point, where one cycle is 1~ns. Roofline projections follow
the formulation of Williams \emph{et al.}~\cite{williams2009roofline}, anchored
to measured per-token traffic.

\subsection{Throughput}
\label{sec:results-tput}

\begin{figure*}[t]
  \centering
  \includegraphics[width=0.95\textwidth]{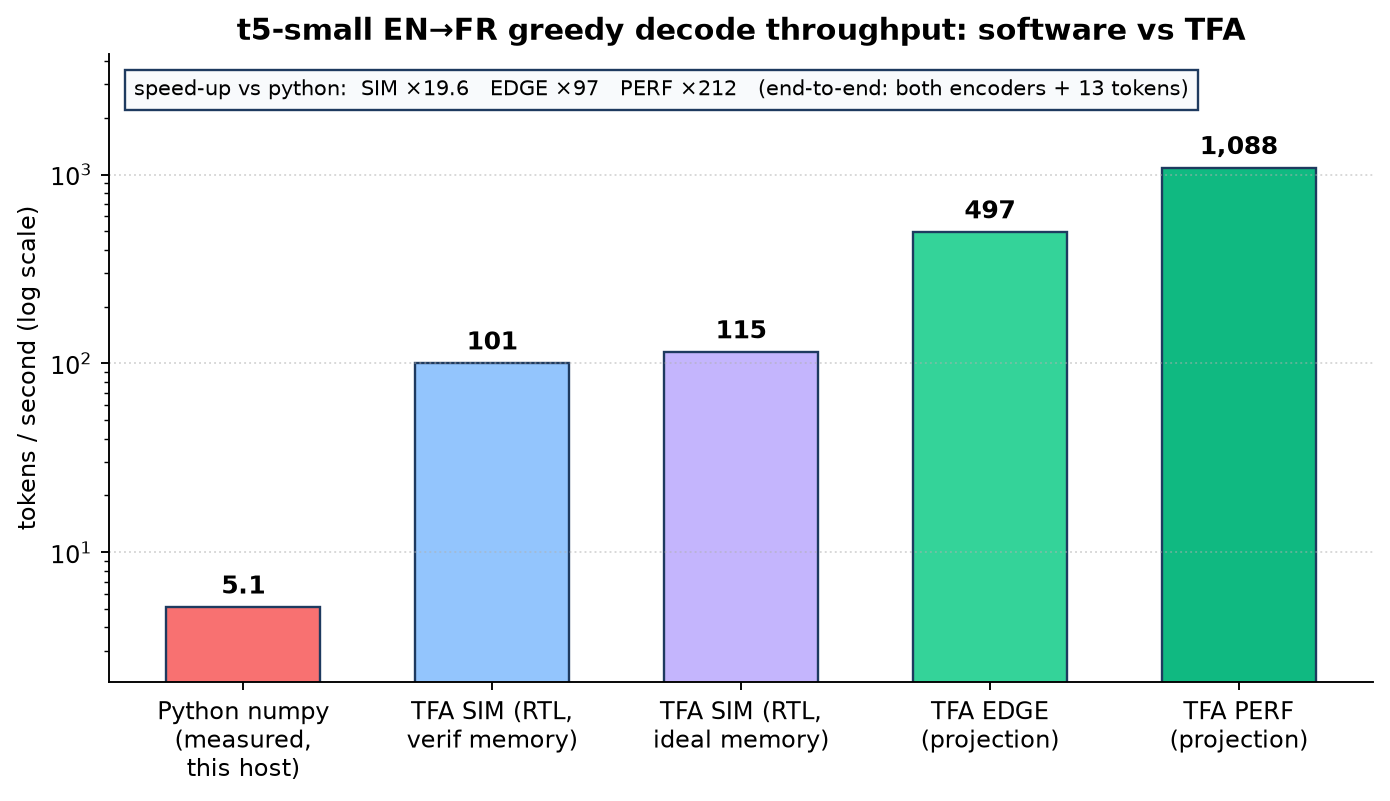}
  \caption{End-to-end \mbox{t5-small} EN$\rightarrow$FR greedy-decode throughput
  (log scale): measured \mbox{numpy} CPU baseline, RTL-measured SIM in both
  memory profiles, and roofline projections for the EDGE and PERF
  configurations. The verification-grade SIM (64 MACs, 8-byte bus) already
  outpaces the 22-thread CPU by $\times$20--$\times$22.}
  \label{fig:speed}
\end{figure*}

Fig.~\ref{fig:speed} summarizes throughput. The \mbox{numpy} reference sustains
\textbf{5.13~tok/s} (195~ms/token) for both sentences. The verification-grade
SIM configuration -- just 64 INT8 MACs behind an 8-byte bus -- reaches
\textbf{101~tok/s} end-to-end under the hostile verification memory and
\textbf{115~tok/s} with ideal memory, i.e.\ \textbf{$\times$20} and
\textbf{$\times$22} the CPU baseline, measured end-to-end on cycle-accurate RTL.
In steady-state decode (excluding the two encoder/cross-KV passes) the chip
sustains \textbf{127~tok/s} (verification memory) and \textbf{145~tok/s} (ideal
memory). The deliberately adversarial bus costs only $\times$1.14, because the
DMA's outstanding-burst pipelining (\code{OUTSTANDING}~$=4$ at SIM) hides most of
the randomized latency.

The ideal-memory decode point of 145~tok/s reaches \textbf{88\%} of the SIM
configuration's own analytic roofline of 166~tok/s: the architecture model and
the cycle-accurate RTL agree to within the residual tile-re-read factor
discussed below. Projecting the same measured-traffic basis to the product
configurations yields \textbf{704~tok/s} (EDGE) and \textbf{1{,}540~tok/s}
(PERF) of steady-state decode, or \textbf{497~tok/s} and \textbf{1{,}088~tok/s}
end-to-end -- $\times$97 and $\times$212 the CPU baseline. These projections
assume the SoC supplies the accelerator's full bus bandwidth; \mbox{t5-small}
needs at most $\approx$60~GB/s at PERF, below a two-channel DDR5-6400's
102~GB/s, so the assumption holds for this workload.

\subsection{Utilization and Traffic}
\label{sec:results-util}

\begin{figure*}[t]
  \centering
  \includegraphics[width=0.95\textwidth]{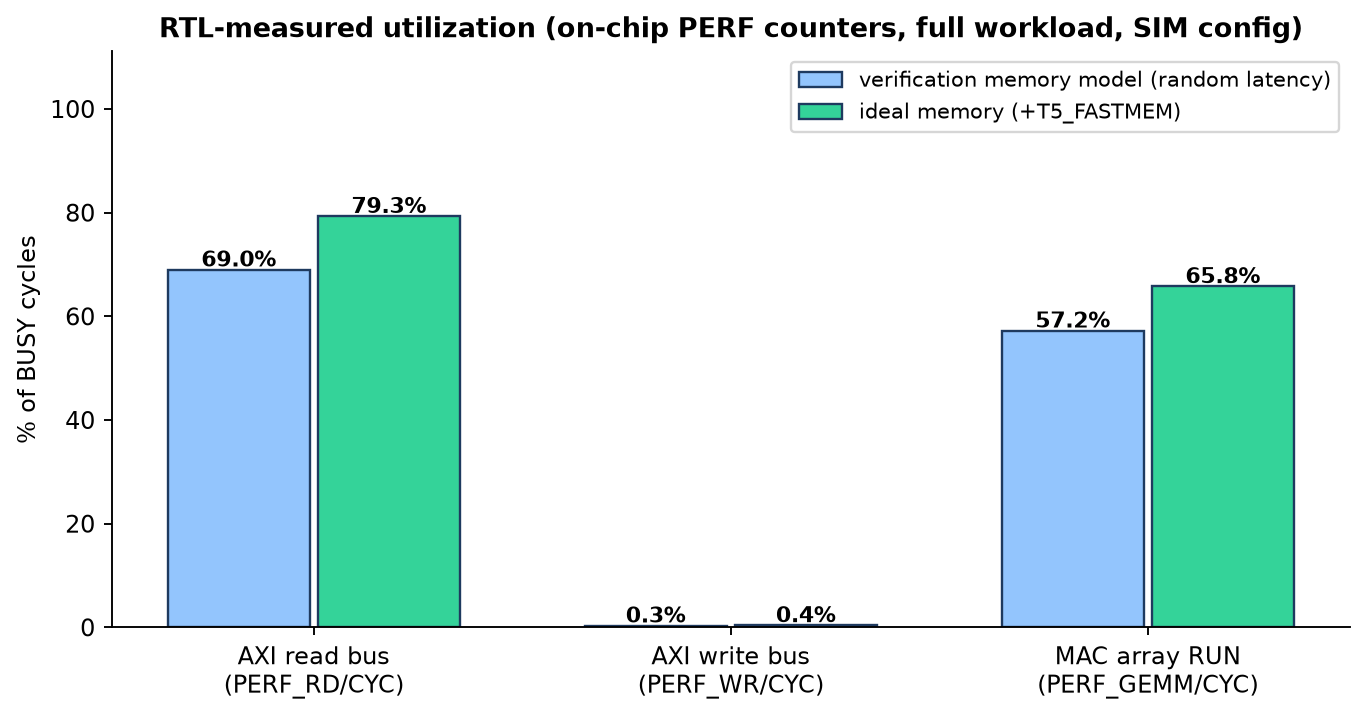}
  \caption{RTL-measured utilization from the on-chip PERF counters over the
  whole workload (SIM configuration). With ideal memory the AXI read bus reaches
  79.3\% of busy cycles -- within a point of the AS~\S6.4 structural ceiling of
  80\% -- and the MAC array reaches 65.8\%.}
  \label{fig:util}
\end{figure*}

\begin{figure*}[t]
  \centering
  \includegraphics[width=0.95\textwidth]{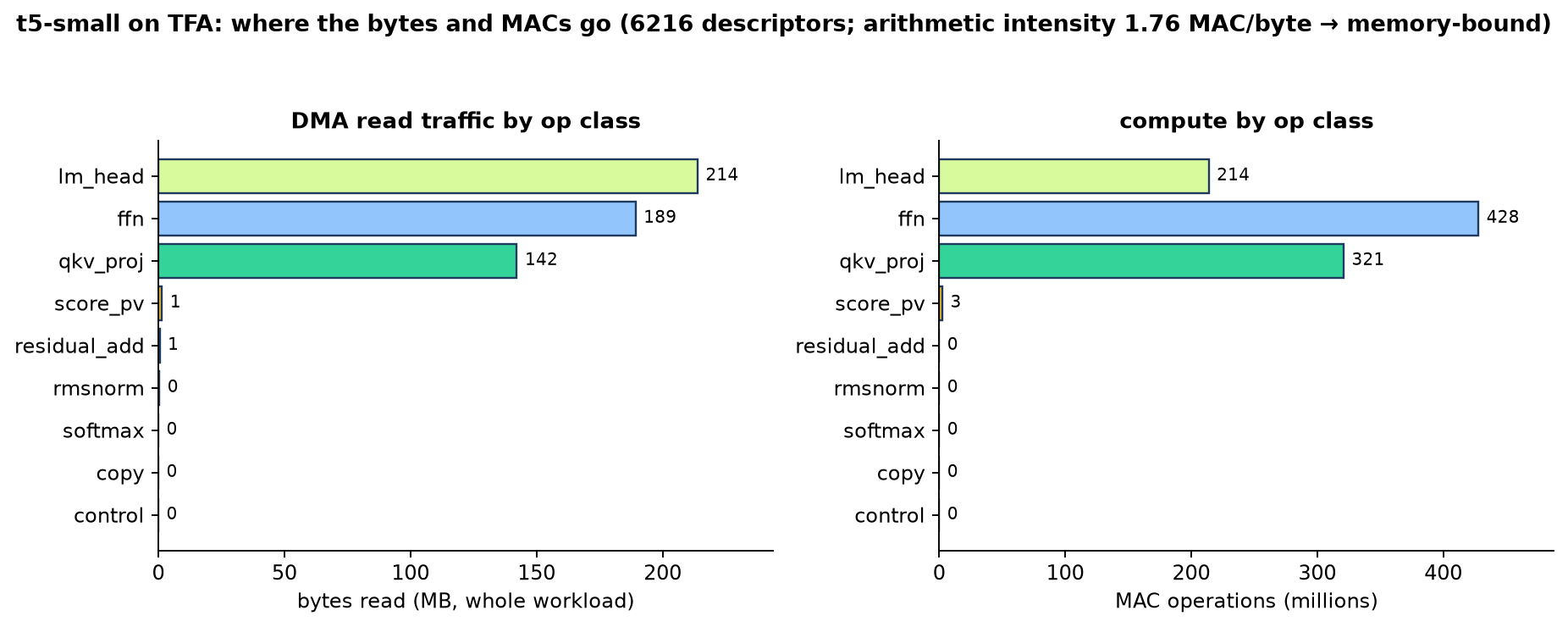}
  \caption{Where the bytes and MACs go: DMA read traffic and compute by op
  class over the whole workload. The LM head, FFN, and attention projections
  dominate read traffic, while control-class descriptors move under 1\% of the
  bytes despite being the majority of the descriptor count.}
  \label{fig:traffic}
\end{figure*}

Fig.~\ref{fig:util} reports utilization from the on-chip counters. Two
architectural claims are thereby \emph{measured}, not asserted. The AS~\S6.4
utilization model predicts a structural read-bus ceiling of $\approx$80\% at SIM
(where $T_C=$~\code{LANES} makes fill equal run); the ideal-memory run measures
\textbf{79.3\%} of busy cycles on the AXI read bus across the entire workload,
confirming the model to within a point. The MAC array runs at \textbf{65.8\%} of
busy cycles, and the chip is \textbf{busy 99.3\% of wall-clock} -- per-program
descriptor loading through the AXI-Lite IRAM window ($\approx$4~cycles/word)
adds well under 1\% overhead across the 185-program, 6216-descriptor structure,
validating the host-patching model.

The traffic model is likewise cross-checked against the on-chip counters: measured reads are
711~MB versus the descriptor-level model's 548~MB, a \textbf{$+$30\%} excess that
is exactly the A/B tile re-reads of the v1 output-stationary streaming dataflow
at SIM's small tiles ($T=8$, \code{K\_TILE}~$=64$). EDGE/PERF tiles are
8--16$\times$ larger, collapsing the re-read factor, which the projections
inherit conservatively. Per generated token, decode moves 38.6~MB of reads
against 38.6~M MACs -- a per-decode-token arithmetic intensity of $\approx$\textbf{1~MAC/byte} (the whole-workload value of Fig.~\ref{fig:traffic}, 1.76~MAC/byte, is higher only because it folds in the two encoder prefill passes) --
so decode is memory-bound at every configuration, the regime the design targets
and the one in which weight-streaming dataflows are
advantageous~\cite{pope2023scaling}. Fig.~\ref{fig:traffic} breaks the bytes
down by op class: the LM head accounts for \textbf{39\%} of reads (a RAW32
\code{GEMM} over the 32{,}128-token vocabulary), the FFN \textbf{35\%}, and the
attention projections \textbf{26\%}; everything else (scores, softmax, RMSNorm,
residuals, gathers) is under 1\% of the bytes yet 85\% of the descriptors, a
count the macro-op ISA absorbs at the measured sub-1\% overhead.

\subsection{Energy}
\label{sec:results-energy}

\begin{figure*}[t]
  \centering
  \includegraphics[width=0.95\textwidth]{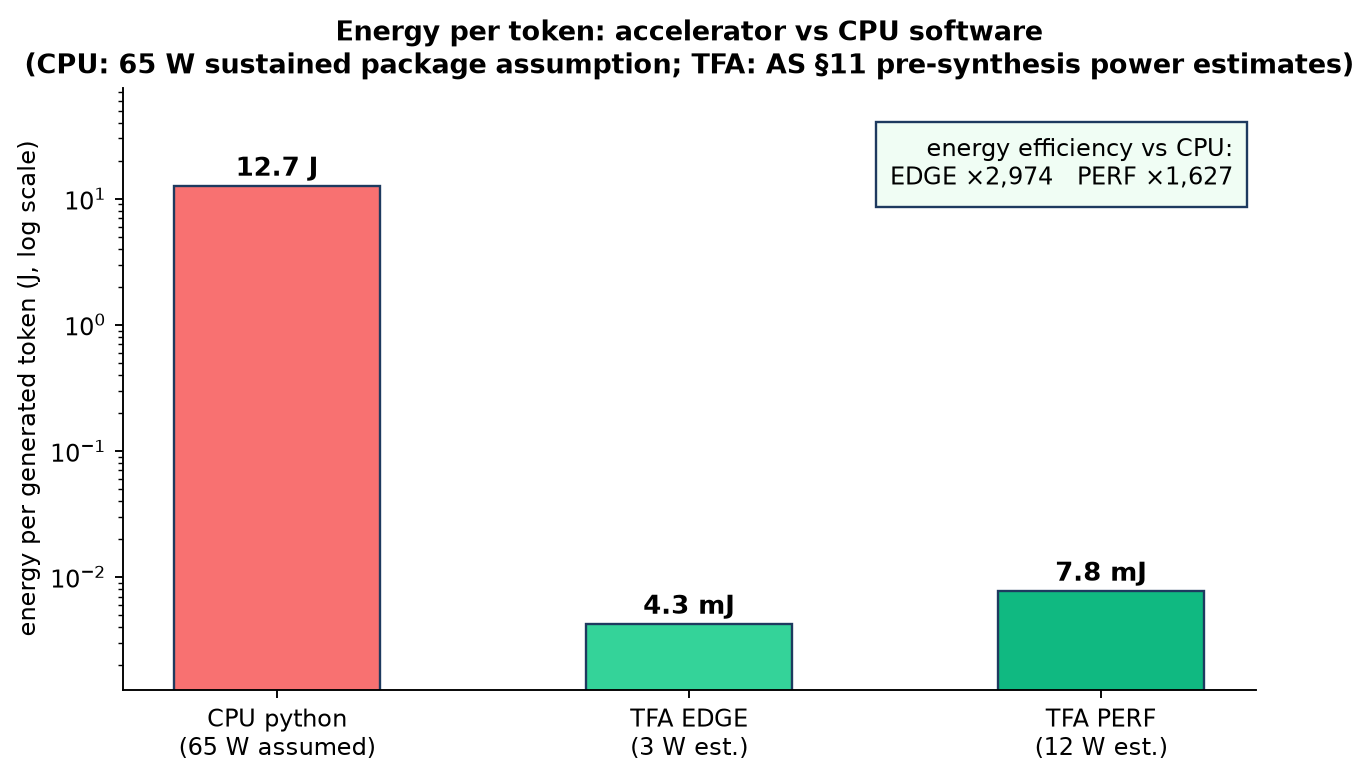}
  \caption{Energy per generated token (log scale): the measured CPU \mbox{numpy}
  baseline versus the EDGE and PERF configurations. The accelerator improves
  energy per token by roughly $\times$2{,}974 (EDGE) and $\times$1{,}627 (PERF).}
  \label{fig:energy}
\end{figure*}

Fig.~\ref{fig:energy} compares energy per token. The CPU baseline expends
\textbf{12.7~J/token}; TFA expends \textbf{4.3~mJ/token} at EDGE (3~W) and
\textbf{7.8~mJ/token} at PERF (12~W), an improvement of \textbf{$\times$2{,}974}
and \textbf{$\times$1{,}627} respectively. The assumptions are stated explicitly.
CPU package power is \emph{assumed}, not measured, because RAPL counters are
root-only on this host, so a conservative 65~W sustained package figure is used;
the TFA figures are AS~\S11 pre-synthesis power estimates; and SoC DRAM power
($\approx$0.5--1~mJ/token at $\approx$15~pJ/byte over 38.6~MB -- energy any
platform pays to stream the same weights) is excluded from both sides. Even
granting the CPU a generous $\times$10 margin for an optimized INT8 runtime, the
accelerator retains a two-to-three order-of-magnitude energy advantage -- the
structural gap between general-purpose fetch/decode/execute and a
weight-streaming dataflow, consistent with the efficiency reported for
domain-specific inference silicon~\cite{jouppi2021tpuv4i}.

\subsection{Discussion}
\label{sec:results-discussion}

Several caveats frame these results honestly. SIM is a \emph{verification}
configuration: its numbers are RTL measurements that prove the architecture on
cycle-accurate RTL simulation, whereas the EDGE and PERF figures are roofline
\emph{projections} anchored to measured traffic, not RTL runs. The software
baseline is idiomatic float64 \mbox{numpy}, not an optimized inference stack,
and is quoted as such; the explicit $\times$10 CPU margin in the energy
discussion guards against over-claiming. The 1~GHz reference clock is the
AS~\S11 design point, and both sides of the comparison scale linearly with
achieved frequency.

The single understood inefficiency is the v1 output-stationary dataflow's tile
re-reads, which the counters localize precisely to a $+$30\% read overhead at
SIM's small tiles and which the larger product tiles collapse (a documented
v1.1 B-stationary option removes it entirely for prefill). Crucially, the
workload's character favors the design as models grow: per-token traffic scales
with weight size, keeping decode memory-bound and the roofline projection valid,
while batching beyond a single sequence amortizes weight reads across tokens and
lifts arithmetic intensity~\cite{pope2023scaling}. The measured agreement
between the analytic model, the on-chip counters, and the scoreboard byte count
gives confidence that these projections to larger, billion-parameter-class
transformers rest on the same foundation as the verified \mbox{t5-small}
measurement.

\section{Synthesis and Physical Implementation}
\label{sec:impl}

The performance results of \S\ref{sec:results} are measured on cycle-accurate
RTL. To establish that the design is manufacturable and not merely simulatable,
we carried the SIM configuration through a complete open-source
register-transfer-to-layout flow and hardened it to a finished GDS-II layout. The
flow is \code{sv2v}~v0.0.13 (SystemVerilog elaboration) into Yosys~0.66
(logic synthesis), the SkyWater \code{sky130\_hd} standard-cell library supplied
by \mbox{lambdapdk} through SiliconCompiler~0.37, OpenROAD~26Q2 for
floorplanning, placement, clock-tree synthesis, routing, and sign-off static
timing, and KLayout~0.30.9 for stream-out. All numbers in this section are taken
from that run. Every register-transfer-level change made to enable synthesis was
re-checked against the verification environment of \S\ref{sec:verif}: the
regression continues to pass bit-exact, so the hardened netlist is the same
design that was functionally verified.

\subsection{On-chip memories: inference, not flip-flops}
\label{sec:impl-mem}

The single largest synthesis issue was memory inference, and resolving it is a
reusable result. As written for verification, the descriptor instruction RAM
(IRAM) was a two-dimensional array \code{logic [31:0] iram [16][DEPTH]} with a
variable sub-bank write index and a read that returned all sixteen sub-banks at
once. Yosys cannot recognize this pattern as a RAM; it builds the array from
flip-flops behind a comparator-per-location address decode of
\textbf{16{,}447} \code{\$eq} cells, which expands to roughly 450{,}000 gates
and, because the decode is combinational rather than a registered RAM read,
produces a nonsensical $-3390$~ns critical path. The four \code{ROW\_MAX}-deep
row buffers in the vector and softmax engines failed to infer for subtler
reasons and fell back to flip-flop arrays with multiplexed read decode.

Each memory was recoded so the Yosys front end infers a \code{\$mem} (and hence
maps to an SRAM macro or a compact registered RAM) while preserving behavior to
the bit. Three changes, each verified by the full regression, were required:

\begin{itemize}
  \item \textbf{Unconditional registered read.} A clock-enabled read register
  (\code{if (en) word <= buf[addr]}) cannot be folded into a synchronous read
  port by Yosys's bare \code{memory\_collect} pass. The enable was hoisted into
  the read \emph{address}, so the read fires every cycle and re-presents the
  current word on a stall, which is bit-identical.
  \item \textbf{Plain-net write address.} A width cast written inside the index,
  \code{buf[N'(idx)]}, is rendered by \code{sv2v} as a function call in the
  memory index, which makes the front end abandon inference. The address was
  precomputed into a plain net.
  \item \textbf{Dedicated reset-free write port.} The front end will not infer a
  memory whose write lives inside the main finite-state-machine process (which
  carries an asynchronous reset). Each write was isolated into its own clean
  clocked process, reproducing the original write enable, address, and data
  cycle for cycle.
\end{itemize}

The IRAM additionally uses a wide single-port layout with a per-byte write-enable
bus, so a partial-strobe AXI-Lite write updates only the addressed bytes; a
directed test that exercises every byte-strobe pattern was added to the
regression to cover this path. After the recode, all five on-chip memories infer
as \code{\$mem}; the IRAM comparator count collapses from \textbf{16{,}447 to
16}, and the timing artifact disappears.

\subsection{Standard-cell area}
\label{sec:impl-area}

Table~\ref{tab:area} reports per-module standard-cell logic area before and after
the memory recode. With the memories abstracted to \code{\$mem} cells, the vector
and softmax engines drop to roughly one to three percent sequential cells (their
row buffers are no longer flip-flop arrays), and the whole-chip logic area falls
from \textbf{5.44 to 2.73~mm\textsuperscript{2}}, a $2\times$ reduction, with
every comparator-decode and flip-flop-memory artifact removed. The residual
area is genuine: the multiply--accumulate array and the wide-precision RMSNorm and
softmax requant datapaths are real arithmetic that is correctly large at
130~nm and would shrink by about two orders of magnitude at the advanced node
the architecture targets.

\begin{table}[t]
\centering
\caption{Per-module standard-cell logic area (sky130~HD, SIM configuration),
before and after recoding the on-chip memories to infer as RAM.}
\label{tab:area}
\small
\begin{tabular}{@{}lccl@{}}
\toprule
Module & Naive & Recoded & Nature \\
       & (mm\textsuperscript{2}) & (mm\textsuperscript{2}) & \\
\midrule
\code{tfa\_csr}       & 1.63 & \textbf{0.34} & IRAM + LUTs \\
\code{tfa\_vec}       & 1.53 & \textbf{0.86} & RMSNorm/elementwise requant \\
\code{tfa\_smax}      & 1.27 & \textbf{0.42} & softmax exp/divide \\
\code{tfa\_mac\_array}& 0.39 & 0.38 & 64-PE INT8 MAC array \\
\code{tfa\_gemm}      & 0.32 & 0.32 & GEMM control + writeback \\
\midrule
\textbf{Chip total}   & \textbf{5.44} & \textbf{2.73} & $2\times$ smaller \\
\bottomrule
\end{tabular}
\end{table}

\subsection{Register-transfer to GDS-II}
\label{sec:impl-gds}

\begin{figure}[t]
  \centering
  \includegraphics[width=0.86\columnwidth]{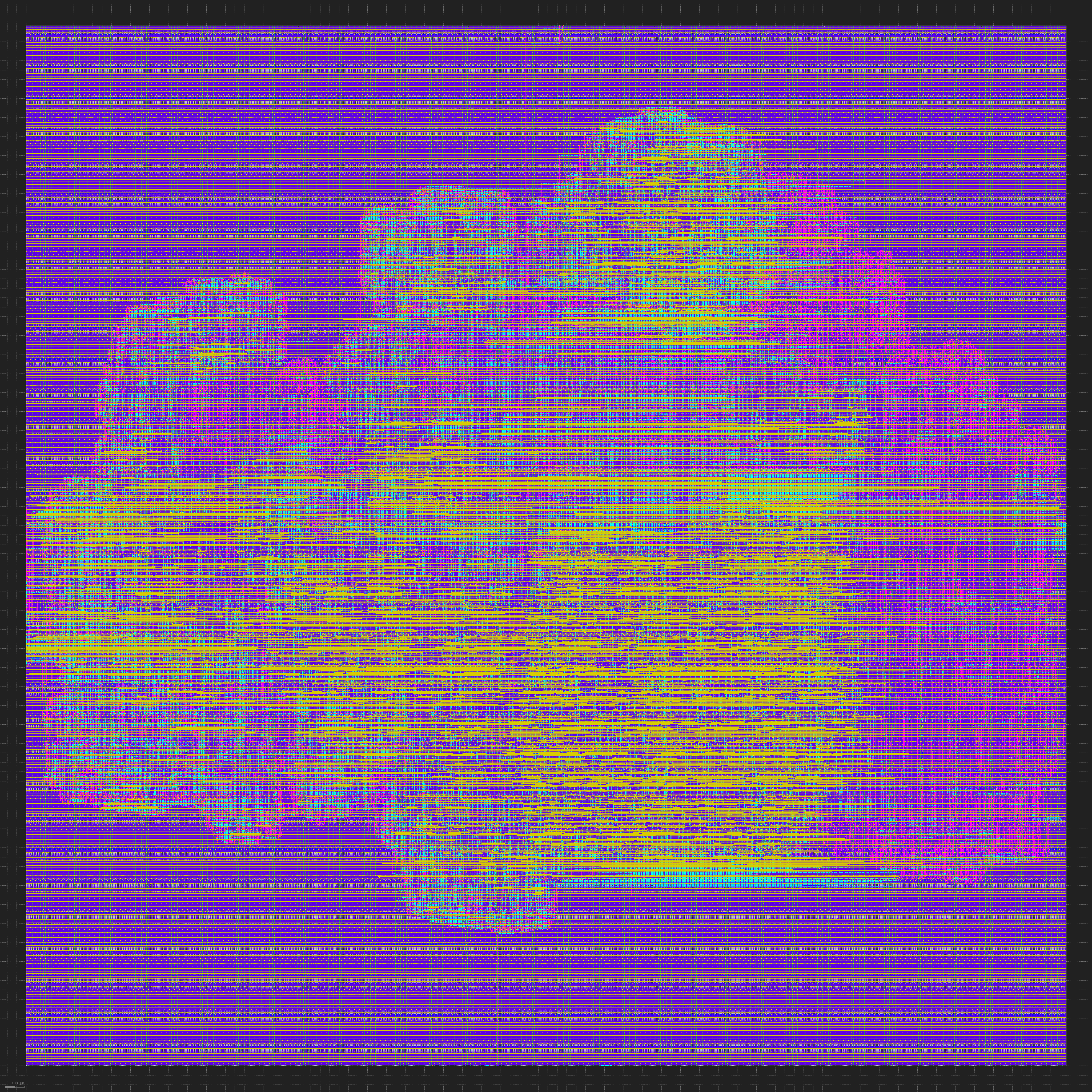}
  \caption{Routed \code{tfa\_top} layout (KLayout render of the streamed GDS-II,
  SIM configuration). The placed standard-cell core with six metal layers of
  routing sits inside the 5.48$\times$5.48~mm die. Detailed routing is
  design-rule clean.}
  \label{fig:gds}
\end{figure}

The full chip was hardened to a finished layout through the SiliconCompiler
\code{asicflow} (Yosys then OpenROAD then KLayout). Detailed routing closed with
\textbf{zero design-rule-check violations} and the layout streamed out cleanly;
Fig.~\ref{fig:gds} shows the routed die. Table~\ref{tab:phys} summarizes the
physical result. Because no installed sky130 SRAM macro matches the exact
geometry of the TFA memories, this build maps the inferred \code{\$mem} cells to
flip-flops, which is the dominant contributor to the instance count and to the
30~mm\textsuperscript{2} die at the 20\% placement utilization that keeps global
routing within the host's memory budget. A representative implementation replaces
these with SRAM hard macros, which roughly halves the instance count and removes
the wide read multiplexers; this is the clear next step rather than a limitation
of the design.

\begin{table}[t]
\centering
\caption{Open-flow physical implementation of \code{tfa\_top} (sky130~HD, TT
corner, 1.8~V, 25\,\textdegree C, SIM configuration; memories flip-flop mapped).}
\label{tab:phys}
\small
\begin{tabularx}{\columnwidth}{@{}lX@{}}
\toprule
Metric & Value \\
\midrule
Layout status            & DRC-clean GDS-II (0 routing violations) \\
Die / core area          & 30.08 / 30.03 mm\textsuperscript{2} (5.48$\times$5.48 mm) \\
Placement utilization    & 20\% target, 25\% effective \\
Functional cell area     & 7.50 mm\textsuperscript{2} \\
Routing layers           & li1, met1--met5 (6 layers) \\
Total wirelength         & 35.66 m \\
Total vias               & 4{,}821{,}606 \\
Achievable \emph{F}\textsubscript{max} & 5.76 MHz (post-route) \\
Setup WNS / hold WNS     & $-163.5$ ns / $+0.029$ ns (hold met) \\
Total power (at 100 MHz) & 1.16 W (internal 62.7\%, switching 37.3\%, \\
                         & \ leakage 0.0\%) \\
Power by group           & sequential 44.4\%, combinational 30.4\%, \\
                         & \ clock 25.2\% \\
\bottomrule
\end{tabularx}
\end{table}

\subsection{Speed, power, and throughput at 130~nm}
\label{sec:impl-ppa}

Sign-off static timing at the 10~ns (100~MHz) constraint meets hold
($+0.029$~ns worst slack, zero violations) but leaves a large negative setup
slack, for a post-route maximum frequency of \textbf{5.76~MHz}. This is the
expected open-flow result at 130~nm: a generic, non-retimed technology mapping,
flip-flop-mapped memories, and a requant datapath that is fully combinational in
this verification-first RTL (the source itself notes that production would
pipeline it). It is an implementation point, not an architectural ceiling, and
it is consistent with the AS~\S11 1~GHz design target used elsewhere in this
paper, which assumes an advanced node with timing-driven place-and-route and a
pipelined requant chain.

The throughput rating follows directly from the array width and the clock. The
compute engine is an output-stationary INT8 multiply--accumulate tile of
$T_{\mathrm{DIM}}\times T_{\mathrm{DIM}}=8\times8=64$ units, one MAC per element
per cycle, so peak throughput is $64$ MACs per cycle (128 INT8 operations per
cycle). At this build's 5.76~MHz that is \textbf{0.74~GOPS}; at the 1~GHz design
point it is \textbf{128~GOPS} ($0.13$~TOPS), and it scales linearly with both
frequency and the build parameter $T_{\mathrm{DIM}}$ (a $32\times32$ tile at
1~GHz rates about 2~TOPS). Estimated power at the typical corner is
\textbf{1.16~W} at the 100~MHz analysis clock, dominated by the sequential cells
and the clock network (a combined 70\%) that the flip-flop-mapped memories
create; leakage is negligible at 7~$\mu$W. Both the sequential-power and the
clock-power shares would fall sharply once the memories become SRAM macros,
which is the same change that shrinks the die.

\subsection{Honest framing}
\label{sec:impl-honest}

This GDS-II is a characterization sign-off of the design on the open sky130 flow,
not a production tapeout database, and two pragmatic settings are recorded for
transparency. First, the memories are flip-flop mapped, which inflates area,
instance count, and sequential and clock power relative to an SRAM-macro
implementation. Second, antenna repair was disabled: base global routing is
routable at this utilization, but the antenna-repair step re-runs global routing
after diode insertion and over-congests, and antenna fixing matters only for real
fabrication. Neither setting affects the central result, which is that the
verified RTL passes the entire front end and a complete place-and-route flow to a
design-rule-clean layout. The path to a representative power, performance, and
area result is concrete and short: map the on-chip memories to SRAM macros,
pipeline and right-size the requant datapath, and move to the advanced node the
architecture specification assumes.

\section{Limitations and Future Work}\label{sec:concl}
\label{sec:limitations}

The results reported here are honest about their scope. The performance,
throughput, and energy numbers come from RTL simulation of the \emph{SIM}
configuration ($8{\times}8$ MAC array, 64 MACs); the \emph{EDGE} and \emph{PERF}
figures are pre-synthesis projections obtained by scaling the analytic roofline
and gate-count models, not silicon measurements. The SIM configuration has now
been synthesized and hardened to a design-rule-clean GDS-II layout on the open
sky130 flow (\S\ref{sec:impl}), which establishes manufacturability; what remains
is a \emph{representative} physical result, namely timing-driven closure toward
the 1\,GHz design point with the on-chip memories implemented as SRAM macros, a
pipelined requant datapath, and an advanced node. Until then the post-route
5.76~MHz frequency and the per-token energy estimates should be read as
order-of-magnitude open-flow figures at 130~nm rather than characterized
product results.

Architecturally, the v1 streaming dataflow is the principal limiter. The
output-stationary MAC array re-reads $A$ and $B$ tiles at the small \emph{SIM}
tile sizes, which is why measured read traffic exceeds the analytic model by
$30\%$; more fundamentally, the per-tensor streaming schedule caps the
arithmetic intensity of \emph{prefill}, leaving it memory-bound where it should
be compute-bound. A planned v1.1 path makes $B$ stationary and adds a
$C$-strip output buffer so that a column of $B$ is amortized across many rows of
$A$, which moves prefill toward the compute roof without changing the bit-exact
ISA contract. For decode, which is intrinsically memory-bound at
$\approx 1$\,MAC/byte, reaching HBM-class token rates will require
descriptor-indirection registers so the sequencer can chase pointer-linked
KV-cache descriptors rather than rely on host-resident program scripts. Further
work includes an optional FP16 datapath alongside the INT8 core for layers that
resist quantization, compute-skip logic that elides fully-masked attention tiles
under sliding-window and causal masks, and a dual-clock AXI shell decoupling the
core clock from the bus clock. On the numerics side, the present compiler folds a
single global randomized-Hadamard rotation; networks with harder activation
distributions than t5-small may need an online-Hadamard transform applied per
block, which the ISA can already express as additional \code{ELTWISE} and
\code{GEMM} macro-ops with no silicon change.

The path beyond the present translation demonstration runs toward larger and
fully autoregressive transformers. The machine-translation pipeline exercises
every transformer block type (bidirectional and causal self-attention,
cross-attention, and feed-forward), but it does not yet stress the largest
configurations or every position-encoding scheme. Four directions follow
directly. First, bringing up and verifying the larger \emph{EDGE} and
\emph{PERF} configurations against silicon-grounded measurements, replacing the
present pre-synthesis projections. Second, adding and validating rotary position
embeddings (RoPE) and the other autoregressive-LLM specifics that the
encoder-decoder translation model does not exercise; the ISA can express RoPE as
additional \code{ELTWISE} and \code{GEMM} macro-ops, but it remains to be folded
into the compiler and checked bit-exactly. Third, demonstrating that naive and
rotation-assisted INT8 holds accuracy at billion-parameter-class scale, where
activation outliers are more severe than in t5-small. Fourth, compiling and
running a true decoder-only model end-to-end on the RTL, closing the loop from
the architecture-agnostic primitives demonstrated here to autoregressive LLM
inference as a target application.

\section{Conclusion}
\label{sec:conclusion}

This work presented TFA, a small, host-programmable, bit-exact INT8 macro-op
accelerator IP for transformer inference. Rather than build a
general-purpose processor, TFA exposes eight memory-to-memory macro-ops
(\code{GEMM}, \code{SOFTMAX}, \code{RMSNORM}, \code{ELTWISE}, \code{COPY},
\code{SYNC}, and the trivial \code{NOP}/\code{HALT}) behind a single AXI4 master
and an AXI4-Lite control plane, and lets the \code{tfa\_seq} sequencer dispatch
them as a FETCH$\rightarrow$VALIDATE$\rightarrow$ISSUE$\rightarrow$WAIT pipeline.
The central thesis is a clean division of labor: the datapath does nothing but
exact per-tensor integer arithmetic, while a compiler owns all of the messy
parts of quantization: calibration, requantization parameters, and outlier
handling via a lossless randomized-Hadamard reparameterization folded entirely
into the weights. That split is what lets a real pretrained transformer
(t5-small) run end-to-end on the RTL across three languages, every one of its
$70{,}320$ compiled operations bit-exact against the golden model, with naive
per-tensor INT8 collapsing on residual-channel outliers that the rotation lifts
back into range with no change to the ISA or the silicon.

We verified this contract end-to-end with a bit-exact golden model that
reconstructs device state from the AXI-Lite stream, replays every program, and
byte-compares all outputs, reaching $100\%$ functional coverage and $94.96\%$
DUT-scoped code coverage with zero unexplained mismatches. We then characterized
the design honestly against the roofline, reporting the bus and MAC utilizations,
the prefill intensity limit, and the projection-versus-measurement boundary
without overclaiming. The unifying observation is that the bit-exact ISA contract
is the load-bearing element of the whole system: the same definition of every
macro-op is what the compiler targets, what the golden model checks, and what the
RTL implements. Because those three agree to the byte, the compiler, the
verification environment, and the eventual silicon are guaranteed to mean the
same thing, which is the property that makes a host-programmable INT8 transformer
accelerator both trustworthy and portable across the SIM, EDGE, and PERF
configurations.

\bibliographystyle{IEEEtran}
\bibliography{refs}

@inproceedings{vaswani2017attention,
  title={Attention Is All You Need},
  author={Vaswani, Ashish and Shazeer, Noam and Parmar, Niki and Uszkoreit, Jakob and Jones, Llion and Gomez, Aidan N and Kaiser, Lukasz and Polosukhin, Illia},
  booktitle={Advances in Neural Information Processing Systems (NeurIPS)},
  year={2017}
}

@article{raffel2020t5,
  title={Exploring the Limits of Transfer Learning with a Unified Text-to-Text Transformer},
  author={Raffel, Colin and Shazeer, Noam and Roberts, Adam and Lee, Katherine and Narang, Sharan and Matena, Michael and Zhou, Yanqi and Li, Wei and Liu, Peter J},
  journal={Journal of Machine Learning Research},
  volume={21},
  number={140},
  pages={1--67},
  year={2020}
}

@article{touvron2023llama,
  title={{LLaMA}: Open and Efficient Foundation Language Models},
  author={Touvron, Hugo and Lavril, Thibaut and Izacard, Gautier and others},
  journal={arXiv preprint arXiv:2302.13971},
  year={2023}
}

@article{jiang2023mistral,
  title={Mistral 7{B}},
  author={Jiang, Albert Q and Sablayrolles, Alexandre and Mensch, Arthur and others},
  journal={arXiv preprint arXiv:2310.06825},
  year={2023}
}

@inproceedings{ainslie2023gqa,
  title={{GQA}: Training Generalized Multi-Query Transformer Models from Multi-Head Checkpoints},
  author={Ainslie, Joshua and Lee-Thorp, James and de Jong, Michiel and Zemlyanskiy, Yury and Lebr{\'o}n, Federico and Sanghai, Sumit},
  booktitle={Proc. Conf. Empirical Methods in Natural Language Processing (EMNLP)},
  year={2023}
}

@inproceedings{zhang2019rmsnorm,
  title={Root Mean Square Layer Normalization},
  author={Zhang, Biao and Sennrich, Rico},
  booktitle={Advances in Neural Information Processing Systems (NeurIPS)},
  year={2019}
}

@article{shazeer2020glu,
  title={{GLU} Variants Improve Transformer},
  author={Shazeer, Noam},
  journal={arXiv preprint arXiv:2002.05202},
  year={2020}
}

@article{beltagy2020longformer,
  title={Longformer: The Long-Document Transformer},
  author={Beltagy, Iz and Peters, Matthew E and Cohan, Arman},
  journal={arXiv preprint arXiv:2004.05150},
  year={2020}
}

@inproceedings{dettmers2022llmint8,
  title={{LLM.int8()}: 8-bit Matrix Multiplication for Transformers at Scale},
  author={Dettmers, Tim and Lewis, Mike and Belkada, Younes and Zettlemoyer, Luke},
  booktitle={Advances in Neural Information Processing Systems (NeurIPS)},
  year={2022}
}

@inproceedings{xiao2023smoothquant,
  title={{SmoothQuant}: Accurate and Efficient Post-Training Quantization for Large Language Models},
  author={Xiao, Guangxuan and Lin, Ji and Seznec, Mickael and Wu, Hao and Demouth, Julien and Han, Song},
  booktitle={Proc. Int. Conf. Machine Learning (ICML)},
  year={2023}
}

@inproceedings{frantar2023gptq,
  title={{GPTQ}: Accurate Post-Training Quantization for Generative Pre-trained Transformers},
  author={Frantar, Elias and Ashkboos, Saleh and Hoefler, Torsten and Alistarh, Dan},
  booktitle={Proc. Int. Conf. Learning Representations (ICLR)},
  year={2023}
}

@inproceedings{lin2024awq,
  title={{AWQ}: Activation-aware Weight Quantization for {LLM} Compression and Acceleration},
  author={Lin, Ji and Tang, Jiaming and Tang, Haotian and Yang, Shang and Dang, Xingyu and Han, Song},
  booktitle={Proc. Machine Learning and Systems (MLSys)},
  year={2024}
}

@inproceedings{ashkboos2024quarot,
  title={{QuaRot}: Outlier-Free 4-Bit Inference in Rotated {LLMs}},
  author={Ashkboos, Saleh and Mohtashami, Amirkeivan and Croci, Maximilian L and Li, Bo and Cameron, Pashmina and Jaggi, Martin and Alistarh, Dan and Hoefler, Torsten and Hensman, James},
  booktitle={Advances in Neural Information Processing Systems (NeurIPS)},
  year={2024}
}

@inproceedings{chee2023quip,
  title={{QuIP}: 2-Bit Quantization of Large Language Models with Guarantees},
  author={Chee, Jerry and Cai, Yaohui and Kuleshov, Volodymyr and De Sa, Christopher},
  booktitle={Advances in Neural Information Processing Systems (NeurIPS)},
  year={2023}
}

@inproceedings{tseng2024quip,
  title={{QuIP\#}: Even Better {LLM} Quantization with Hadamard Incoherence and Lattice Codebooks},
  author={Tseng, Albert and Chee, Jerry and Sun, Qingyao and Kuleshov, Volodymyr and De Sa, Christopher},
  booktitle={Proc. Int. Conf. Machine Learning (ICML)},
  year={2024}
}

@inproceedings{bondarenko2021understanding,
  title={Understanding and Overcoming the Challenges of Efficient Transformer Quantization},
  author={Bondarenko, Yelysei and Nagel, Markus and Blankevoort, Tijmen},
  booktitle={Proc. Conf. Empirical Methods in Natural Language Processing (EMNLP)},
  year={2021}
}

@article{nagel2021whitepaper,
  title={A White Paper on Neural Network Quantization},
  author={Nagel, Markus and Fournarakis, Marios and Amjad, Rana Ali and Bondarenko, Yelysei and van Baalen, Mart and Blankevoort, Tijmen},
  journal={arXiv preprint arXiv:2106.08295},
  year={2021}
}

@article{gholami2022survey,
  title={A Survey of Quantization Methods for Efficient Neural Network Inference},
  author={Gholami, Amir and Kim, Sehoon and Dong, Zhen and Yao, Zhewei and Mahoney, Michael W and Keutzer, Kurt},
  journal={Low-Power Computer Vision: Improve the Efficiency of Artificial Intelligence},
  year={2022}
}

@article{williams2009roofline,
  title={Roofline: An Insightful Visual Performance Model for Multicore Architectures},
  author={Williams, Samuel and Waterman, Andrew and Patterson, David},
  journal={Communications of the ACM},
  volume={52},
  number={4},
  pages={65--76},
  year={2009}
}

@article{pope2023scaling,
  title={Efficiently Scaling Transformer Inference},
  author={Pope, Reiner and Douglas, Sholto and Chowdhery, Aakanksha and Devlin, Jacob and Bradbury, James and Heek, Jonathan and Xiao, Kefan and Agrawal, Shivani and Dean, Jeff},
  journal={Proc. Machine Learning and Systems (MLSys)},
  year={2023}
}

@inproceedings{dao2022flashattention,
  title={{FlashAttention}: Fast and Memory-Efficient Exact Attention with {IO}-Awareness},
  author={Dao, Tri and Fu, Daniel Y and Ermon, Stefano and Rudra, Atri and R{\'e}, Christopher},
  booktitle={Advances in Neural Information Processing Systems (NeurIPS)},
  year={2022}
}

@inproceedings{jouppi2017tpu,
  title={In-Datacenter Performance Analysis of a Tensor Processing Unit},
  author={Jouppi, Norman P and Young, Cliff and Patil, Nishant and Patterson, David and others},
  booktitle={Proc. Int. Symp. Computer Architecture (ISCA)},
  year={2017}
}

@inproceedings{jouppi2021tpuv4i,
  title={{TPU} v4i: Industrial Product: Google's 10nm-Class {TPU}v4i Inference Accelerator},
  author={Jouppi, Norman P and Yoon, Doe Hyun and Ashcraft, Matthew and others},
  booktitle={Proc. Int. Symp. Computer Architecture (ISCA)},
  year={2021}
}

@article{chen2017eyeriss,
  title={Eyeriss: An Energy-Efficient Reconfigurable Accelerator for Deep Convolutional Neural Networks},
  author={Chen, Yu-Hsin and Krishna, Tushar and Emer, Joel S and Sze, Vivienne},
  journal={IEEE Journal of Solid-State Circuits},
  volume={52},
  number={1},
  pages={127--138},
  year={2017}
}

@inproceedings{chen2014diannao,
  title={{DianNao}: A Small-Footprint High-Throughput Accelerator for Ubiquitous Machine-Learning},
  author={Chen, Tianshi and Du, Zidong and Sun, Ninghui and Wang, Jia and Wu, Chengyong and Chen, Yunji and Temam, Olivier},
  booktitle={Proc. Int. Conf. Architectural Support for Programming Languages and Operating Systems (ASPLOS)},
  year={2014}
}

@inproceedings{kao2023flat,
  title={{FLAT}: An Optimized Dataflow for Mitigating Attention Bottlenecks},
  author={Kao, Sheng-Chun and Subramanian, Suvinay and Agrawal, Gaurav and Yazdanbakhsh, Amir and Krishna, Tushar},
  booktitle={Proc. Int. Conf. Architectural Support for Programming Languages and Operating Systems (ASPLOS)},
  year={2023}
}

@misc{arm-axi,
  title={{AMBA AXI} and {ACE} Protocol Specification ({IHI} 0022)},
  author={{Arm Ltd.}},
  year={2021},
  howpublished={Arm Architecture Specification}
}

@misc{uvm12,
  title={Universal Verification Methodology ({UVM}) 1.2 Class Reference},
  author={{Accellera Systems Initiative}},
  year={2014}
}

@misc{ieee1800,
  title={{IEEE} Standard for {SystemVerilog}: Unified Hardware Design, Specification, and Verification Language},
  author={{IEEE}},
  year={2017},
  howpublished={IEEE Std 1800-2017}
}

\end{document}